\documentclass[prd,nofootinbib,twocolumn,superscriptaddress,preprintnumbers,longbibliography]{revtex4-1}

\usepackage{amsfonts, amssymb,amsmath}
\usepackage{graphics, graphicx}
\usepackage{booktabs, tabularx, makecell, multirow}
\usepackage{xcolor} 
\usepackage{afterpage}
\usepackage[utf8]{inputenc}
\usepackage{url}
\usepackage{hyperref}

\hypersetup{colorlinks=true
,urlcolor=blue
,anchorcolor=blue
,citecolor=blue
,filecolor=blue
,linkcolor=blue
,menucolor=blue
,linktocpage=true
,pdfproducer=medialab
,pdfa=true}

\renewcommand{\arraystretch}{1.8}
\newcolumntype{C}[1]{>{\centering\let\newline\\\arraybackslash\hspace{0pt}}m{#1}}

\newcommand {\be} {\begin {equation}}
\newcommand {\ee} {\end {equation}} 
\newcommand{\es}[2] {\begin{equation} \label{#1} \begin{split} #2 \end{split} \end{equation}}

\newcommand\Tstrut{\rule{0pt}{2.6ex}}         
\newcommand\Bstrut{\rule[-0.9ex]{0pt}{0pt}}   

\begin{document}

\title{Characterizing the Nature of the Unresolved Point Sources in the Galactic Center:\\
\vspace{0.07in}
An Assessment of Systematic Uncertainties
}

\preprint{LCTP-19-20}

\author{Laura J. Chang}
\affiliation{Department of Physics, Princeton University, Princeton, NJ 08544, USA}

\author{Siddharth Mishra-Sharma}
\affiliation{Center for Cosmology and Particle Physics, Department of Physics, New York University, New York, NY 10003, USA}

\author{Mariangela Lisanti}
\affiliation{Department of Physics, Princeton University, Princeton, NJ 08544, USA}

\author{Malte Buschmann}
\affiliation{Leinweber Center for Theoretical Physics, Department of Physics, University of Michigan, Ann Arbor, MI 48109, USA}

\author{Nicholas L. Rodd}
\affiliation{Berkeley Center for Theoretical Physics, University of California, Berkeley, CA 94720, USA}
\affiliation{Theoretical Physics Group, Lawrence Berkeley National Laboratory, Berkeley, CA 94720, USA}

\author{Benjamin R. Safdi}
\affiliation{Leinweber Center for Theoretical Physics, Department of Physics, University of Michigan, Ann Arbor, MI 48109, USA}

\date{\today}

\begin{abstract}
The Galactic Center Excess (GCE) of GeV gamma rays can be explained as a signal of annihilating dark matter or of emission from unresolved astrophysical sources, such as millisecond pulsars.  Evidence for the latter is provided by a statistical procedure---referred to as Non-Poissonian Template Fitting (NPTF)---that distinguishes the smooth distribution of photons expected for dark matter annihilation from a ``clumpy'' photon distribution expected for point sources. In this paper, we perform an extensive study of the NPTF on simulated data, exploring its ability to recover the flux and luminosity function of unresolved sources at the Galactic Center.  When astrophysical background emission is perfectly modeled, we find that the NPTF successfully distinguishes between the dark matter and point source hypotheses when either component makes up the entirety of the GCE.  When the GCE is a mixture of dark matter and point sources, the NPTF may fail to reconstruct the correct contribution of each component.  These results are related to the fact that in the ultra-faint limit, a population of unresolved point sources is exactly degenerate with Poissonian emission.  We further study the impact of mismodeling the  Galactic diffuse backgrounds, finding that while a dark matter signal could be attributed to point sources in some outlying cases for the scenarios we consider, the significance of a true point source signal remains robust.  Our work enables us to comment on a recent study by Leane and Slatyer~(2019) that questions prior NPTF conclusions because the method does not recover an artificial dark matter signal injected on actual \emph{Fermi} data.  We  demonstrate  that  the  failure  of  the NPTF to extract an artificial dark matter signal can be natural when point sources are present in the data---with the effect further exacerbated by the presence of diffuse mismodeling---and does not on its own  invalidate the conclusions of the NPTF analysis in the Inner Galaxy.  
\end{abstract}
\maketitle

\section{Introduction}  
\label{sec:intro}

The observed excess of GeV gamma-rays at the center of the Milky Way has withstood many tests over the course of the last decade~\cite{Goodenough:2009gk,Hooper:2010mq,Boyarsky:2010dr, Hooper:2011ti,Abazajian:2012pn, Hooper:2013rwa, Gordon:2013vta, Huang:2013pda,Macias:2013vya,Abazajian:2014fta,Daylan:2014rsa, Zhou:2014lva,Calore:2014nla,Calore:2014xka,Abazajian:2014hsa, TheFermi-LAT:2015kwa, Linden:2016rcf, Karwin:2016tsw}. Referred to as the Galactic Center Excess (GCE), the energy spectrum and morphology of the excess as observed by the \emph{Fermi}~Large Area Telescope~\cite{Atwood:2009ez} are consistent with a signal of dark matter (DM) annihilation~\cite{Gordon:2013vta, Calore:2014nla,Calore:2014xka, Daylan:2014rsa, Karwin:2016tsw}.  However, astrophysical sources---such as a population of unresolved millisecond pulsars---may also explain the signal~\cite{Abazajian:2010zy,Hooper:2013nhl, Mirabal:2013rba, Calore:2014oga, Cholis:2014lta, Petrovic:2014xra,Yuan:2014yda, Abazajian:2014fta, OLeary:2015qpx,Ploeg:2017vai,Eckner:2017oul,Cholis:2014noa,Bartels:2018xom}. Characterizing the nature of these potential sources, either through direct discovery or indirect statistical tests, is of paramount importance in establishing the viability of the DM hypothesis.

Two separate but complementary studies have argued for evidence of unresolved point sources (PSs) at the Galactic Center.  The first method, referred to as Non-Poissonian Template Fitting (NPTF), used the statistics of fluctuations in photon counts to demonstrate evidence for an unresolved PS population in the Inner Galaxy~\cite{Lee:2015fea}.  The second study performed a wavelet decomposition of the gamma-ray sky and found evidence of small-scale structure consistent with a population of unresolved PSs rather than smooth emission from DM~\cite{Bartels:2015aea}.  Since then, the case for unresolved PSs has continued to be strengthened by studies suggesting that the shape of the excess is correlated with the stellar overdensity in the Galactic bulge and the nuclear stellar bulge, a scenario strongly preferred over spherically-symmetric emission from DM annihilation~\cite{Macias:2016nev,Macias:2019omb,Bartels:2017vsx}.  

The exact nature of these unresolved PSs continues to remain a mystery, however.  Both the NPTF and wavelet methods are only sensitive to the spatial distribution of PSs and, in the case of the NPTF, their luminosity function, but are otherwise model-independent.  These sources can be actual astrophysical PSs, such as millisecond pulsars~\cite{Brandt:2015ula,FermiLAT:2017yoi,Ploeg:2017vai,Eckner:2017oul}, residual structure due to mismodeling of cosmic-ray background emission~\cite{Carlson:2015ona,Carlson:2016iis}, or even DM substructure~\cite{Clark:2016mbb}.  The statistical analyses are themselves agnostic to these possibilities.  In lieu of a direct discovery of these sources~\cite{Calore:2015bsx}, we can only hope to infer their properties indirectly.  

In this paper, we study the ability of the NPTF method to characterize the flux contribution of PSs to the GCE as well as their source-count distribution.  A source-count distribution describes the number of sources of a given flux, and therefore encodes information about the relative number of bright and faint sources in a population.  When applied to spatially binned (pixelated) photon count data, the NPTF procedure distinguishes PSs from smooth Poissonian emission based on the number of photons that each PS produces, distributed across a number of pixels due to the finite spatial resolution of the telescope.  A population of PSs will typically yield more ``hot" and ``cold" pixels relative to a smooth Poissonian component.  

Using simulated data, we consider scenarios where the GCE is comprised entirely of DM or PSs, as well as cases where the GCE flux is divided between the two.  In each case, we study how reliably the NPTF recovers the correct composition of the GCE.   Our work here builds on previous studies of the NPTF on simulated data~\cite{Lee:2015fea}, which only considered PSs with a source-count distribution matching that recovered on data.  This empirically-motivated distribution described a population of reasonably bright unresolved sources.  We now present a more systematic study on simulated data that carefully considers different source-count functions, focusing on cases where more ultra-faint sources are present, as is expected for a population of millisecond pulsars (MSPs)~\cite{Hooper:2013nhl,Petrovic:2014xra,Cholis:2014lta,Cholis:2014noa,Eckner:2017oul,Ploeg:2017vai, Bartels:2018xom}. This enables us to characterize any potential biases of the statistical method that can shift the recovered source-count function away from its true distribution.

When the backgrounds are perfectly modeled, we find that the NPTF always accurately identifies the origin of the signal in the case where the GCE consists entirely of DM.\footnote{Note that, throughout this work, when we refer to the NPTF we also implicitly refer to the standard parametrization of the source-count distribution and the associated priors, which are described in Sec.~\ref{sec:NPTF}.  It is possible that different parametrizations of these distributions within the framework of the NPTF would give different results to those presented here.}  If, instead, 100\% of the GCE consists of PSs, then some fraction of the PS flux can be misattributed to DM.  When the GCE is comprised of contributions from both DM and PSs, the NPTF can misidentify the flux as belonging entirely to DM or entirely to PSs.  The challenge associated with reconstructing the correct fraction of DM when PSs are also present in the data stems from the basic fact that in the ultra-faint limit, PSs are exactly degenerate with smooth Poissonian emission.  This degeneracy can lead to biases when inferring the proportion of PSs and DM preferred by the data, which we explore in detail.  These biases are tempered for a population of (relatively) bright unresolved sources,  as they are easier to distinguish from smooth Poissonian emission with the same spatial distribution.  This is the scenario that had been studied previously in Ref.~\cite{Lee:2015fea}.

When the Galactic diffuse backgrounds are mismodeled, as expected in any analysis on actual \emph{Fermi} data,  additional challenges arise. We mock up this scenario by creating simulated data with one diffuse model and analyzing it with a different diffuse model, using this setup to show that bright residuals from mismodeling can be absorbed as point sources in the analysis.  This may explain why the source-count distribution recovered on data in Ref.~\cite{Lee:2015fea} is different from that expected for MSPs.  When 100\% of the GCE flux is in DM, the statistical preference for PSs is significantly reduced relative to the strong preference recorded when the GCE flux is entirely in PSs.  However, we identify some instances where the DM can be misidentified as PSs with reasonable statistical confidence. For the particular pair of diffuse models we use, we find that the significance for PSs varies strongly depending on whether the ``correct" or ``incorrect" model is assumed in the NPTF analysis, a strong indication that the NPTF is picking up residuals from diffuse mismodeling as PSs. These findings motivate a detailed study of ways to mitigate the effects of diffuse mismodeling in NPTF analyses on data. 
We will present these results  separately in a companion paper~\cite{companion}. 

Lastly, our work enables us to comment on a recent study that draws doubt on the PS interpretation of the GCE~\cite{Leane:2019xiy}.  In reaching this conclusion, the authors inject an artificial DM signal on the \emph{Fermi} data, pass it through the NPTF pipeline, and find that the injected DM signal is misattributed to PS flux.  In general, signal recovery tests on data can be quite valuable---as we have shown in separate studies on \emph{Fermi} data~\cite{Lisanti:2017qlb,Chang:2018bpt}.  However, interpreting the results must be done with great care, especially in the case when a true signal (either DM or PSs) is present in the data. We demonstrate using simulated data that, within the current NPTF framework, misattribution of injected DM flux to PSs can be a natural consequence of the fact that smooth Poissonian emission is exactly degenerate with a population of ultra-faint PSs;  from the perspective of the NPTF, the artificially injected DM signal can either be its own separate Poissonian contribution, or it can be flux that ``fills in" the ultra-faint end of the source-count distribution for PSs with the same spatial distribution as the DM.\footnote{In principle, one can construct an analysis framework that is indiscriminate between the DM and PS hypotheses---rather than attribute the flux to one component or the other---in the degenerate regime. However, this is beyond the scope of the work presented here.}  It is thus unsurprising that the NPTF does not recover the injected DM signal if PSs are already present in the data.  We also show that the recovery of the injected DM signal can be significantly worsened (\emph{i.e.}, an appreciable fraction of the DM signal is misattributed in more realizations) in the presence of diffuse mismodeling, an irreducible effect on real \emph{Fermi} data. Additionally, on the real data, the injected signal recovery could further be compounded by complications from diffuse mismodeling that are not captured by the simulations studied here, or by other populations of unmodeled PSs. We show that such misattribution of an injected DM signal does not by itself point to issues with an NPTF analysis of the underlying map that does not contain an injected signal, and thus conclude that the signal injection tests performed on data in Ref.~\cite{Leane:2019xiy} are not by themselves indicative of an issue with the original NPTF analysis.

We take a pedagogical approach in this work in order to help the reader build intuition for interpreting the output of an NPTF analysis.   The paper is organized as follows.  In Sec.~\ref{sec:NPTF}, we review the basics of the NPTF procedure focusing specifically on the importance of the source-count function for the PSs and its interpretation.  We emphasize where the source-count function becomes potentially degenerate with Poissonian emission, a crucial point for any NPTF study claiming to set constraints on the flux contribution of DM and PSs at the Galactic Center.  In Secs.~\ref{sec:anatomy}--\ref{sec:mismodeling}, we present the NPTF tests on simulated data. Throughout, we emphasize the significance for these results in interpreting the results of signal injection tests on the \emph{Fermi} data.  We conclude in Sec.~\ref{sec:conclusions}. Appendix~\ref{sec:supplementary_figs} includes supplementary figures that further illustrate the points of the main text. Appendix~\ref{sec:residuals} discusses the residuals from diffuse mismodeling. Appendix~\ref{sec:cutoff} explores the effects of forcing the source-count function to zero below the flux near which the DM and PS contributions become degenerate.

\section{Statistical Methodology}  
\label{sec:NPTF}

This work uses simulated data to better characterize the ability of the NPTF procedure to recover the  properties of the unresolved GCE PSs.  We will start from simple maps that only contain PSs, and build up to include diffuse emission, DM, and other non-PS components. In this way, we will clearly see how the recovery of the PS and DM fractions is affected as the simulated maps become increasingly more realistic. This section  reviews the NPTF procedure and describes how the maps are made. 

\subsection{NPTF Procedure}
\label{sec:NPTF_procedure}

We assume that the data can be described by a set of different gamma-ray components, each with its own specified spatial distribution.  Each component is modeled by a ``template'' that traces its spatial morphology.  Some of the templates in the study model smooth Poissonian emission, while others trace populations of unresolved PSs that are described by non-Poissonian statistics. 
Consider a spatially binned data map $d$ that consists of $n_p$ photon counts in pixel $p$.  For a given model $\cal M$ with free parameters $\boldsymbol{\theta}$, the likelihood function is defined as
\es{eq:likelihood}{
p(d \, | \, \boldsymbol{\theta}, \mathcal{M}) = \prod_p \, p_{n_p}^{(p)}(\boldsymbol{\theta}) \,,}
where $p_{n_p}^{(p)}(\boldsymbol{\theta})$ is the probability of observing $n_p$ photons in pixel $p$ for the assumed model.  In the Poissonian case, the templates---which are spatially binned in the same way as the data---predict the mean expected number of counts $\mu_p(\boldsymbol{\theta})$ in pixel $p$:
\es{eq:mu}{
\mu_p(\boldsymbol{\theta})=\sum_{l}\mu_{p,l}(\boldsymbol{\theta}) \,,}
where $l$ is the index over Poissonian templates. $\mu_p(\boldsymbol{\theta})$ is fully specified by the overall normalizations of the templates. 
In this case, $p_{n_p}^{(p)}(\boldsymbol{\theta})$ in Eq.~\ref{eq:likelihood} is simply given by the Poisson probability of observing $n_p$ photons given the expected number of counts:
\es{eq:poiss_prob}{
p_{n_p}^{(p)}(\boldsymbol{\theta})=\frac{\mu_p^{n_p}(\boldsymbol{\theta})}{n_p!}e^{-\mu_p(\boldsymbol{\theta})} \,.}

When modeling unresolved PSs, however, $p_{n_p}^{(p)}(\boldsymbol{\theta})$ is non-Poissonian.  The reason for this is that one must first ask what the probability is that a PS is in pixel $p$ and then ask what the probability is that it contributes $n_p$ photons to the data (modulo corrections for a finite point-spread function, which will be discussed below).   
 
When modeling the non-Poissonian templates, an essential input is the flux distribution of the sources.  To aid the calculation, this is typically parameterized as a multiply-broken power law.  In the first iteration of the NPTF method from Ref.~\cite{Lee:2015fea}, a singly-broken power law was used, but additional breaks can allow for greater flexibility in the recovery of the underlying PS flux distribution. In this work, we  use a two-break model to describe how the number of sources $N$ is distributed with photon count $S$:
\es{eq:sourcecount2break}{
\frac{dN}{dS} = A^\text{PS} \,\begin{cases} 
 \left(  \frac{S}{S_{b,2}}\right)^{-n_3} &  S < S_{b,2}\\ 
\left( \frac{S}{S_{b,2}}\right)^{-n_2}  & S_{b,2} \leq S< S_{b,1}  \\ 
\left( \frac{S_{b,1}}{S_{b,2}}\right)^{-n_2} \left( \frac{S}{S_{b,1}}\right)^{-n_1}   & S_{b,1} \leq S  \\ 
 \end{cases} \,,}
where $S_{b,1...2}$ are the breaks, $n_{1..3}$ denote the power-law indices, and $A^\text{PS}$ is the overall normalization.  The photon count $S$ is related to the flux $F$ through the equation $S = \langle\epsilon\rangle F$, where $\langle\epsilon\rangle\sim6.59\times 10^{10}$\,cm$^{2}$\ is the mean exposure per pixel for the dataset under consideration.  Note that, for computational simplicity, we consider a flat exposure map with the value in every pixel equal to the mean \emph{Fermi}-LAT exposure in the relevant energy range.  The effect of non-uniform exposure can be corrected using the procedure described in Ref.~\cite{Mishra-Sharma:2016gis}, and would not affect the conclusions of our study.  In general, such corrections require that Eq.~\ref{eq:sourcecount2break} be written in terms of flux, with the translation to counts occurring on a pixel-by-pixel basis.  

It is important to emphasize that the shape of the flux distribution is a critical assumption of the method, and an inherent systematic uncertainty. The choice of the doubly broken power law is useful as it provides sufficient freedom to capture known features in the distribution.  For example, the upper break ($S_{b,1}$) corresponds roughly to the threshold of resolved sources, when they are masked.  For \emph{Fermi}, we take this to be the threshold for the third source catalog (3FGL)~\cite{Acero:2015hja}.\footnote{Although the fourth source catalog has recently become available~\cite{Fermi-LAT:2019yla}, we use the 3FGL in our study to motivate comparison with previous work. Using the updated catalog would not affect the conclusions presented here since we restrict ourselves to studying simulated data.}  The lower break ($S_{b,2}$)  corresponds to the region where the method starts to lose sensitivity.\footnote{When this is not the case, an additional break is often useful as it allows the model to capture features in the source-count distribution at intermediate fluxes, as was done in Ref.~\cite{Lisanti:2016jub}.}  As discussed further below, ultra-faint sources are inherently degenerate with smooth  Poissonian emission.  In this low-flux regime, we expect that the NPTF analysis will struggle to distinguish PSs from smooth emission. 

The NPTF method was discussed in depth in Refs.~\cite{Lee:2014mza, Lee:2015fea, Mishra-Sharma:2016gis}, and we refer the reader to those works for a full review and technical details of algorithms used. Here, we provide basic pertinent information relevant to this study.  The NPTF likelihood is most conveniently cast in the language of probability generating functions, also known as moment generating functions, following Ref.~\cite{Malyshev:2011zi}. For a discrete probability distribution $p_k$, with $k=0,1,2,\ldots$, the generating function  is defined as
\es{eq:prob}{
P(t) \equiv \sum_{k=0}^{\infty} \, p_k \, t^k \,,}
from which the probabilities can be recovered by taking successive derivatives:
\es{eq:deriv}{
p_k = \frac{1}{k!} \left. \frac{d^k P(t)}{dt^k} \right|_{t=0} \,.
}
The generating function for a Poissonian template is given by
\es{eq:P-PGF}{
P_{\rm P}(t; {\boldsymbol{\theta}}) = \prod_p \text{exp}\left[ \mu_p( {\boldsymbol{\theta}}) (t - 1) \right]  \,.}
The generating function for a PS template takes the more complicated form
\es{eq:NP-PGF}{
P_{\rm NP}(t; {\boldsymbol{\theta}}) = \prod_p \exp \left[ \sum_{m=1}^{\infty} x_{p,m}( {\boldsymbol{\theta}}) ( t^m - 1) \right] \,,}
where
\es{eq:xm-def}{
x_{p,m}( {\boldsymbol{\theta}}) =\int_0^{\infty} dS \, \frac{dN_p}{dS}(S;{\boldsymbol{\theta}}) \int_0^1 df \, \rho(f) \, \frac{\left(fS\right)^m}{m!} \, e^{-fS} \,.}
The $x_{p,m}$ can be interpreted as the average number of PSs contributing $m$ photons in expectation within the pixel $p$, given the pixel-dependent source-count distribution $dN_p(S;{\boldsymbol{\theta}})/dS$.  
When $m=1$, the functional form of Eq.~\ref{eq:NP-PGF} reduces to that of Eq.~\ref{eq:P-PGF} (see Ref.~\cite{Mishra-Sharma:2016gis} for more details).  This demonstrates that a Poissonian component (such as DM or diffuse emission) can be thought of as a population of single-photon sources with the same spatial distribution. Using the property that the generating function of a sum of several random variables is the product of the individual corresponding generating functions, we can write the total generating function in our case as the product of the separate Poissonian and non-Poissonian contributions.  

For a ``spatially-averaged'' source-count distribution, \emph{e.g.} Eq.~\ref{eq:sourcecount2break}, an overall pixel-dependent prefactor in $dN_p(S;{\boldsymbol{\theta}})/dS$ modulates the expected number of PSs in each pixel $p$ following the assumed spatial distribution of PSs specified by the template. Additionally, $\rho(f)$ is a function that describes the distribution of flux fractions among pixels, accounting for photon ``leakage'' due to the finite point-spread function (PSF) of the instrument.  By definition $\rho(f) \, df$ equals the number of pixels that, on average, contain between $f$ and $f+ df$ of the flux from a PS; the distribution is normalized such that $ \int_0^1 df \, f \, \rho(f) = 1$. In the absence of this effect, the PSF is a $\delta$-function and $\rho(f) = \delta(f-1)$. For a given PSF model, $\rho(f)$ is obtained through a Monte Carlo procedure. For more details on Eqs.~\ref{eq:P-PGF}--\ref{eq:xm-def} and the NPTF algorithm generally, as well as details of instrument PSF and exposure (\emph{i.e.}, scanning strategy) correction, see Ref.~\cite{Mishra-Sharma:2016gis}.

We use the public code \texttt{NPTFit}~\cite{Mishra-Sharma:2016gis} to implement the NPTF procedure. This is interfaced with \texttt{MultiNest}~\cite{Feroz:2008xx,Buchner:2014nha} (with \texttt{nlive}=500), which implements the nested sampling algorithm~\cite{Feroz:2013hea,Feroz:2007kg,skilling2006}, to efficiently scan the (potentially multi-modal) posterior parameter space associated with the Poissonian normalizations and non-Poissonian source-count parameters in a Bayesian framework. This necessitates a specific choice of prior probabilities on each parameter, summarized in Table~\ref{tab:priors}.  

\begin{table}[tb]
\footnotesize
\begin{center}
\begin{tabular}{C{2cm}C{2cm} | C{2cm}C{2cm}}
\renewcommand{\arraystretch}{1}
\textbf{Parameter}	 & \textbf{Prior}  & \textbf{Parameter}	&  \textbf{Prior}   \Tstrut\Bstrut	\\   
\Xhline{3\arrayrulewidth}
$A_\text{diff}$  & [0,\,20]  & $\log_{10}S_{b,1}$ & [0.5,\,2.0] \\
$A_\text{iso}$ & [0,\,2]  &  $\log_{10}S_{b,2}$ & [-2.0,\,0.5]  \Tstrut\Bstrut \\ 
$A_\text{bub}$  & [0,\,2]  & $n_1$ & [2.05, 15] \Tstrut\Bstrut \\ 
$A_\text{3FGL}$& [0,\,2] &  $n_2$ & [-3.95, 2.95]   \Tstrut\Bstrut \\
$\log_{10}A_\text{NFW}$& [-5,\,2] &  $n_3$ & [-10, 0.95]     \Tstrut\Bstrut \\
$\log_{10}A_\text{NFW}^{\text{PS}}$  & [-10,\,5] &   & \Tstrut\Bstrut \\
\end{tabular}
\end{center}
\caption{Fiducial priors assumed for the NPTF analyses in this work.  The left column lists the priors for the template normalizations.  From top to bottom: diffuse, isotropic, \emph{Fermi} bubbles, \emph{Fermi} 3FGL sources, NFW dark matter, and NFW point sources. The right column lists the priors on the source-count parameters, as in Eq.~\ref{eq:sourcecount2break}. A flat prior distribution between the specified parameter ranges is assumed.}
\label{tab:priors}
\end{table}  

In all of the results presented and described in this paper, we have discarded scans for which the upper break $S_{b,1}$ of the NFW PS template is peaked at the lower edge of its prior range.\footnote{In practice, we quantify this convergence test by histogramming the NFW PS $S_{b,1}$ posterior for a single run into 50 log-spaced bins from $10^{0.5}$ to $10^{2}$, and require that either the counts in the first bin do not exceed 0.4 times the maximum counts or the counts in the last bin are no lower than 0.2 times the maximum counts. The former criterion ensures that the peak of the distribution is not pushed against the lower prior edge, and the latter allows for a posterior distribution that is unconverged over the prior range, as is expected for \emph{e.g.} the case of no PS contribution. Example triangle plots of scans that pass these criteria can be found in Figs.~\ref{fig:corner_pass_PS} and~\ref{fig:corner_pass_DM}; a failing example can be found in Fig.~\ref{fig:corner_fail}.} This is because in such cases, the recovered source-count function peaks in precisely the flux regime where the PSs are degenerate with smooth Poissonian emission.  Any such source-count function recovered on data would be suspicious as it would suggest a population of ultra-faint sources that is concentrated in the regime where the NPTF method loses sensitivity. Fig.~\ref{fig:corner_fail} shows the triangle plot from an example of a discarded scan. In Appendix~\ref{sec:cutoff}, we consider what happens when the source-count function is forced to zero below the flux near which the method loses sensitivity for the aforementioned reason.  We find that imposing such a flux cutoff reduces the number of cases that would be removed using this procedure.   

\subsection{Simulated Data Maps}
The region of interest (ROI) in our analysis is $|b|\geq2^\circ,\,r<30^\circ$, and we mask the resolved \emph{Fermi} 3FGL PSs~\cite{Acero:2015hja} at a $0.8^\circ$ radius. This is essentially the same setup as was used in Refs.~\cite{Lee:2015fea,Leane:2019xiy}.
We use the datasets and templates from Ref.~\cite{rodd_nicholas_safdi_siddharth_2016} (packaged with Ref.~\cite{Mishra-Sharma:2016gis}) to create our simulated maps. The data corresponds to 413 weeks of \emph{Fermi}-LAT (Large Area Telescope)  Pass 8 data collected between August 4, 2008 and July 7, 2016. The top quarter of photons in the energy range 2--20 GeV by quality of PSF reconstruction (corresponding to PSF3 event type) in the event class \texttt{ULTRACLEANVETO} are used. The recommended quality cuts are applied, corresponding to zenith angle less than 90$^\circ$, \texttt{LAT\_CONFIG} = 1, and \texttt{DATA\_QUAL} $> 0.1$.\footnote{\url{https://fermi.gsfc.nasa.gov/ssc/data/analysis/documentation/Cicerone/Cicerone_Data_Exploration/Data_preparation.html}} The maps are spatially binned using HEALPix~\cite{Gorski:2004by} with \texttt{nside} = 128.

We build up the simulated maps as a combination of Poissonian and PS contributions, as necessary. A PS population is completely specified by a spatial distribution and a source-count distribution, in our case parameterized with Eq.~\ref{eq:sourcecount2break}.  We draw the fluxes of simulated PSs  from the source-count function, using it as a probability density. The total number of simulated PSs is determined by the desired total flux contribution from the PS population. We then put the simulated PSs down on a higher-resolution HEALPix map with \texttt{nside} = 2048, using the template describing the spatial distribution of PSs as a probability density informing the locations of PSs. The PS map is smoothed with the \emph{Fermi} PSF at 2 GeV, modeled as a linear combination of King functions,\footnote{\url{https://fermi.gsfc.nasa.gov/ssc/data/analysis/documentation/Cicerone/Cicerone_LAT_IRFs/IRF_PSF.html}} and downgraded to the baseline \texttt{nside} = 128.  A Poisson realization of this downgraded map then represents a single Monte Carlo simulated map.

In this work, we model the known astrophysical emission as a sum of Poissonian templates, which include, \emph{e.g.}, \emph{(i)}~the Galactic diffuse emission, described by the \emph{Fermi} \texttt{gll\_iem\_v02\_P6\_V11\_DIFFUSE} (\texttt{p6v11}) model\footnote{\url{https://fermi.gsfc.nasa.gov/ssc/data/access/lat/ring_for_FSSC_final4.pdf}} \emph{(ii)}~the \emph{Fermi} bubbles~\cite{2010ApJ...724.1044S}, \emph{(iii)}~isotropic emission, and \emph{{(iv)}}~resolved \emph{Fermi} 3FGL PSs~\cite{Acero:2015hja}. For each template, the normalization is the best fit obtained with a traditional Poissonian template fit on \emph{Fermi} data in the ROI. The final set of maps is obtained by creating a linear combination of a (sub)set of these templates as a baseline, and subsequently Poisson fluctuating to obtain multiple Monte Carlo realizations. In addition, whenever applicable, we model the DM(PS) GCE emission as a Poissonian(non-Poissonian) template following the line-of-sight integrated square of an NFW distribution~\cite{Navarro:1995iw}. We perform a separate Poissonian template fit including the astrophysical emission templates \emph{(i)}$-$\emph{(iv)}, with the addition of the Poissonian NFW template, to determine the best-fit GCE flux. In Sec.~\ref{sec:anatomy}, we will start with just a simulated PS contribution accounting for the entirety of the GCE flux and build up to increasingly more realistic scenarios by adding astrophysical background templates.  We then consider more complicated scenarios where the GCE might consist of flux contributions from PSs and DM. Approaching our study from a pedagogical angle, we will also initially consider the Galactic diffuse emission as the sole tracer of astrophysical emission (and use a single Poissonian template, the \emph{Fermi} \texttt{p6v11} model, to describe it), then incorporate the effect of additional Poissonian templates later.

\section{Anatomy of a Source-Count Function}
\label{sec:anatomy}

\begin{figure*}[t]
\centering
\includegraphics[width=3.5in]{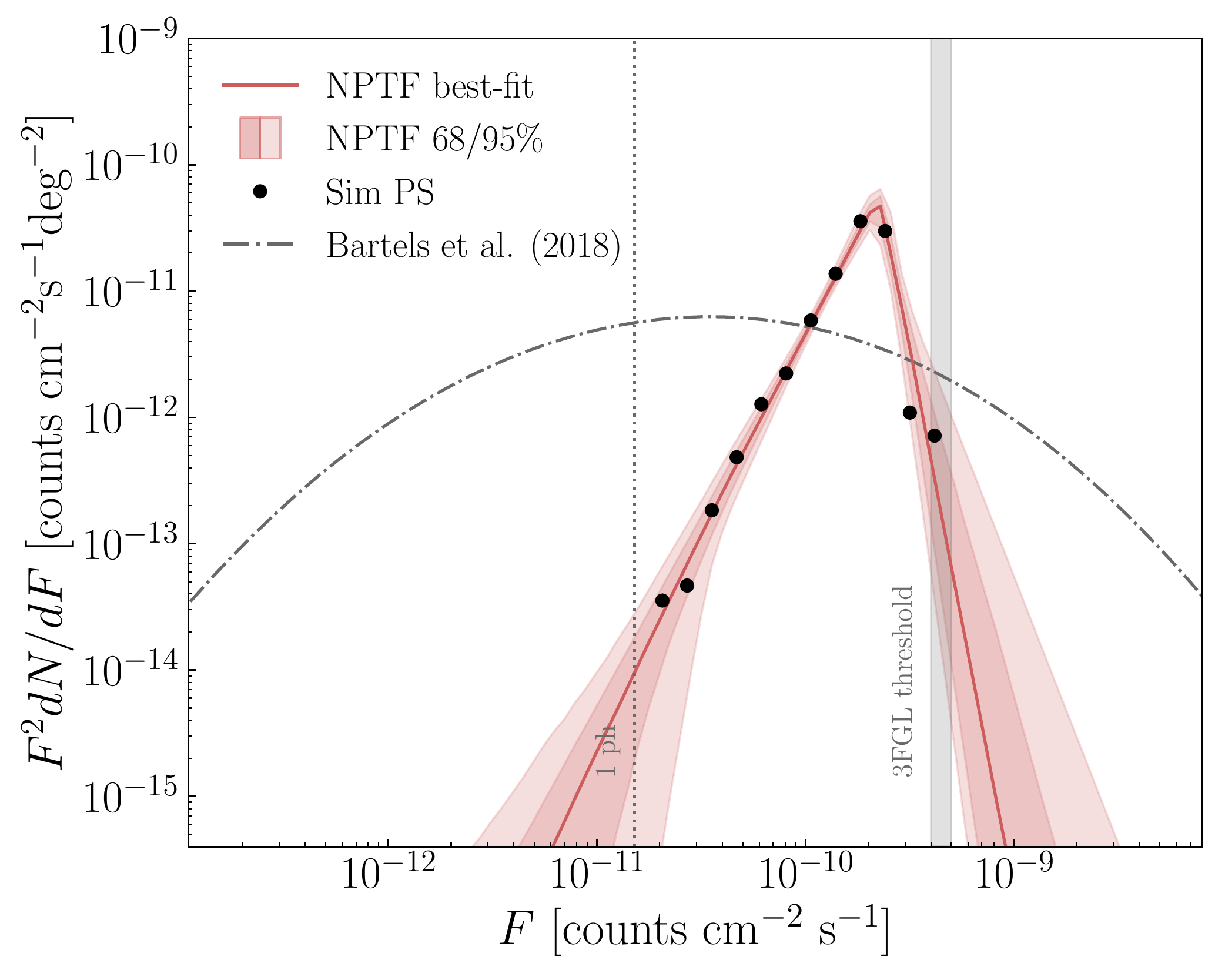} 
\includegraphics[width=3.5in]{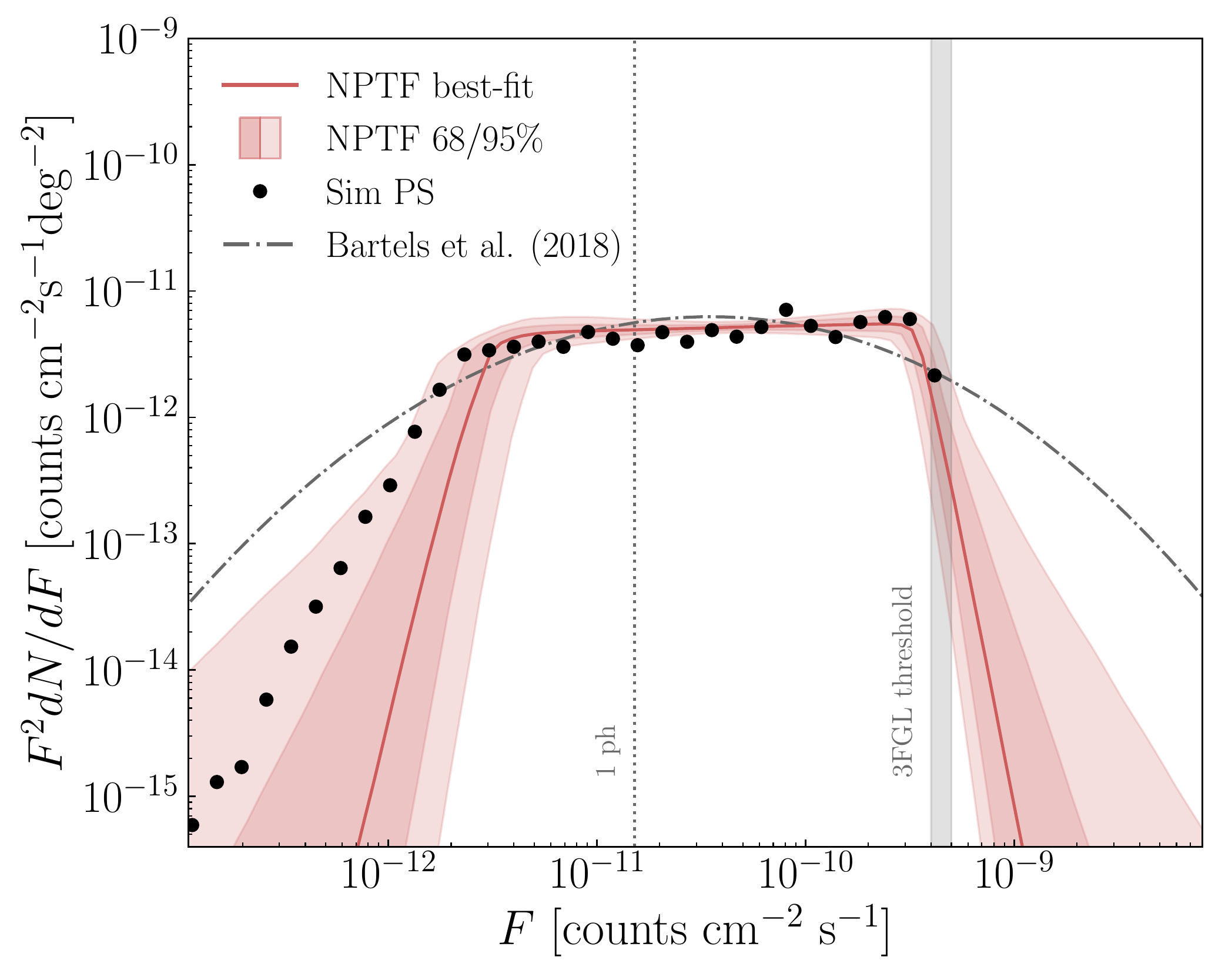} 
\caption{Recovery of two benchmark source populations, both normalized to account for 100\% of the GCE flux.  The black points correspond to the source-count distribution of the simulated sources with a hard (left) and soft (right) flux distribution.  At this stage, the simulated maps contain no contributions other than from the PSs, and the sources themselves are not PSF-smoothed.  We run the NPTF on these maps using three templates: \emph{(i)} NFW PS, \emph{(ii)} NFW DM, and \emph{(iii)} Galactic diffuse emission.  The solid red line indicates the best-fit source-count function for the PSs, with the bands corresponding to the 68 and 95\% confidence interval.  We assume that all sources above the 3FGL threshold of $F \sim 4$--$5 \times 10^{-10}$~counts~cm$^{-2}$~s$^{-1}$ would be resolved and thus choose, by design, source-count functions that are suppressed above this threshold flux.  The vertical gray dotted line indicates the flux corresponding to $\sim1$~ph.  Above this flux, the NPTF successfully recovers the source-count distributions in both cases.  Below $\sim1$~ph, the uncertainty on the recovered source-count function increases dramatically for the soft population, as this is the regime where the statistical method loses sensitivity.  In both cases, the total flux of the PSs recovered by the NPTF matches the true value. We also show for reference the millisecond pulsar (MSP) source-count distribution derived using the luminosity function for disk MSPs obtained in Ref.~\cite{Bartels:2018xom} and assuming the energy spectrum found in Ref.~\cite{Cholis:2014noa}, normalized to account for 100\% of the GCE flux (gray dash-dotted line). The soft source-count distribution is a reasonable approximation to the MSP source-count distribution, while the hard source-count distribution significantly underpredicts the number of dim PSs.}
\label{fig:0DM100PS}
\end{figure*}

When studying the ability of the NPTF to distinguish PSs from smooth emission at the Galactic Center, one must know something about the properties of those sources---both their spatial and flux distribution---as well as the average photon count per pixel that is expected for other gamma-ray sources that could be degenerate with the PS signal.  In this work, we will only consider PSs whose emission traces the square of an NFW distribution.\footnote{In practice, we treat the line-of-sight integrated NFW squared map as the number density distribution of the PSs, from which we draw the positions of simulated sources.}  This is intended to match the  spatial distribution of the GCE flux as characterized in Ref.~\cite{Daylan:2014rsa, Calore:2014xka}.\footnote{The study could also be repeated assuming a bulge-shaped template (as in Ref.~\cite{Macias:2016nev,Macias:2019omb,Bartels:2017vsx}) for the DM and PSs.  As the results of this work are mostly driven by the fact that the DM and PSs share the same spatial distribution, we do not expect that the overall conclusions would be significantly altered in this case.  However, the finer details of the recovered fit parameters would likely change.}

\begin{figure*}[t]
\centering
\includegraphics[width=0.9\textwidth]{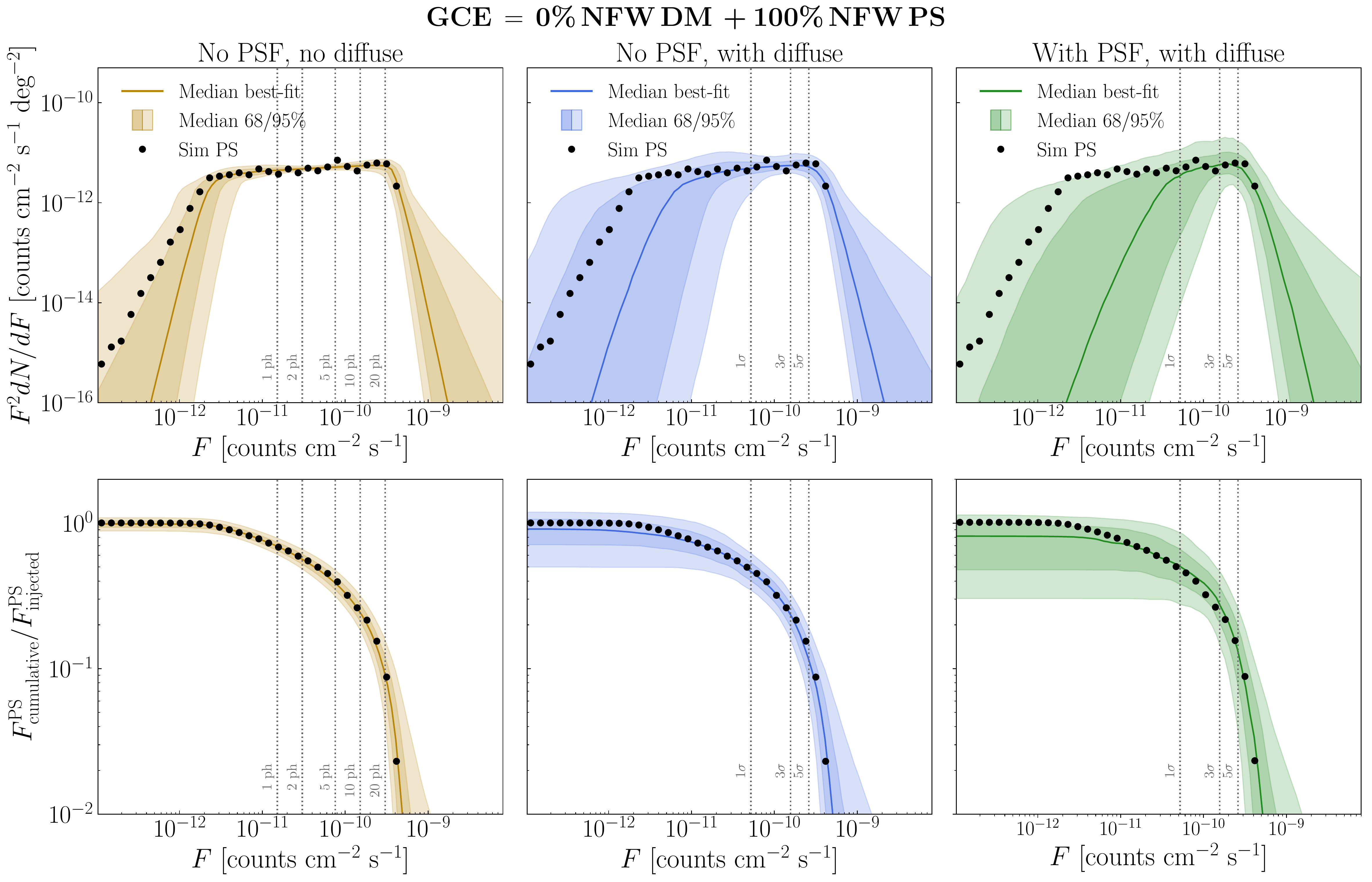}
\caption{Differential source-count distributions (top panels) and cumulative flux distributions, integrated above a given threshold flux (bottom panels), shown for simulations with PSs following a soft source-count distribution, accounting for 100\% of the GCE (no DM contribution).  We run the NPTF on these maps using three templates: \emph{(i)}~NFW PS, \emph{(ii)}~NFW DM, and \emph{(iii)}~Galactic diffuse emission.  Black points indicate the true simulated distributions while solid lines indicate the median best-fit distributions recovered over 100 Monte Carlo realizations of the simulated data maps.  The bands show the median 68 and 95\% confidence bands over these 100 iterations.  In order, the results are shown without including PSF effects in the simulation or NPTF analysis and without Galactic diffuse emission in the simulated data (left column); without PSF effects but including Galactic diffuse emission (middle column); and finally, accounting for PSF effects while also including Galactic diffuse emission (right column). In the left column, the dotted lines designate the photon counts associated with a given flux; in the middle and right columns, the lines designate the approximate significance of a source with flux $F$.  Note that the 3FGL threshold corresponds roughly to $\sim 5\sigma$ sources.  We observe that PS recovery at low fluxes gets progressively worse as Galactic diffuse emission is added into the simulated data maps and PSF effects are included.  The individual flux posteriors for a random subset of 50 out of the 100 realizations summarized in the right panel are provided in Fig.~\ref{fig:0DM_100PS_fplot}. All subsequent figures in this paper include Galactic diffuse emission in the simulated data maps, as well as PSF effects in both the simulated data and the NPTF analysis.}
\label{fig:0DM100PS_fullpanel}
\end{figure*}

We consider two benchmarks for the PS flux distribution: a hard source-count distribution where the distribution of PSs is peaked towards high fluxes, and a softer source-count distribution with a larger number of faint sources. The hard source-count function is generated with a singly-broken power law, with parameters: $S_{b,1}\approx15,\,n_1\approx9.5,\,n_2\approx-1$.  This benchmark is motivated by previous NPTF studies of the Galactic Center and roughly matches the function recovered in the data~\cite{Lee:2015fea}.\footnote{When using a two-break power law in the fit to data, an essentially equivalent function is returned.}  The soft source-count distribution is generated with a two-break power law, with parameters: $S_{b,1}\approx22,\,S_{b,2}\approx0.2,\,n_1=10,\,n_2=1.9,\,n_3=-0.8$ and is motivated by the luminosity function of MSPs estimated in the literature~\cite{Bartels:2018xom,Cholis:2014lta,Cholis:2014noa,Eckner:2017oul,Hooper:2013nhl,Petrovic:2014xra,Ploeg:2017vai}, which tend to be softer than that inferred in Ref.~\cite{Lee:2015fea}. The black points in Fig.~\ref{fig:0DM100PS} illustrate these two cases for simulated maps that only include PSs (and no other contributions), for the hard~(left) and soft~(right) source-count functions.  The spatially averaged 3FGL flux threshold is approximately 4--5$\times 10^{-10}$~counts~cm$^{-2}$~s$^{-1}$---\emph{i.e.,} we assume that all sources above this flux value would be resolved by \emph{Fermi}.\footnote{This threshold was conservatively estimated as the flux corresponding to the peak of the source-count distribution $F^2\,dN/dF$ associated with the sources in the 3FGL catalog in the region of interest.}  We also show the line corresponding to $\sim1$~ph.  Because we can think of Poissonian emission as a combination of single-photon sources, this represents the approximate flux boundary below which Poissonian and non-Poissonian emission are fundamentally degenerate.  In the case of the hard (soft) source-count function, about 0.02~(34)\% of the emission falls below the $\sim 1$~ph line. 

Running the NPTF pipeline on these simulated maps, we can test how well the analysis recovers the properties of the simulated sources in the very simple case where the map consists only of NFW PSs.  We include three templates in the model:  \emph{(i)}~NFW PSs, \emph{(ii)}~NFW DM, and \emph{(iii)}~Galactic diffuse emission. This will allow us to verify that the PS emission is predominantly picked up by the appropriate non-Poissonian template, and characterize any possible degeneracies with the other two templates.  Figure~\ref{fig:0DM100PS} provides the best-fit source-count function (solid red line) that is recovered by the NPTF analysis for the map with hard~(left) and soft~(right) sources; the red bands span the 68 and 95\% containment.  Above the $\sim1$~ph threshold, the source-count function is recovered exactly for both benchmark scenarios.  Below this threshold, the uncertainty on the recovered source-count function increases for the soft sources, as this is the regime where the sources are so ultra-faint that their photon counts are effectively Poissonian. 

As a point of reference, we also show in Fig.~\ref{fig:0DM100PS} the source-count distribution derived using the median (log-normally parameterized) disk MSP luminosity function obtained in Ref.~\cite{Bartels:2018xom}, assuming the MSP energy spectrum found in Ref.~\cite{Cholis:2014noa} and an NFW-squared spatial distribution for the MSP population. We see that our simple model of soft PSs is a reasonable approximation to the MSP scenario, while the hard source-count distribution motivated by previous NPTF studies significantly underpredicts the number of dim sources. 

Next, we test the source-count recovery over several realizations of the simulated data maps.  In particular, we re-run the analysis for 100 Monte Carlo iterations of each map, and take the best-fit source-count function from each run.  The solid yellow line in the top-left panel of Fig.~\ref{fig:0DM100PS_fullpanel} shows the median best-fit over these separate iterations for the soft source-count function.  The yellow bands show the median 68 and 95\% containment regions.  Importantly, these uncertainty bands are fundamentally different than those shown in Fig.~\ref{fig:0DM100PS}, as they summarize the average uncertainty over multiple realizations of the simulated map.  Comparing Fig.~\ref{fig:0DM100PS} (right) with the top-left panel of Fig.~\ref{fig:0DM100PS_fullpanel}, we see that the specific case shown in Fig.~\ref{fig:0DM100PS} is not an outlier over the the distribution of many realizations.

The bottom-left panel of Fig.~\ref{fig:0DM100PS_fullpanel} shows the cumulative PS flux above a given threshold, compared to the total injected PS flux, as a function of decreasing flux threshold.  The black points represent the true distribution, and the yellow bands show the results recovered by the NPTF.  The cumulative flux distribution is useful for inferring the fraction of the injected PS flux that is correctly recovered by the NPTF procedure.  Ideally, the fractional cumulative flux distribution should track the true distribution, asymptoting to unity at low fluxes.  Deviations from unity indicate that the PS template has over/under-absorbed flux relative to the truth.  

We now add Galactic diffuse emission to the simulated maps.  Specifically, we include Poissonian emission from the \texttt{p6v11} model in addition to the NFW PSs and repeat the NPTF studies described above.  The median recovered source-count distribution is provided in the middle column of Fig.~\ref{fig:0DM100PS_fullpanel}, in blue.   The vertical dotted lines denote the approximate significance ($\sigma$) of the PSs at a given flux $F$.  To estimate the significance, we calculate $\sigma = S/\sqrt{B}$, where $S$ and $B$ are respectively the average photon count per pixel for the signal and the diffuse background in the ROI (in our case the pixel size roughly matches the extent of the PSF).  We provide the lines for $\{1, 3, 5\}\sigma$.  The 3FGL threshold corresponds roughly to a significance of $5\sigma$ for each source.  

The presence of diffuse emission results in a greater spread of the recovered source-count distribution relative to the scenario with no diffuse emission.  In this case, the median source-count distribution recovered by the NPTF analysis resembles a harder population of sources.  As already discussed, a population of ultra-faint sources becomes indistinguishable from Poisson emission, so we expect the uncertainties to grow at low fluxes.  This occurs for sources with fluxes $\lesssim 4$--$5\times10^{-11}$~counts~cm$^{-2}$~s$^{-1}$.  While the recovered distribution is consistent with truth at the $\sim 95\%$ level at these low fluxes, the median best-fit is systematically lower than the true source-count distribution, which indicates that the analysis is biased against PS recovery (we observe this even in the absence of diffuse emission).  While a  comprehensive study of the source of this bias is beyond the scope of this work, we comment that it may be related to the parameterization of the source-count function as a multiply-broken power law and/or the choice of priors. 

Lastly, we repeat the same tests, but now smear the simulated map with the \emph{Fermi} PSF function at 2\,GeV---modeled as a double King function---and perform the NPTF with the correction described by Eq.~\ref{eq:xm-def}. The results are summarized in the right column of Fig.~\ref{fig:0DM100PS_fullpanel}, in green.  We see that the inference with the PSF is further biased against recovery of PSs at the faint end of the source-count distribution. This is intuitively expected, as the inclusion of the PSF exacerbates the challenges with recovering the faintest sources.  When the PS flux is underestimated, the flux is picked up by the DM template or the diffuse template.  The amount of flux that is erroneously attributed to these other templates can vary considerably between different Monte Carlo iterations of the map, as demonstrated in Fig.~\ref{fig:0DM_100PS_fplot}.

The degeneracy between ultra-faint sources and Galactic diffuse emission is primarily due to the fact that the total flux of the diffuse emission is $\gtrsim 30$ times greater than the GCE flux in this ROI. As a means of testing this hypothesis, we ran an analysis scaling down the diffuse emission flux by an order of magnitude.  In this case, the recovery of the PS source-count function is significantly improved relative to the fiducial scenario presented in the right panel of Fig.~\ref{fig:0DM100PS_fullpanel}. As a separate test, we replaced the diffuse emission with isotropic emission (of equivalent flux magnitude) in the map. In this latter case, the recovered source-count distributions look very similar to the right panel of Fig.~\ref{fig:0DM100PS_fullpanel}.
We therefore conclude that the degeneracy between ultra-faint unresolved sources and the diffuse emission is primarily driven by the brightness, rather than the morphology, of the diffuse emission. This degeneracy is fundamental to the analysis method in this region of the sky. We further note that the diffuse emission is spatially structured towards the Galactic Center region due to gas-correlated emission, which, if not modeled properly, can potentially lead to residual clusters of ``hot" or ``cold" pixels, as one would also get from a population of PSs. We discuss the effects of diffuse mismodeling in more detail in Sec.~\ref{sec:mismodeling}.

We only show the results of these tests for the soft source-count distribution.  In general, we find that the hard source-count distribution is unaffected by the types of degeneracies we see here, primarily because there are more unresolved sources with high photon counts that are more easily distinguishable from diffuse background emission.

\begin{figure*}[t]
\centering
\includegraphics[width=0.9\textwidth]{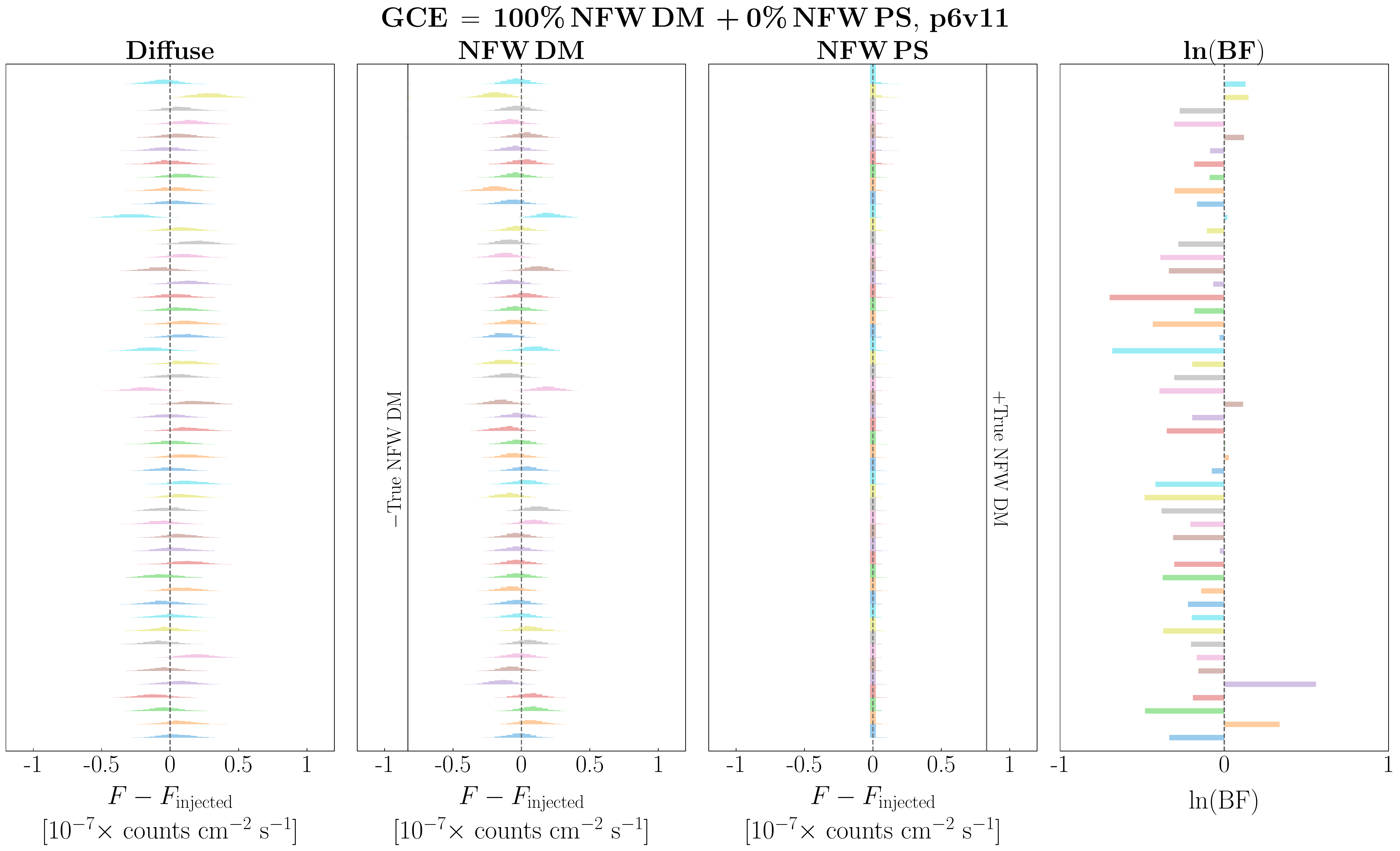} 
\caption{Comparison of the flux posterior (relative to the true injected flux) for the Galactic diffuse, NFW DM, and NFW PS components. The last column shows the Bayes factors (BFs), characterizing the statistical preference of a model of the data that includes NFW PSs over a model that does not include them, for each realization shown. These results pertain to maps where 100\% of the GCE is accounted for by DM and there are no PSs present in the simulated data.  We run the NPTF on these maps using three templates: \emph{(i)}~NFW PS, \emph{(ii)}~NFW DM, and \emph{(iii)}~Galactic diffuse emission.  Each row in the figure represents a different Monte Carlo iteration of the map.  In cases where the flux distributions are recovered exactly, the posteriors are centered at zero.  Importantly, we see that when there are no unresolved PSs in the map, the NPTF analysis does not erroneously attribute the DM to PSs. For this example, the simulated map is made using the \texttt{p6v11} model, and the same template is used in the analysis---this represents the case where there is no diffuse mismodeling. As expected, the $\ln\mathrm{(BF)}$ is negative for the majority of realizations, pointing to a preference for a model without PSs. These are a random subset of 50 out of the 100 iterations shown in the left half of Fig.~\ref{fig:BF_dif}.}
\label{fig:100DM_0PS_fplot}
\end{figure*}

\section{Dark Matter and the GCE}
Having introduced the NPTF procedure and demonstrated how well it works in recovering soft and hard source-counts functions in simulated data, we now begin to test the method on more complex simulated maps.  Specifically, this section will explore what happens as the relative amount of DM and PS flux contributing to the GCE varies, and the ability of the method to accurately recover these flux combinations.  All examples in this section include modeling of the diffuse background emission and PSF effects, both in the construction of the simulated maps and in the analysis.  For now, we assume that there is no diffuse mismodeling.

\subsection{Dark Matter-Only GCE} 
\label{sec:DMrecovery}
We begin by considering the case where the entirety of the GCE is comprised of DM; the simulated maps consist of a DM signal accounting for 100\% of the GCE flux, as well as \texttt{p6v11} diffuse background emission.  We generate 100 different Monte Carlo realizations of this map and run the NPTF on each realization using the following three templates: NFW DM, NFW PS, and Galactic diffuse emission.  Figure~\ref{fig:100DM_0PS_fplot} shows (for a random subset of 50 out of the 100 scans) the flux posteriors for the three separate components, centered around the true injected value.  The last column shows the Bayes factors (BFs) for each realization, which quantifies the statistical preference of a model of the simulated data that includes NFW PSs over a model that does not include them. Each row in the figure corresponds to a different Monte Carlo iteration.  The posterior distribution for the NFW PS template is sharply peaked at zero in every case, meaning that the analysis recovers no PSs.  The recovered DM flux is always close to the injected amount, albeit with some spread due to degeneracy with the diffuse emission.  Importantly, however, a non-zero DM flux is recovered in all cases (the vertical line at ``$-$True NFW DM" in the second panel denotes where the flux posteriors would lie if zero DM flux were recovered), and there is no statistical preference for an NFW PS population based on the Bayes factor, as should be the case.

\subsection{Fractional Dark Matter Recovery}

\begin{figure*}[t]
\centering{}
\includegraphics[width=0.9\textwidth]{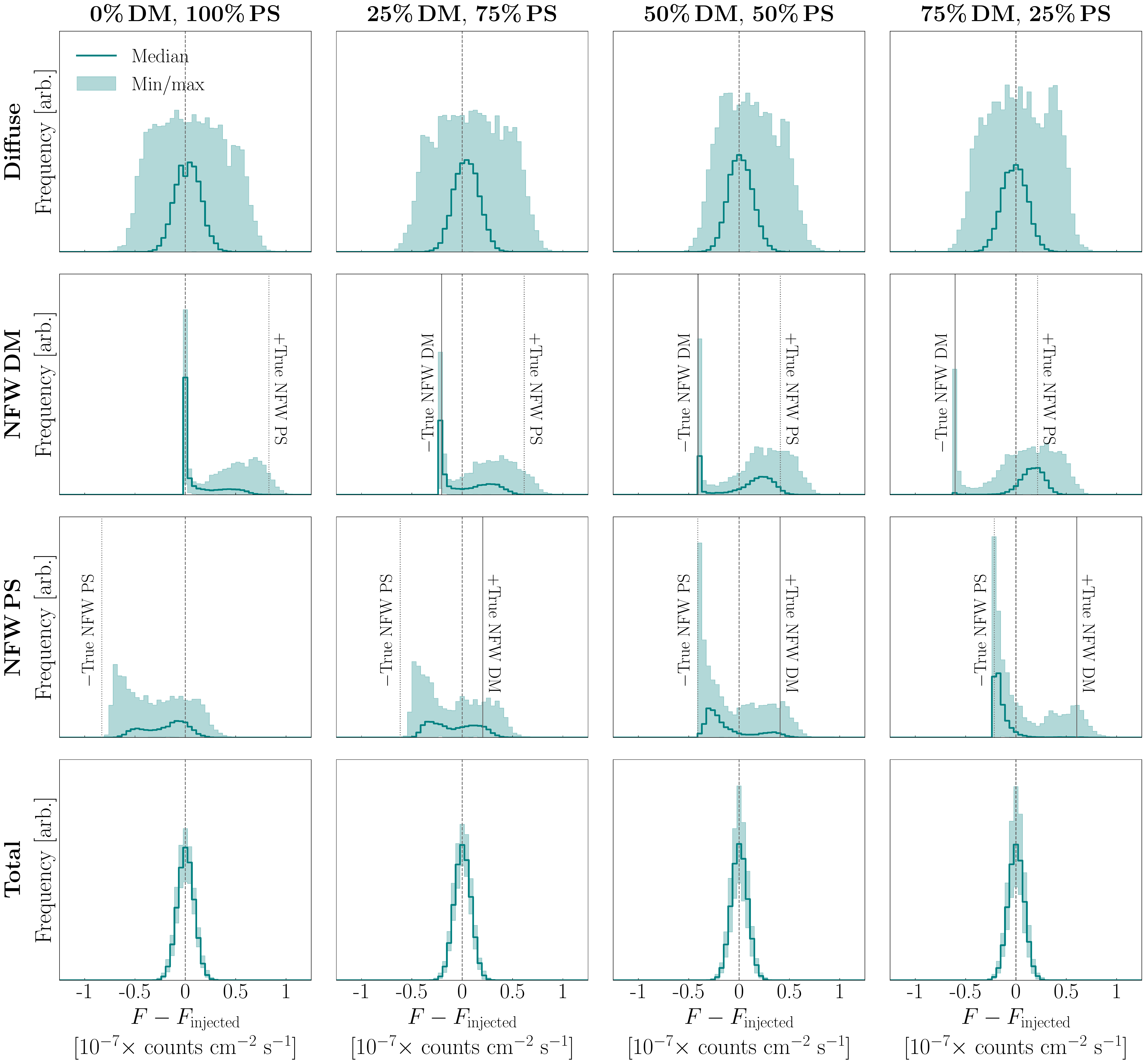}
\caption{Comparison of the flux posterior (relative to the true injected flux) for the Galactic diffuse, NFW DM, and NFW PS components, as well as the total flux, varying the relative fraction of the GCE accounted for by DM and PS.  These results pertain specifically to the soft source-count function.  We consider the four cases where the GCE is 100\% PS (first column), 25\% DM and 75\% PS (second column), 50\% DM and 50\% PS (third column), and 75\% DM and 25\% PS (fourth column). We run the NPTF on these maps using three templates: \emph{(i)} NFW PS, \emph{(ii)} NFW DM, and \emph{(iii)} Galactic diffuse emission. In each panel, the solid line represents the median and the shaded region spans the minimum and maximum value in a given flux bin, across 100 Monte Carlo iterations. In each case where there are contributions from both DM and PSs, there is a probability that the DM signal is absorbed by the PS template, as evidenced by the peaks at ``$-$True NFW DM'' in the second row of each of the right three panels; the probability of the PS signal being absorbed by the NFW DM template increases with increased DM fraction, as evidenced by the increasingly large peak at ``$-$True NFW PS'' in the third row of each of the right three panels.  The corresponding plot for the hard source-count distribution is provided in the Appendix as Fig.~\ref{fig:25_75_fullpanel_hardSC}.}
\label{fig:25_75_fullpanel}
\end{figure*}

We now consider the case where the simulated maps include Galactic diffuse emission, NFW DM, and NFW PSs---but vary the relative fraction of the total GCE flux that comes from DM and PSs. We specifically consider 25/75\%, 50/50\%, and 75/25\% splits as examples. In each case, we run the NPTF analysis to test the recovered flux of each component and the details of the best-fit source-count function.  We use the standard three templates: \emph{(i)}~NFW PSs, \emph{(ii)}~NFW DM, and \emph{(iii)}~Galactic diffuse emission.

Figure~\ref{fig:25_75_fullpanel} summarizes the range of possibilities that occur when varying the DM/PS contributions.  As a point of comparison, the left-most column shows the results for the case where the GCE is comprised of 100\% soft PSs (and no DM). For each of the 100 Monte Carlo realizations of the map, we obtain the posterior distributions for the Galactic diffuse, NFW DM, and NFW PS components, similar to what is shown in Fig.~\ref{fig:100DM_0PS_fplot}.  The median posterior distribution for each component is indicated by the solid line in each panel of Fig.~\ref{fig:25_75_fullpanel}; the shaded bands denote the maximum and minimum value obtained in any given flux bin over the separate map realizations.  In this case, the median PS flux posterior is peaked at its true injected value, but there is a tail towards lower fluxes where some of the emission is instead absorbed by the DM template.  We see this explicitly as the tail in the DM posterior extending towards the `$+$True NFW PS' line.  This speaks to the inherent degeneracy between the ultra-faint PSs and truly Poissonian emission.  

It is notable that the effects of this degeneracy are evident in this case, but not when the GCE consists of 100\% DM (Fig.~\ref{fig:100DM_0PS_fplot}).  When there is only DM present, there are not enough bright pixels to give the PS template any statistical advantage in the fit. However, in the opposite scenario where the GCE is 100\% PS, the PS template can still pull the statistical weight of fitting the comparatively brighter unresolved sources, while the fainter sources are either attributed to the PS or DM template.  Fig.~\ref{fig:25_75_fullpanel_hardSC} shows the corresponding figure for the hard source-count function.  As expected,  when there are fewer ultra-faint sources present (in the 100\% PS GCE case), the probability that the DM template picks up the PS flux is reduced.   

The second column of Fig.~\ref{fig:25_75_fullpanel} illustrates the case where 25\% of the GCE flux originates from DM and the remaining 75\% comes from PSs.  In this case, the median DM posterior is no longer peaked at the true injected value.  The peak at `$-$True NFW DM' indicates that the entirety of the DM flux is not recovered most of the time---the flux is instead absorbed by the PS template, whose posterior distribution now has a tail extending to larger fluxes and encompasses the `$+$True NFW DM' line.  However, a wide range of possibilities remains viable, depending on the Monte Carlo realization of the map; this includes the possibility that the PS flux is underestimated and is instead absorbed by the DM template.

As the relative amount of DM to PS flux in the GCE increases, this trend reverses.  In particular, it becomes increasingly more likely that the PS flux is underestimated and incorrectly absorbed by the DM template.  This can be seen in Fig.~\ref{fig:25_75_fullpanel}: going from left to right, the median PS flux posterior becomes increasingly peaked towards `$-$True NFW PS,' while the median DM flux posterior becomes increasingly peaked towards `$+$True NFW PS.'  Although the average 75/25\% DM/PS map clearly follows this behavior, there are still cases where the entirety of the DM flux is absorbed by the PS template.  These cases are more unlikely, but still viable.

\begin{figure*}[p]
\centering
\includegraphics[width=0.9\textwidth]{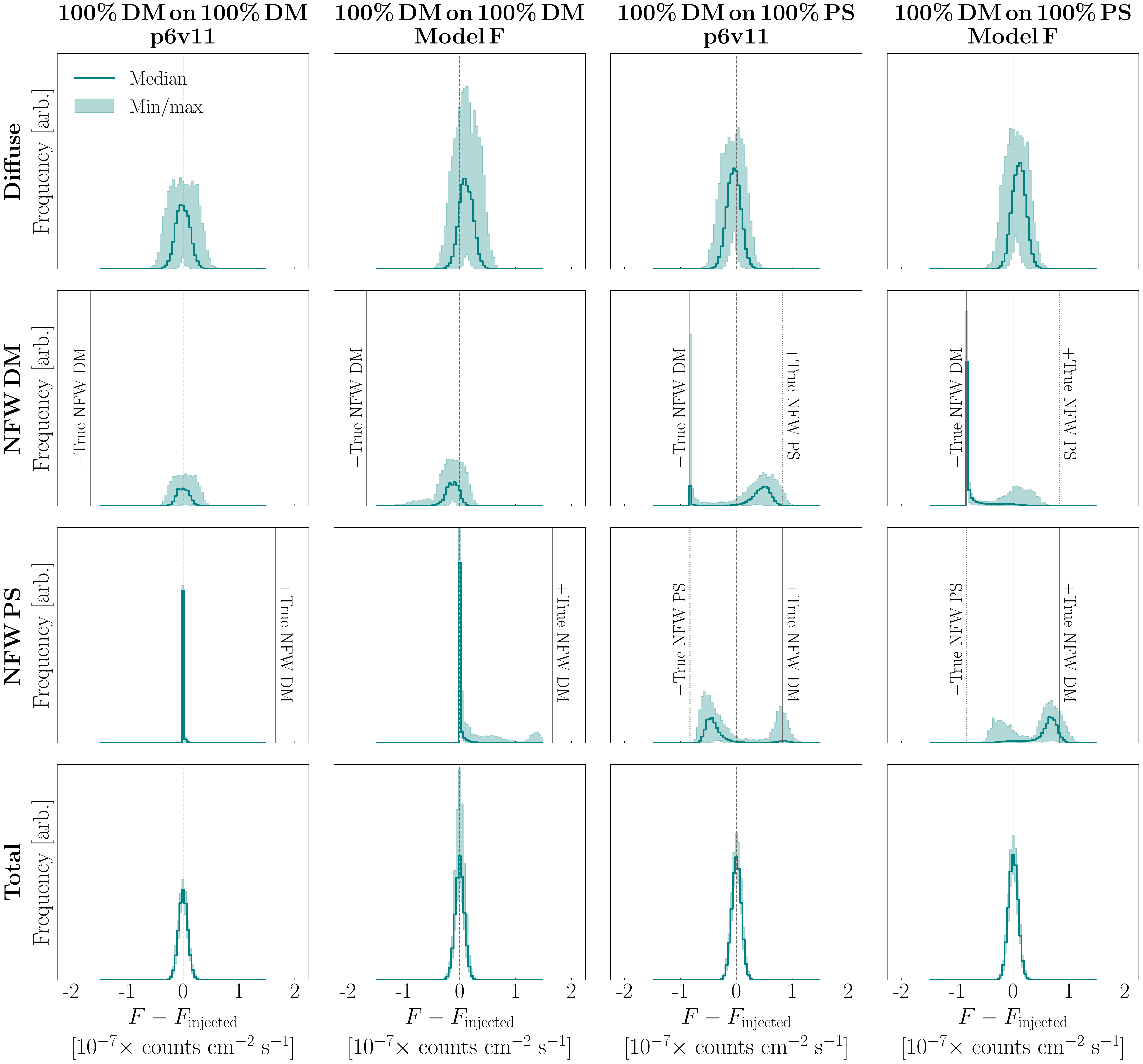} 
\caption{Same as Fig.~\ref{fig:25_75_fullpanel}, but with mock data constructed by injecting an additional GCE-strength DM signal on top of an existing GCE signal. Each panel shows the median (solid lines) and minimum/maximum (shaded regions) value in a given flux bin over 100 Monte Carlo iterations. The simulated data maps consist of 10 distinct ``base'' maps in which the GCE is entirely accounted for by DM (left two columns) or PSs (right two columns), onto which the additional DM signal is injected. The injected DM signal is Poisson fluctuated to generate 10 realizations for each base case. All of the simulated data maps are made using the \texttt{p6v11} diffuse model; to explore the effects of diffuse mismodeling, the second and fourth columns are analyzed using a Galactic diffuse template that is based on Model~F from Ref.~\cite{Calore:2014xka}. Importantly, when injecting the additional DM signal onto an existing GCE-strength DM signal, the analysis is well-behaved and there is no significant misattribution of flux between the DM and PS components, even in the presence of diffuse mismodeling. When injecting the additional signal onto an existing PS signal, there is confusion between the two components and the flux posteriors for the DM and PS templates become bimodal. Notably, in the absence of diffuse mismodeling, the amount by which the median PS flux posterior is shifted negative matches the true total flux contributed by ultrafaint sources below the $\sim1\sigma$ significance threshold. In the presence of diffuse mismodeling, the DM signal is typically absorbed by the PS template, as evidenced by the median DM posterior peaking at ``$-$True NFW DM" in the fourth column. These results clearly demonstrate that an artificial DM signal injected on the data may fail to be recovered by the NPTF if point sources are already present, and especially when the diffuse emission is mismodeled, thereby providing a simple possible explanation for the results presented in Ref.~\cite{Leane:2019xiy}. The corresponding plot for the hard source-count distribution is provided in Fig.~\ref{fig:signal_on_signal_hardSC}.}
\label{fig:signal_on_signal}
\end{figure*}

When the unresolved PSs have a hard flux distribution, as illustrated in Fig.~\ref{fig:25_75_fullpanel_hardSC}, these trends continue to hold, but are less pronounced.  In particular, the PS flux is never fully absorbed by the DM template, even in scenarios where the DM constitutes the majority of the GCE.  Such behavior makes sense as it is more difficult for a collection of bright sources to fake a Poissonian DM signal.

To summarize, we find that, in the absence of diffuse mismodeling: 
\begin{itemize}
    \item The NPTF analysis never misattributes DM as PSs in the case where the GCE is 100\% DM.  
    \item When the GCE is 100\% PSs, some fraction of the PS flux can be misattributed to DM.  This depends sensitively on the flux distribution of the PSs, and is exacerbated for cases where there are more ultra-faint sources.
    \item When the GCE flux is split between DM and PSs, it is possible that the emission is  misattributed between the two templates.  In particular, when the DM accounts for a minority of the emission, it can be entirely absorbed by the PS template.  As the DM fraction increases, this behavior reverses, with the DM template preferentially absorbing the PS contribution.  This effect is again exacerbated for the soft source-count function.
    \item In all instances where the GCE flux is split between DM and PSs, we find cases---regardless of whether the relative DM/PS flux contribution is 25/75, 50/50, or 75/25\%---where the entirety of the DM flux is misattributed to PSs.  This becomes increasingly rare as the relative DM contribution to the GCE increases, but can still occur.  We emphasize, however, that this never occurs when the GCE consists entirely of DM.
\end{itemize}
These results pertain specifically to the case where there is no mismodeling of the Galactic diffuse emission.  We will consider the implications of diffuse mismodeling in the following section.

\subsection{Signal Injection Tests on Monte Carlo}
\label{sec:siginjection}

The study performed by Ref.~\cite{Leane:2019xiy} showed that injecting an artificial DM signal on data results in the signal being misattributed to PSs in the NPTF analysis.  When interpreting results from such signal injection tests, it is important to consider potential subtleties that may arise from injecting an artificial DM signal into data that itself contains a signal---the GCE. We explore on simulated datasets the NPTF recovery of a GCE-strength DM signal injected on top of an existing GCE signal, in the cases where the GCE is entirely accounted for by DM or by NFW PSs. 

In the case where the GCE is 100\% DM, we choose 10 of the Monte Carlo realizations from Fig.~\ref{fig:100DM_0PS_fplot} as the base ``data" maps onto which we inject an additional GCE-strength DM signal. The base maps are chosen to bracket a range of the possibilities depicted in Fig.~\ref{fig:100DM_0PS_fplot}. We inject 10 different Monte Carlo realizations of the additional DM signal onto each of the base maps, resulting in a total of 100 composite maps. We show the results across the 100 maps in the first column of Fig.~\ref{fig:signal_on_signal}. In this case, the DM signal is never absorbed by the PS template. We note that analyzing 10 Monte Carlo iterations for each base map is sufficient, because the variations in the recovered fluxes are dominated by differences in the base maps themselves rather than differences in Poisson realizations of the injected signal.

The results are quite different in the case where the GCE is 100\% soft PSs. As we have already seen, a population of soft PSs can be more challenging to distinguish from DM or Galactic diffuse emission in Inner Galaxy.  This confusion can be exacerbated as the total Poissonian flux increases and becomes spatially correlated with the PS distribution. Indeed, this is precisely what we see on simulation, as shown in the third column of Fig.~\ref{fig:signal_on_signal}. Similarly to the previous case, we choose 10 base maps spanning a range of possibilities, and inject 10 separate Monte Carlo realizations of an additional DM signal onto of each of the base maps. There is considerable spread in the posterior distributions---in some cases, no DM is recovered and all the flux is entirely attributed to PSs; in other cases, a large fraction of the PS flux is attributed to DM. On average, there is a higher probability of the latter occurring. The bimodality of the NFW DM and NFW PS flux posteriors is exacerbated compared to the base case with no additional injected signal, shown in the first column of Fig.~\ref{fig:25_75_fullpanel}.  In that case, the recovered DM and PS fluxes are correct the majority of the time.  However, for the signal injection test, the recovered DM and PS fluxes are almost always incorrect. Additionally, the spread in signal-on-signal case is larger than that of Fig.~\ref{fig:25_75_fullpanel} (left panel)---note that the $x$-axes are different between the two figure panels.  The corresponding results for the hard PSs can be found in the left column of Fig.~\ref{fig:signal_on_signal_hardSC}.  In this scenario, the spread in the DM and PS posteriors is smaller---in particular, the PS flux is never absorbed by the DM template, while the DM flux still may be absorbed by the PS template.

\begin{figure*}[t]
\centering
\includegraphics[width=0.65\textwidth]{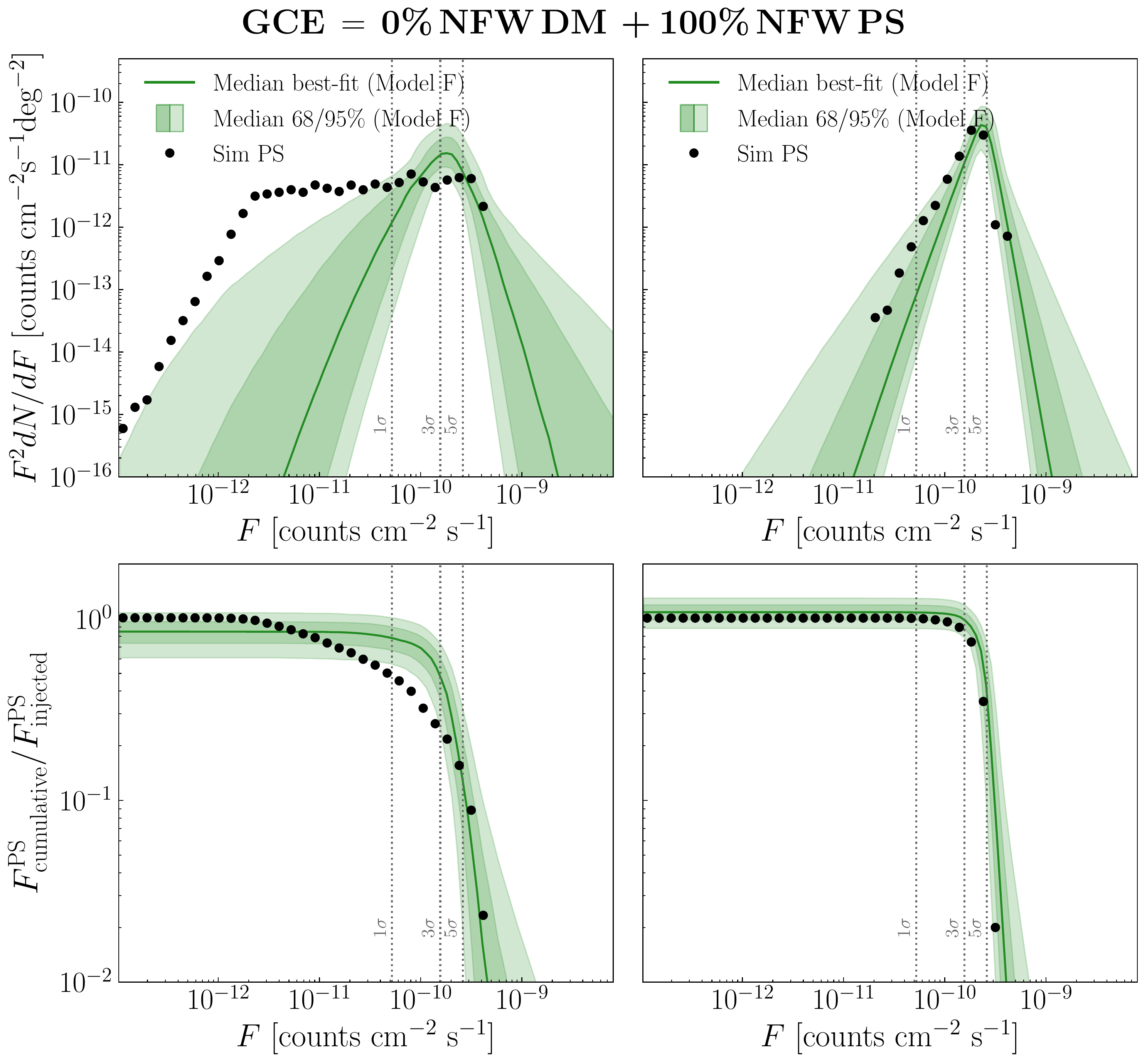} 
\caption{Same as the right-most panel of Fig.~\ref{fig:0DM100PS_fullpanel}, but analyzed using Model F rather than the \texttt{p6v11} model, which was used to generate the simulated data maps.  Results are shown for the soft PSs (left panels) as well as the hard PSs (right panels).  Even at the $\sim95$\% level over 100 Monte Carlo iterations, the recovered source-count function fails to capture the low-flux sources for the soft PSs. The cumulative flux distribution for the soft PSs shows that in the presence of diffuse mismodeling, the recovered flux is consistent with the true injected flux down to $\sim2\times 10^{-10}$~counts~cm$^{-2}$~s$^{-1}$, but is in excess for fluxes between $\sim 1$--$2\times 10^{-10}$~counts~cm$^{-2}$~s$^{-1}$. In the case of the hard source-count distribution, the recovered function reliably captures the input at the $\sim95$\% level. The individual flux posteriors for a random subset of 50 out of the 100 runs shown in the left panels are provided in Fig.~\ref{fig:0DM_100PS_fplot_modelF}.  For comparison, we also provide (in Fig.~\ref{fig:100DM_0PS_fplot_modelF}) the flux posteriors for the case where the GCE consists of 100\% DM and there is diffuse mismodeling. }
\label{fig:100PS_modelF}
\end{figure*}

The results presented in this section demonstrate that signal injection tests on the GCE can be biased, even when the NPTF works robustly on the actual underlying data with no artificial signals present. This bias can arise from the fact that the injected DM is degenerate with ultra-faint PSs. If a soft PS population is already present in the data, the fit may not be penalized by absorbing injected DM flux into the PS template.  Signal injection tests therefore yield less information than they would in the absence of this degeneracy.  Indeed, we see that when a DM signal is injected on a map that already contains a population of PSs at the Galactic Center, the NPTF may naturally absorb this injected signal into the PS template.  When the GCE consists entirely of DM, on the other hand, the signal injection test is well-behaved.

\section{Diffuse Mismodeling} 
\label{sec:mismodeling}

Thus far, the Poissonian templates included in the NPTF analyses have perfectly described the astrophysical backgrounds in the data (up to Poisson noise).  In a realistic setting however, the spatial morphology of the Galactic diffuse emission is rather poorly constrained.  As a result, the templates used to model the diffuse emission describe the actual underlying background with uncertainty far exceeding the level of Poisson noise. This raises the possibility of, \emph{e.g.}, spurious residuals in the data that could mimic a PS signal even when actual astrophysical sources, such as MSPs, are not present. Conversely, it could be possible for a PS signal to be absorbed into the mismodeled diffuse background, resulting in the true flux and source-count distribution not being properly reconstructed. 

In this section, we explore both these effects and comment on how recovery of DM and PS signals in the Inner Galaxy could be affected by mismodeling of the Galactic diffuse emission.  We mock up the effect of diffuse mismodeling by analyzing the same simulated data that we have used so far (created using the \emph{Fermi} \texttt{p6v11} diffuse model) with an alternate diffuse model. In particular, we use Model~F, which was found to be the best-fit to Inner Galaxy data out of the models considered in Ref.~\cite{Calore:2014xka}. We explore the impact of diffuse mismodeling on PS and DM recovery in turn.  

\subsection{Point Source Signal Recovery}
\label{sec:ps_mismodeling}

\begin{figure*}[t]
\centering
\includegraphics[width=0.9\textwidth]{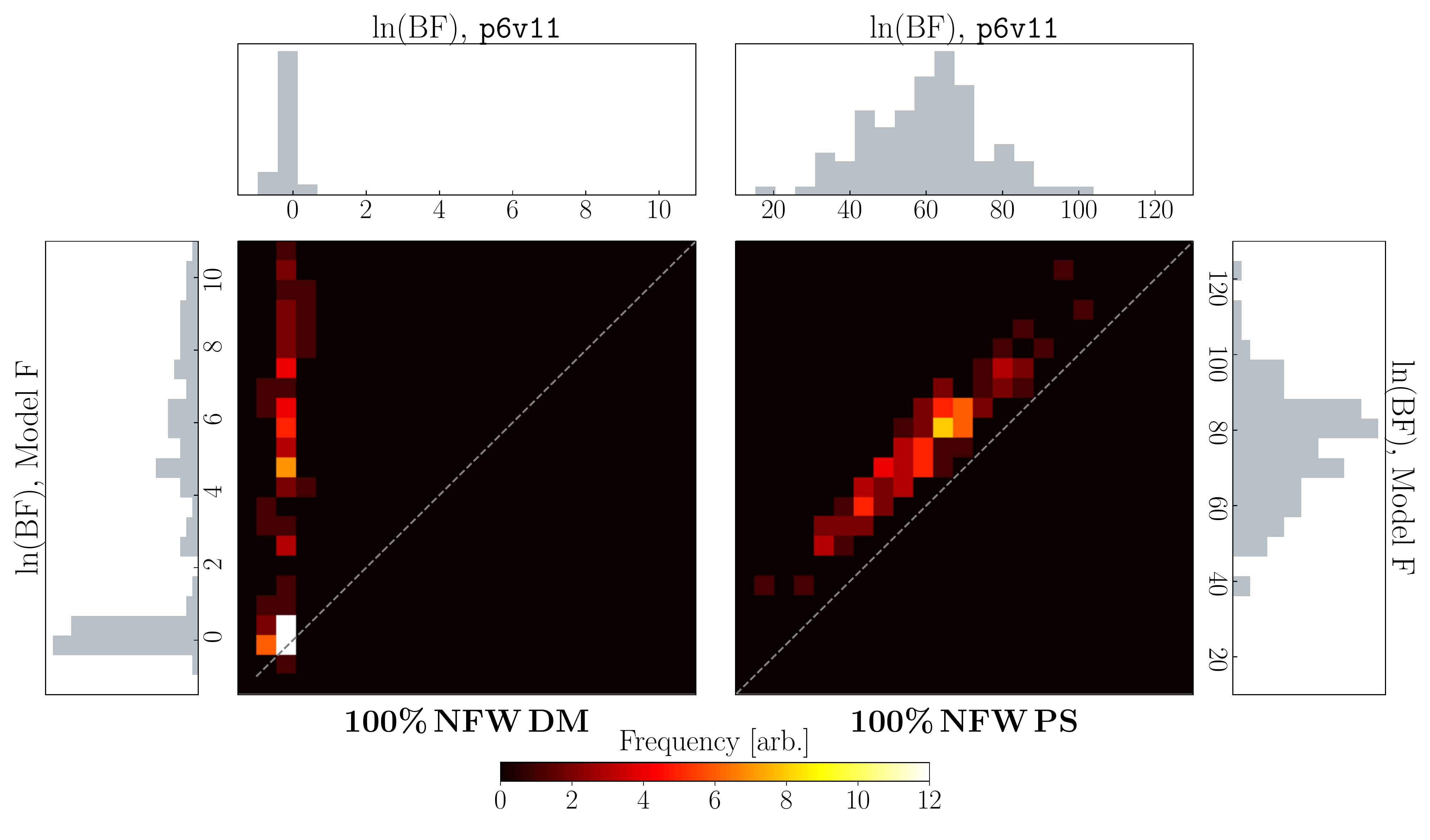} 
\caption{Bayes factors (BFs) characterizing the statistical preference for a model of the data that includes NFW PSs over a model that does not include them. The mock data consists of the GCE, accounted for by either 100\% NFW DM (left) or by 100\% NFW PS following the soft source-count distribution (right), and diffuse emission modeled by \texttt{p6v11}. Results are shown for 100 Monte Carlo iterations. We run the NPTF on these maps using three templates: \emph{(i)}~NFW PS, \emph{(ii)}~NFW DM, and \emph{(iii)}~Galactic diffuse emission.  The BFs recovered using the \texttt{p6v11} model for the diffuse template are shown along the horizontal axes, while the BFs recovered using the Model~F template are shown along the vertical axes. (Note the difference in scale for the axes between the left and right halves of the figure.)  Even in the presence of diffuse mismodeling, the NPTF robustly picks up evidence for PSs when they constitute 100\% of the GCE.  When the GCE is 100\% DM, the evidence for PSs is always negligible in the \texttt{p6v11} case; in the Model F case, while there is a strong peak around $\ln(\text{BF})\sim 0$, there are 35 iterations in which the diffuse mismodeling leads to residuals that are picked up as PSs with $5< \ln(\text{BF}) \lesssim 13$.  The significance of these detections is still smaller than the BF range when true PSs are actually present.  We emphasize that while the relative values of the BFs are useful in comparing the different tests studied here, their overall scale should not be compared to any results on \emph{Fermi} data as these maps are not intended to closely model an actual data realization.}
\label{fig:BF_dif}
\end{figure*}

\begin{figure*}[t]
\centering
\includegraphics[width=0.9\textwidth]{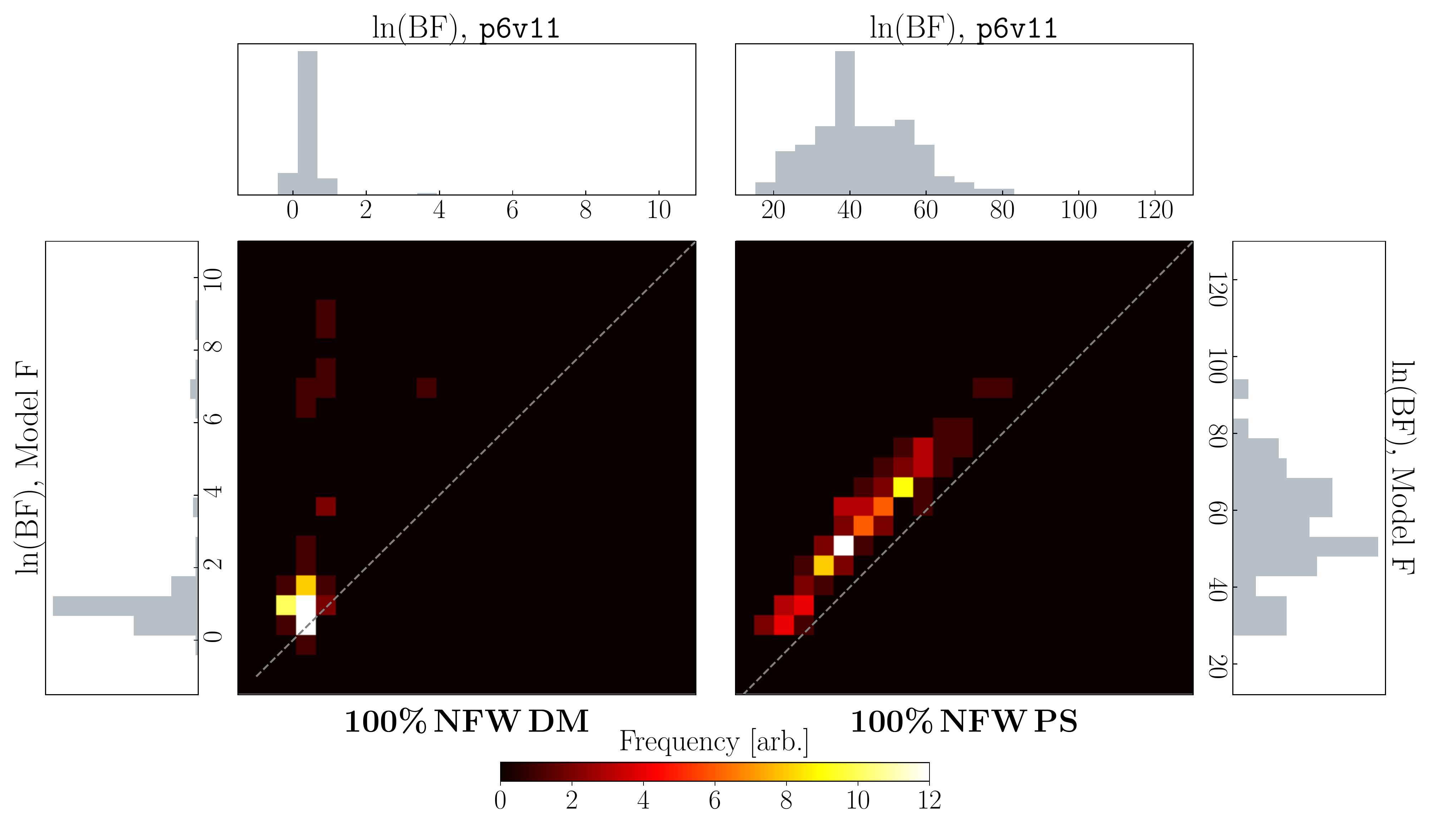} 
\caption{Same as Fig.~\ref{fig:BF_dif}, but where the mock data consists of, in addition to the DM(PS) signal: \texttt{p6v11} diffuse emission, emission from the \emph{Fermi} bubbles~\cite{2010ApJ...724.1044S}, isotropic emission, and emission from 3FGL sources~\cite{Acero:2015hja}.  The left(right) panel shows results where the GCE is 100\% DM(soft PSs).  The NPTF analysis now also includes templates to model the isotropic, 3FGL, and bubbles emission.  Compared to what we see in Fig.~\ref{fig:BF_dif}, the overall scale of the BFs is lower.  Additionally, in the 100\% DM case with diffuse mismodeling, the distribution is much more strongly peaked near a low value of $\ln (\text{BF})\sim 1$, and the number of iterations for which residuals are picked up as PSs with $5< \ln(\text{BF}) \lesssim 13$ is reduced to 7 (compared to 35 cases in Fig.~\ref{fig:BF_dif}). We emphasize again that while the relative values of the BFs are useful in comparing the different tests studied here, their overall scale should not be compared to any results on \emph{Fermi} data as the simple setup is not intended to accurately model the analysis on real data.}
\label{fig:BF_all_bkg}
\end{figure*}

We redo the analysis presented in Sec.~\ref{sec:anatomy}, but now use diffuse Model F to analyze the simulated maps.  To build the Model~F template, we obtain the best-fit normalizations of the gas (Bremsstrahlung and $\pi^0$ decay) and Inverse Compton components on data, and then sum them together.  Therefore, the Model~F template is associated with a single fit parameter (\emph{i.e.}, its overall normalization), just like the \texttt{p6v11} template that we used previously.  The primary differences between the two lie in the specific assumptions made regarding the gas and IC models, as described in Ref.~\cite{Calore:2014xka} (see also Ref.~\cite{Ackermann:2012pya}).  

The results are presented in Fig.~\ref{fig:100PS_modelF}, in analogy to the right-most panel of Fig.~\ref{fig:0DM100PS_fullpanel}.  For the soft PSs (left panels), the lower-flux biasing effects seen in Sec.~\ref{sec:anatomy} are exacerbated in the presence of diffuse mismodeling, leading in general to a steeper downturn in the source-count function towards lower fluxes. Additionally, an excess in the recovered PS flux is observed in the flux regime of $\sim$1--2$\times10^{-10}$~counts~cm$^{-2}$~s$^{-1}$ as a consequence of the clumpy residuals present due to mismodeling. In the case of the hard PSs (right panels), the PS recovery is unaffected by the presence of the diffuse mismodeling and is accurate, to within 95\% confidence, within the entire flux range. We note that in the absence of diffuse mismodeling, the median recovered source-count distribution is accurate over the full flux range.  Interestingly, the recovered source-count functions in the left and right panels are remarkably similar.  These results demonstrate that diffuse mismodeling can make a genuinely soft population of sources ``fake" a harder population as the diffuse residuals can mimic bright PSs.  This could provide one explanation for why the best-fit source-count function recovered by the NPTF in the Inner Galaxy~\cite{Lee:2015fea}---which is similar to the hard source-count function modeled here---differs from \emph{e.g.} the MSP expectation.

Despite mischaracterizing the soft source-count distribution, the preference for a PS population remains robust in the face of mismodeling. This is quantified in the right half of Fig.~\ref{fig:BF_dif}, which shows the distribution of Bayes factors (BFs) in preference of a model including NFW PSs over a model without them, for 100 Monte Carlo realizations. This is illustrated as a heatmap, with the BFs along the horizontal axis corresponding to those obtained when the simulations are analyzed with the ``correct'' diffuse model (\texttt{p6v11}), and those along the vertical axis corresponding to analyses with the alternative diffuse model (Model F). Projected distributions are shown along both axes.  Note that the overall scale of these BFs should not be compared directly to any results on actual \emph{Fermi} data, as the setup here is very simple and is not intended to accurately represent the real data. The main take-away from these figures is the relative differences in BFs for the tests presented in this section.

Preference for a PS population remains robust in either case, with $\ln(\mathrm{BF})\gtrsim 20$.  A stronger preference for PSs is seen in the case of diffuse mismodeling, as additional residuals are also picked up by the NFW PS template. There is a tight correlation between the BFs obtained from the \texttt{p6v11} and Model F analyses, bolstering the fact that, in the 100\% PS scenario, the preference for PSs comes from the true underlying PS population rather than as a consequence of the diffuse mismodeling on its own.

The analogous results with the addition of other astrophysical background components (resolved 3FGL points sources, isotropic emission and emission from the \emph{Fermi} bubbles) and corresponding Poissonian templates are shown in the right half of Fig.~\ref{fig:BF_all_bkg}. The same overall conclusion is seen to hold, with the typical BFs in preference for a model with PSs now being somewhat tempered, as expected in the presence of additional smooth background emission. We emphasize once more that the overall scale of these BFs should not be compared directly to any results on actual \emph{Fermi} data, as we are not accounting for additional PS populations that may be present in the real data (such as disk-correlated sources), and the degree of mismodeling we study here could be different than that in an analysis on the real data.

We conclude from these tests that, with the degree of diffuse mismodeling that we have considered (\texttt{p6v11} vs Model F), it is unlikely that a true underlying PS population would be mischaracterized as DM.

To gain a sense of how comparable the degree of mismodeling studied here is to that from a typical analysis on real data, we compare the residuals from our Monte Carlo analyses (created with diffuse model \texttt{p6v11} and analyzed with diffuse Model F) to the residuals from fitting the \texttt{p6v11} or Model F diffuse templates to data. In all cases, we include Poissonian templates to account for emission from the $\emph{Fermi}$ bubbles~\cite{2010ApJ...724.1044S}, isotropic emission, as well as emission from the resolved 3FGL point sources~\cite{Acero:2015hja}. We show residual sky maps and a histogram of residual counts in App.~\ref{sec:residuals}. In our region of interest, the median and $16^\mathrm{th}/84^\mathrm{th}$ percentile magnitudes of residuals are (in photon counts per pixel): $2.40^{+3.07}_{-1.70}$ for \texttt{p6v11} fit to data, $2.38^{+3.00}_{-1.68}$ for Model F fit to data, and $2.27^{+2.80}_{-1.59}$ (median values over 100 realizations) for Model F fit to \texttt{p6v11} simulated data, respectively. The degree of mismodeling we simulate is slightly less than but roughly commensurate with that on real data, which suggests that our simulated mismodeling provides a reasonable proxy. However, we emphasize that the magnitude of residuals gives only a rough comparison, and that there will be additional differences in mismodeling on the real data versus on our simulated data. For example, the spatial distribution of the residuals could be very different, which could have implications on the results of the NPTF analysis.

\subsection{Dark Matter Signal Recovery}
\label{sec:dm_mismodeling}

The effect of mismodeling on the recovery of a DM signal can be somewhat more subtle. In particular, whether DM recovery is successful or not can be strongly affected by the specifics of a given Poisson realization. This is expected, since the diffuse emission accounts for a large fraction of the flux in the Inner Galaxy, and large Poisson fluctuations combined with the effects of mismodeling can ``fake'' PS-like statistics. 

\begin{figure*}[t]
\centering
\includegraphics[width=0.65\textwidth]{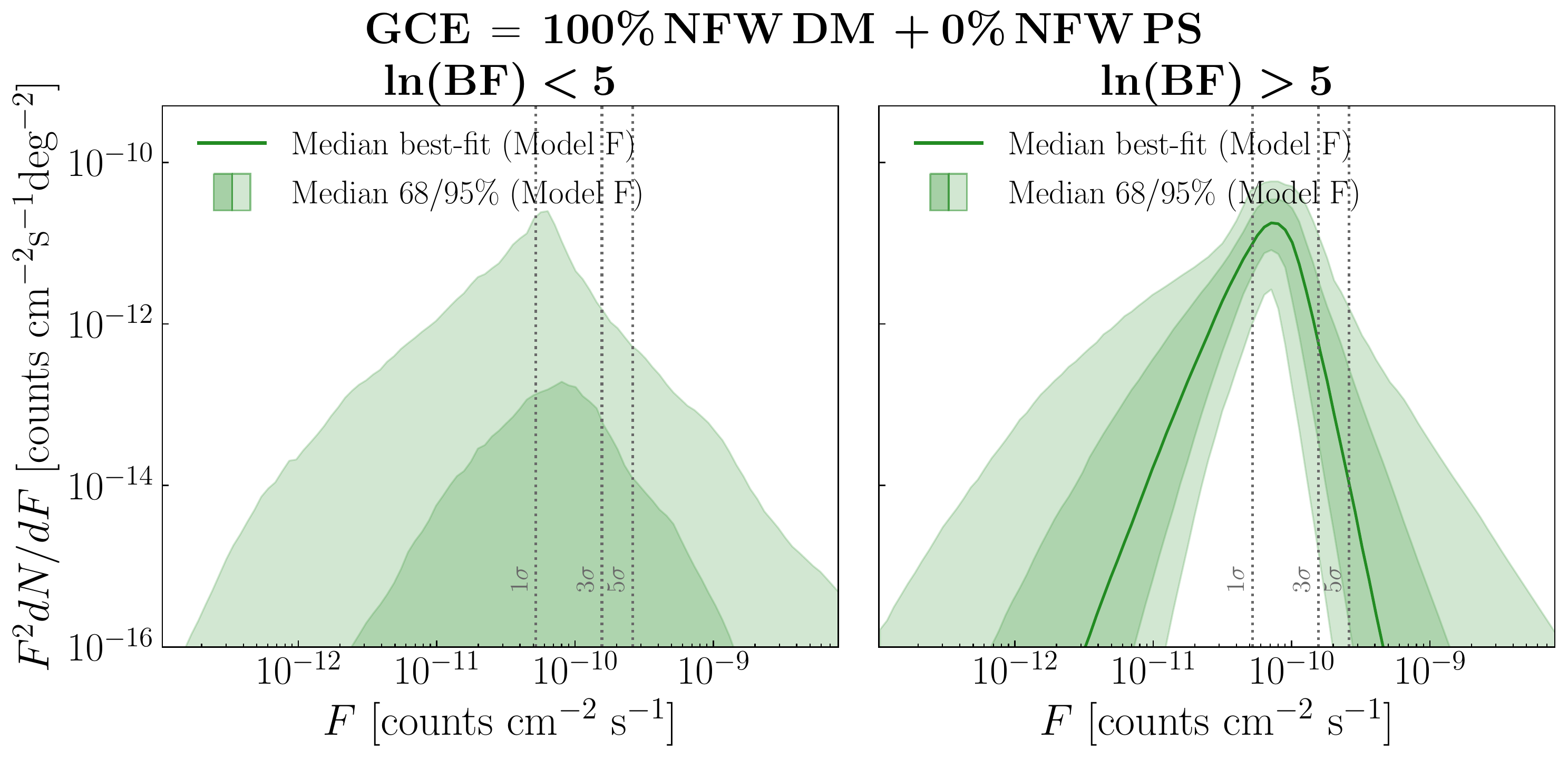} 
\caption{Differential source-count distributions in the presence of diffuse mismodeling. The simulated data consists of the GCE, which is entirely accounted for by DM, and Galactic diffuse emission (corresponding to the left panel of Fig.~\ref{fig:BF_dif}). We run the NPTF on these maps using three templates: \emph{(i)} NFW PS, \emph{(ii)} NFW DM, and \emph{(iii)} Galactic diffuse emission. The simulated maps are generated using the \texttt{p6v11} model and analyzed using Model F. We separately show the results for iterations with $\ln(\text{BF})<5$ (left panel) and iterations with $\ln(\text{BF})>5$ (right panel). In each case, the solid green line is the median best-fit distribution, and the bands denote the median 68/95\% confidence intervals, recovered over 100 Monte Carlo realizations. In the cases with $\ln(\text{BF})<5$, the typical best-fit source-count function is suppressed relative to the examples we have considered thus far, and does not look like the source-count function recovered in the NPTF analysis on \emph{Fermi} data~\cite{Lee:2015fea}; in the outlying cases with $\ln(\text{BF})>5$, which correspond to the tail of the distribution shown in the leftmost panel of Fig.~\ref{fig:BF_dif}, the median recovered source-count function resembles the hard PS population in the left panel of Fig.~\ref{fig:0DM100PS}. Note that when adding in additional Poissonian contributions to the mock data, as in Fig.~\ref{fig:BF_all_bkg}, the number of instances where $\ln(\text{BF})>5$ are significantly reduced.}
\label{fig:DM_SC}
\end{figure*}

Similarly to Sec.~\ref{sec:DMrecovery}, we consider the case where the GCE consists of 100\% DM. The corresponding BFs in preference of a model with PSs over a model without them are shown in the left halves of Fig.~\ref{fig:BF_dif} (with only diffuse background emission) and Fig.~\ref{fig:BF_all_bkg} (with additional Poissonian background components).  When the simulations are analyzed with the ``correct'' \texttt{p6v11} diffuse model, the BFs are always small, peaking around $\ln(\text{BF})\sim 0$ and never showing significant preference for a PS population. When the diffuse emission is mismodeled, the $\ln(\text{BF})$ still peaks near 0, but there is a tail in the distribution extending up to $\ln(\text{BF})\sim 10$. These cases correspond to realizations where mismodeled residuals conspire to mimic a PS-like population in the data.  The number of instances where the 100\% DM case yields a $\ln(\text{BF}) \gtrsim 5$ in the presence of diffuse mismodeling is reduced when additional backgrounds are included in the map. This reduction may be due to the presence of additional Poissonian components in the model that may absorb the residuals. 

Figure~\ref{fig:DM_SC} shows the differential source-count distributions for the NFW~PS template in simulations where the DM accounts for 100\% of the GCE flux (no PS contribution), and there is diffuse mismodeling. Note that, for this example, we use the simulated data corresponding to Fig.~\ref{fig:BF_dif} as opposed to Fig.~\ref{fig:BF_all_bkg} simply because there are more runs with $\ln\text{(BF)} > 5$; as we have seen, the inclusion of the additional Poissonian backgrounds decreases the frequency of such anomalously large BFs.  In each panel of Fig.~\ref{fig:DM_SC}, the solid green line is the median best-fit distribution recovered over 100 Monte Carlo realizations, and the bands denote the median 68/95\% confidence intervals. As there are no PSs in the simulated map, the NFW~PS template should not pick up significant flux, which is confirmed by the peak near $\ln\text{(BF)}\sim1$ in Fig.~\ref{fig:BF_dif}. We show in the left panel that correspondingly, for iterations with $\ln\text{(BF)}<5$, the average best-fit source-count distribution is suppressed relative to the examples we have considered thus far, and certainly does not look like the source-count function recovered in the NPTF analysis on \emph{Fermi} data~\cite{Lee:2015fea}. The 95\% containment band, however, does encompass distributions that resemble the hard PS population shown in the left panel of Fig.~\ref{fig:0DM100PS}. In the right panel, we show the anomalous cases where the BF in preference for PSs falls in the range $\ln(\text{BF}) \sim$ 5--13. In this case, the typical recovered source-count distribution does resemble that of the hard PS population. 

As we have shown, there are some instances where a DM signal can be mischaracterized as a PS population when the diffuse emission is mismodeled.  While this only happens for a subset of realizations when diffuse Model F is used to analyze a map created using the \texttt{p6v11} diffuse model, it is plausible that a differences in mismodeling from what we have considered could lead to more consistent mischaracterization of the DM signal.  It is therefore important when performing the NPTF analysis on the \emph{Fermi} data to vary over the template(s) for the Galactic diffuse emission. In addition, it is crucial to compare the obtained BFs with those expected from corresponding simulations. This is because for different diffuse models (with potentially different degrees of freedom), the interplay between the diffuse model with other components could lead to different expected BFs for the same underlying PS population. If the set of diffuse models span the range of viable possibilities, then it is plausible that the recovered BFs would more consistently favor the PS interpration across different models if PSs are in fact present in the data, compared to the scenario in which the GCE is truly DM---in that case, there would likely be more variation in the recovered BFs because the results would be more sensitive to the specifics of the residuals from mismodeling.  A version of this test was performed in the original NPTF study of the GCE~\cite{Lee:2015fea}, and a preference for PSs was consistently recovered over the different diffuse models studied.  An updated analysis focusing on mitigating the effects of diffuse mismodeling in the NPTF procedure in preparation~\cite{companion}, and is summarized more fully in the Conclusions.

\subsection{Signal Injection Tests}

\begin{figure*}[t]
\includegraphics[width=0.65\textwidth]{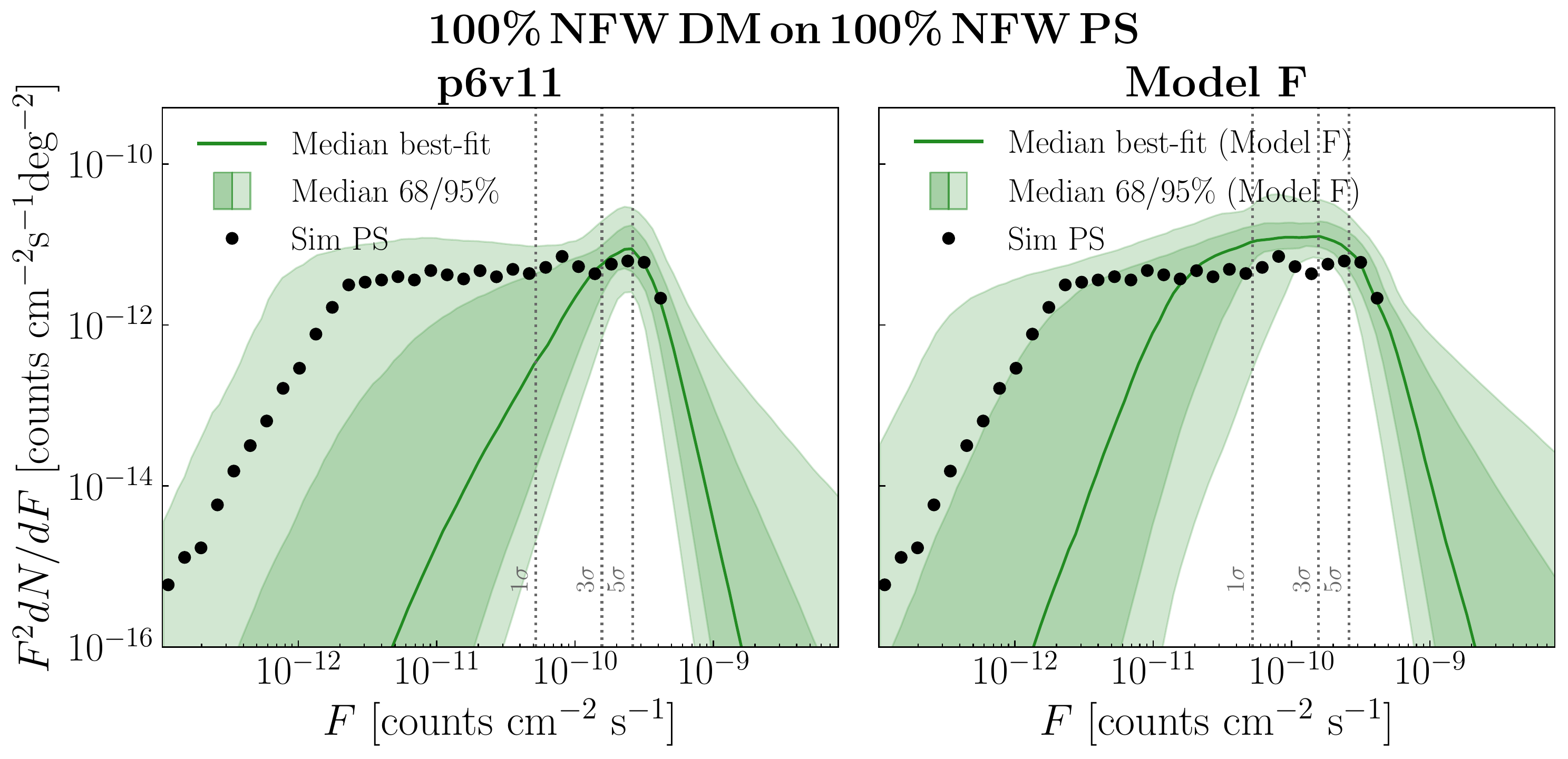} 
\caption{Differential source-count distributions recovered when the simulated data consists of the GCE, comprised entirely of soft PSs, with an additional injected GCE-strength DM signal. We run the NPTF on these maps using three templates: \emph{(i)}~NFW PS, \emph{(ii)}~NFW DM, and \emph{(iii)}~Galactic diffuse emission. The \texttt{p6v11} model is used to generate the Galactic diffuse emission in the simulated data, and the analysis is performed using the ``correct" diffuse model (\texttt{p6v11}, left panel) or the ``incorrect" diffuse model (Model F, right panel). In each case, the solid green line is the median best-fit distribution, and the bands denote the median 68/95\% confidence intervals, recovered over 100 Monte Carlo realizations. The left(right) panel corresponds to the third(fourth) column of Fig.~\ref{fig:signal_on_signal}. These results can be compared with the cases without the additional DM injection, shown in the top right panel of  Fig.~\ref{fig:0DM100PS_fullpanel} for \texttt{p6v11} and the top left panel of Fig.~\ref{fig:100PS_modelF} for Model F. In the analysis with diffuse model \texttt{p6v11}, a significant portion of the PS flux is typically absorbed by the DM template (correspondingly, the source-count distribution is suppressed in the $\sim5\times10^{-11}$--$2\times10^{-10}$~counts~cm$^{-2}$~s$^{-1}$ flux range, relative to Fig.~\ref{fig:0DM100PS_fullpanel}, top right panel). This is similar to the observed misattribution of injected DM flux to PSs in Ref.~\cite{Leane:2019xiy}. On the other hand, when the data is analyzed with the diffuse Model F, the injected DM is consistently absorbed into the PS model (correspondingly, the source-count distribution captures excess flux in the $\sim10^{-11}$--$10^{-10}$~counts~cm$^{-2}$~s$^{-1}$ range, relative to Fig.~\ref{fig:100PS_modelF}, top left panel).}
\label{fig:DM_on_PS_p6_modelF}
\end{figure*}

Lastly, we revisit the signal injection tests discussed in Sec.~\ref{sec:siginjection}, in the presence of diffuse mismodeling. Like before, we consider the cases where we have injected an additional DM signal (with flux equivalent to the GCE) onto simulated data maps in which the GCE is comprised  entirely of either DM or soft PSs.  The \texttt{p6v11} model was used to generate the Galactic emission in the simulated maps, but we now repeat the NPTF analysis using Model~F for the diffuse template instead.  The flux posteriors for 100 Monte Carlo iterations are shown in the second and fourth columns of Fig.~\ref{fig:signal_on_signal}.

When the GCE consists of 100\% DM, the NPTF almost always recovers the correct DM flux (up to a small offset between the diffuse and DM components), including the injected contribution~(second column, Fig.~\ref{fig:signal_on_signal}).  Additionally, the analysis finds on average that there are no PSs.  This result clearly demonstrates that the additional injected DM signal is recovered when there are no PSs present in the data, even if the diffuse emission is mismodeled to the extent that we consider.

In contrast, when the artificial DM signal is injected onto a map where the GCE consists of soft PSs, there is more variation in the results. In particular, when the diffuse emission is accurately modeled, the NPTF on average recovers all of the injected DM flux and additionally absorbs the total flux contributed by ultrafaint PSs (below the $\sim1\sigma$ significance threshold) into the DM template (third column, Fig.~\ref{fig:signal_on_signal}). On the other hand, when the diffuse emission is mismodeled, the NPTF consistently absorbs the injected DM flux into the PS template (fourth column, Fig.~\ref{fig:signal_on_signal}). When the PS population is hard, as shown in Fig.~\ref{fig:signal_on_signal_hardSC}, the injected DM is almost always reconstructed as PSs, especially when analyzed using the Model~F template. These tests demonstrate that the additional DM photons that are injected into the map can conspire with the residuals from diffuse mismodeling to look like a population of PSs.  As a result, the signal injection test fails and the injected DM flux is not correctly reconstructed.

This point is further emphasized in Fig.~\ref{fig:DM_on_PS_p6_modelF}, which shows the differential source-count distributions corresponding to the third and fourth columns of Fig.~\ref{fig:signal_on_signal}. The left panel shows the result using the ``correct'' \texttt{p6v11} diffuse model, and the right panel shows the result using the ``incorrect'' diffuse Model F. These can be directly compared to the recovered source-count distributions in the absence of the additional DM injection, shown in Fig.~\ref{fig:0DM100PS_fullpanel} (top right panel) for \texttt{p6v11} and Fig.~\ref{fig:100PS_modelF} (top left panel) for Model F. It can be seen that, when the diffuse emission is mismodeled using Model F, the injected DM is consistently absorbed by the PS model---evident from the fact that the analysis consistently recovers more PS flux in the $\sim10^{-11}$--$10^{-10}$~counts~cm$^{-2}$~s$^{-1}$ range, compared to the case without the injected DM signal. This is similar to the observed misattribution of injected DM flux to the PS model in Ref.~\cite{Leane:2019xiy}, and is in contrast to the analysis with the ``correct" \texttt{p6v11} diffuse model, where on average, a significant portion of the PS flux gets absorbed by the DM template---correspondingly, the recovered source-count distribution is suppressed in the $\sim5\times10^{-11}$--$2\times10^{-10}$~counts~cm$^{-2}$~s$^{-1}$ flux range, compared to the case without the injected DM signal.

Similarly to~\cite{Leane:2019xiy}, we have also explored the effect of injecting even brighter DM signals (200\% and 300\% of the GCE flux) onto soft PSs constituting the GCE. These results are presented in Fig.~\ref{fig:more_signal_on_signal}. In the absence of diffuse mismodeling (first and third columns), the injection of brighter DM signals mitigates the bimodality of the posterior flux distributions, but the typical results are largely unchanged from the case where the injected DM is 100\% of the GCE flux (third column, Fig.~\ref{fig:signal_on_signal}). When diffuse mismodeling is present (second and fourth columns), the injection of increasingly bright DM signals reduces the probability that the injected DM signal is absorbed by the PS template in the NPTF analysis.  The effects of diffuse mismodeling on signal injection tests are likely to be further exacerbated when analyzing the real data. Understanding how to mitigate these effects on the data warrants a dedicated study, which we will present separately in a companion paper~\cite{companion}.

We emphasize that whether or not the signal injection test fails has no bearing on the validity of the NPTF analysis on the original dataset.  Indeed, we find that the signal injection test fails most spectacularly when PSs are already present in the simulated data and---as we see in Figs.~\ref{fig:BF_dif} and~\ref{fig:BF_all_bkg}---the NPTF analysis finds strong preference for PSs in these cases (with no injected DM), as it should.

\section{Conclusions}
\label{sec:conclusions}

In this paper, we performed a systematic study of the Non-Poissonian Template Fitting (NPTF) method on Monte Carlo data, focusing on its ability to distinguish between the DM and PS origins of the \emph{Fermi} Galactic Center Excess (GCE).  Our primary conclusions are as follows:
\begin{itemize}
\item When the Galactic diffuse backgrounds are perfectly modeled, the NPTF  accurately identifies a GCE that is comprised entirely of DM.  If the GCE is 100\% PSs, then some of the PS flux may be misattributed to DM.  When the GCE is split between DM and PSs, then the NPTF can struggle to identify the correct fluxes of each, especially when the PSs are relatively soft.  These challenges arise from the fact that PSs are exactly degenerate with smooth Poissonian emission in the ultra-faint limit.  
\item Assuming no diffuse mismodeling, we find that, when the GCE is 100\% PSs, the source-count distribution recovered by the NPTF in the Inner Galaxy accurately characterizes the underlying flux distribution of PSs down to a per-source significance of $\sim1\sigma$ while being potentially biased at lower fluxes.  This bias is especially true when the PS population is characterized by a soft source-count distribution with a large number of ultra-faint PSs. 
\item Evidence for a PS population can still be robustly recovered when the Galactic diffuse emission is mismodeled, at least for the one particular (albeit representative) case we considered.  However, the  residuals from the mismodeling can make the recovered source-count function appear to be brighter than it truly is.  This may suggest that the best-fit source-count function in the Inner Galaxy NPTF analysis of Ref.~\cite{Lee:2015fea} is not necessarily indicative of the true distribution for the underlying PS population.  This may potentially explain why the empirical distribution, which is peaked close to the 3FGL threshold, does not resemble models of the MSP luminosity function.  
\item In the presence of the diffuse mismodeling we consider, the NPTF almost always correctly identifies a GCE consisting entirely of DM.  Correspondingly, the Bayes factors in preference for PSs are typically not significant and the recovered source-count functions for the NFW~PS template are suppressed.  In a small fraction of cases, however, we do find that the NPTF can erroneously show evidence for PSs.  This preference is never as strong as what we find when PSs are truly present, and is particularly sensitive to the choice of diffuse template (and corresponding residuals) used in the study.  One way to test that such effects are not driving the preference for PSs on data is to simply rerun the NPTF analysis using a variety of different diffuse templates. In doing so, it is also important to compare the inferred Bayes factors to their expected values from simulation.   
\end{itemize}

Our work also allows us to comment on the results presented in Ref.~\cite{Leane:2019xiy},
which found that the NPTF can misattribute an artificial DM signal injected onto the \emph{Fermi} data to PSs.  We have mocked up such signal injection tests on simulated data maps, injecting an additional DM signal (with the same flux as the GCE) onto maps where the GCE is comprised entirely of DM or of hard/soft PSs.  We conclude that, at least to the extent we have tested: 
\begin{itemize}
\item When an additional DM signal is injected onto a map where the GCE is comprised entirely of DM, the NPTF correctly reconstructs the total (original + injected) DM flux.  This remains true in the majority of iterations even when the Galactic diffuse emission is mismodeled. 
\item When an additional DM signal is injected onto a map where the GCE is comprised entirely of PSs, there is often confusion between the DM and PS components. This can arise from the fact that the DM signal is degenerate with the ultra-faint PSs. In particular, in the presence of diffuse mismodeling, the injected DM signal is consistently reconstructed as PS flux.
\item When an additional DM signal, 2--3 times brighter than the GCE, is injected onto a map where the GCE is comprised entirely of PSs, the confusion between the DM and PS components can be somewhat mitigated. In particular, in the presence of diffuse mismodeling, the probability that the injected DM signal gets reconstructed as PS flux is reduced as the brightness of the injected DM signal is increased.
\end{itemize}

Our results demonstrate that the failure of the NPTF to extract an injected DM signal can be natural in the presence of PSs in the data, particularly in the presence of diffuse mismodeling.  Additionally, whether the signal injection tests succeed or fail is not an accurate diagnostic of the NPTF analysis on the original dataset (without the injected signal).  Indeed, we find that in cases where the signal injection tests fail, the NPTF accurately recovers the PSs on the original dataset.  Our findings demonstrate that great care must be taken when interpreting the results of DM signal injection tests on data.   

Throughout this work, we have only considered the case where the DM and PSs both trace the NFW profile, as opposed to different spatial distributions.  Ref.~\cite{Leane:2019xiy} considered a  scenario where there is a population of unresolved PSs that trace the \emph{Fermi} bubbles---a proof-of-principle example as there is no evidence for such sources in data.  However, in such cases where the DM and PSs follow different spatial morphologies, there are additional handles that may be used to discriminate the DM and PS hypotheses, such as simply changing the ROI.  Such possibilities will be addressed in more detail in a companion paper~\cite{companion}.    

In conclusion, this paper provides a systematic assessment of the NPTF method on simulated data.  Taking a pedagogical approach, we highlight the cases where the NPTF method works robustly, and cases where the output may be biased.  These results provide important context for interpreting the results of the  NPTF studies on actual data. In a companion paper~\cite{companion}, we will revisit the NPTF analysis of the Inner Galaxy in the \emph{Fermi} data, exploring how the results vary with the region of study as well as with the choice of diffuse emission model. We will also present a novel method that uses a spherical harmonic decomposition of the diffuse model to help lessen the effects of large-scale mismodeling.  Taken together with the discussion presented in this paper, these tests are designed to reduce the systematic uncertainties and biases associated with the NPTF analysis and strengthen the conclusions drawn from such studies on data.

\section{Acknowledgements}
\label{sec:acknowledgements}
We thank P.~Fox, R.~Leane, S.~Murgia, K.~Perez, T.~Slatyer, T.~Tait, K.~Van~Tilburg, and N.~Weiner for useful conversations. LJC is supported by a Paul \& Daisy Soros Fellowship and an NSF Graduate Research Fellowship under Grant Number DGE-1656466. ML is supported by the DOE under Award Number DESC0007968 and the Cottrell Scholar Program through the Research Corporation for Science Advancement. SM is partially supported by the NSF CAREER grant PHY-1554858 and NSF grant PHY-1620727. NLR is supported by the Miller Institute for Basic Research in Science at the University of California, Berkeley. MB and BRS are supported by the DOE Early Career Grant DESC0019225. This work was performed in part at the Aspen Center for Physics, which is supported by National Science Foundation grant PHY-1607611.  The work presented in this paper was performed on computational resources managed and supported by Princeton Research Computing, a consortium of groups including the Princeton Institute for Computational Science and Engineering (PICSciE) and the Office of Information Technology's High Performance Computing Center and Visualization Laboratory at Princeton University. This research made use of the  \texttt{healpy}~\cite{2005ApJ...622..759G}, \texttt{IPython}~\cite{PER-GRA:2007}, \texttt{matplotlib}~\cite{Hunter:2007}, \texttt{mpmath}~\cite{mpmath}, \texttt{NumPy}~\cite{numpy:2011}, \texttt{pandas}~\cite{pandas:2010}, \texttt{SciPy}~\cite{Jones:2001ab}, and \texttt{corner}~\cite{corner} software packages.

\bibliographystyle{apsrev4-1}
\bibliography{nptf_systematics}

\clearpage
\appendix
\onecolumngrid

\renewcommand{\thefigure}{A\arabic{figure}}
\renewcommand{\theHfigure}{A\arabic{figure}}
\setcounter{equation}{0}
\setcounter{figure}{0}
\setcounter{table}{0}
\setcounter{section}{0}
\makeatletter
\renewcommand{\theequation}{A\arabic{equation}}
\renewcommand{\thefigure}{A\arabic{figure}}
\renewcommand{\thetable}{A\arabic{table}}

\section{Supplementary Figures}
\label{sec:supplementary_figs}
\begin{figure*}[h]
\centering{}
\includegraphics[width=0.9\textwidth]{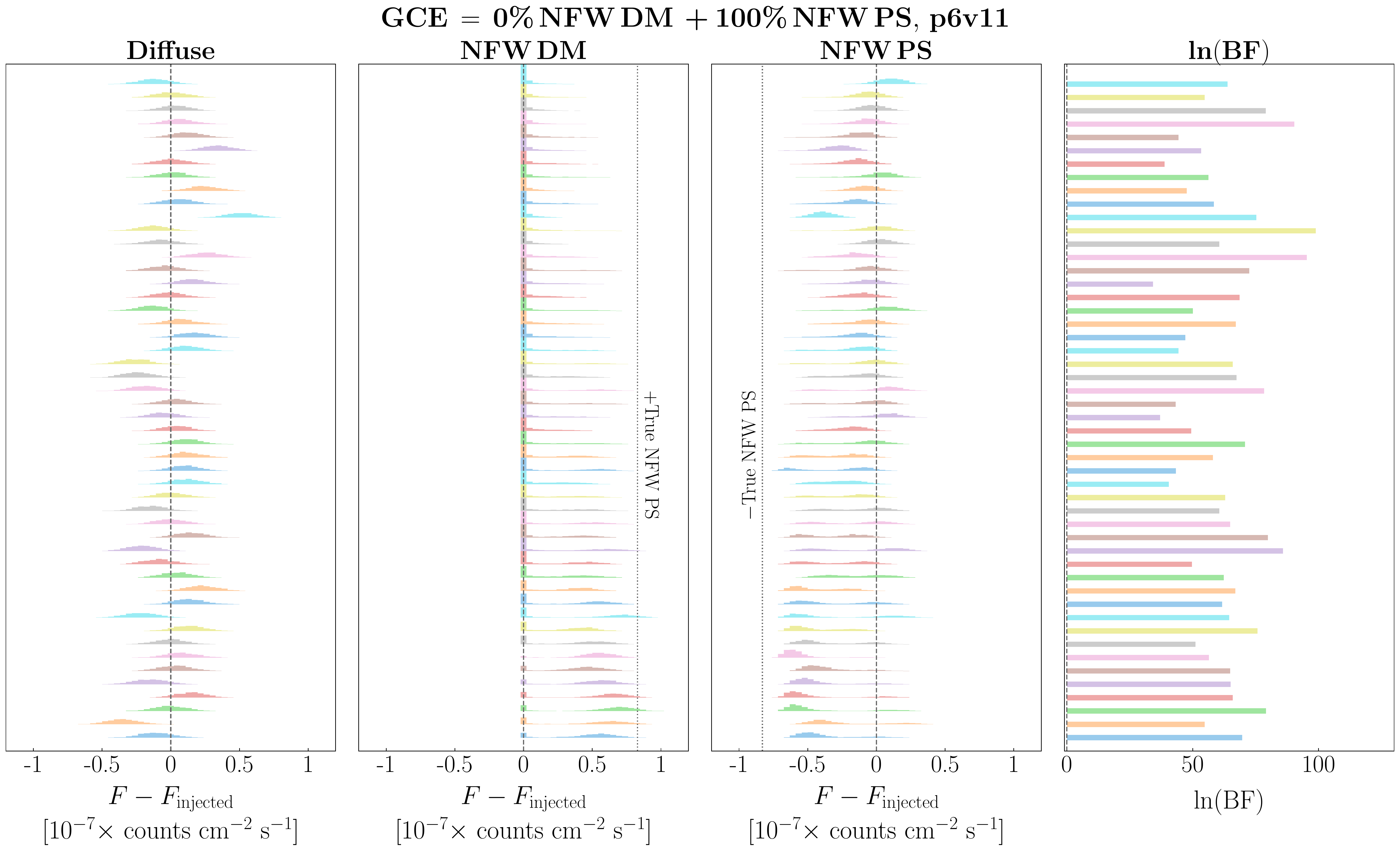}
\caption{Comparison of the flux posterior (relative to the true injected value) for the Galactic diffuse, NFW DM, and NFW PS components. The last column shows the Bayes factors (BFs), characterizing the statistical preference of a model of the data that includes NFW PSs over a model that does not include them. These results pertain to maps where 100\% of the GCE is accounted for by soft PSs (corresponding to the right-most panel of Fig.~\ref{fig:0DM100PS_fullpanel}).  We run the NPTF on these maps using three templates: \emph{(i)} NFW PS, \emph{(ii)} NFW DM, and \emph{(iii)} Galactic diffuse emission (\texttt{p6v11} model).  Each row in the figure represents a different Monte Carlo iteration of the map. These are a random subset of 50 out of the 100 iterations shown in the left half of Fig.~\ref{fig:BF_dif}.  In cases where the PS flux is underestimated, it is typically picked up by the NFW~DM template. Even when the PS flux is underestimated, decisive evidence for a PS population is still seen based on the Bayes factors. We emphasize that the overall scale of the BFs should not be compared to any results on \emph{Fermi} data as these maps are not intended to closely model an actual data realization.}
\label{fig:0DM_100PS_fplot}
\end{figure*}

\begin{figure*}[h]
\centering{}
\includegraphics[width=0.9\textwidth]{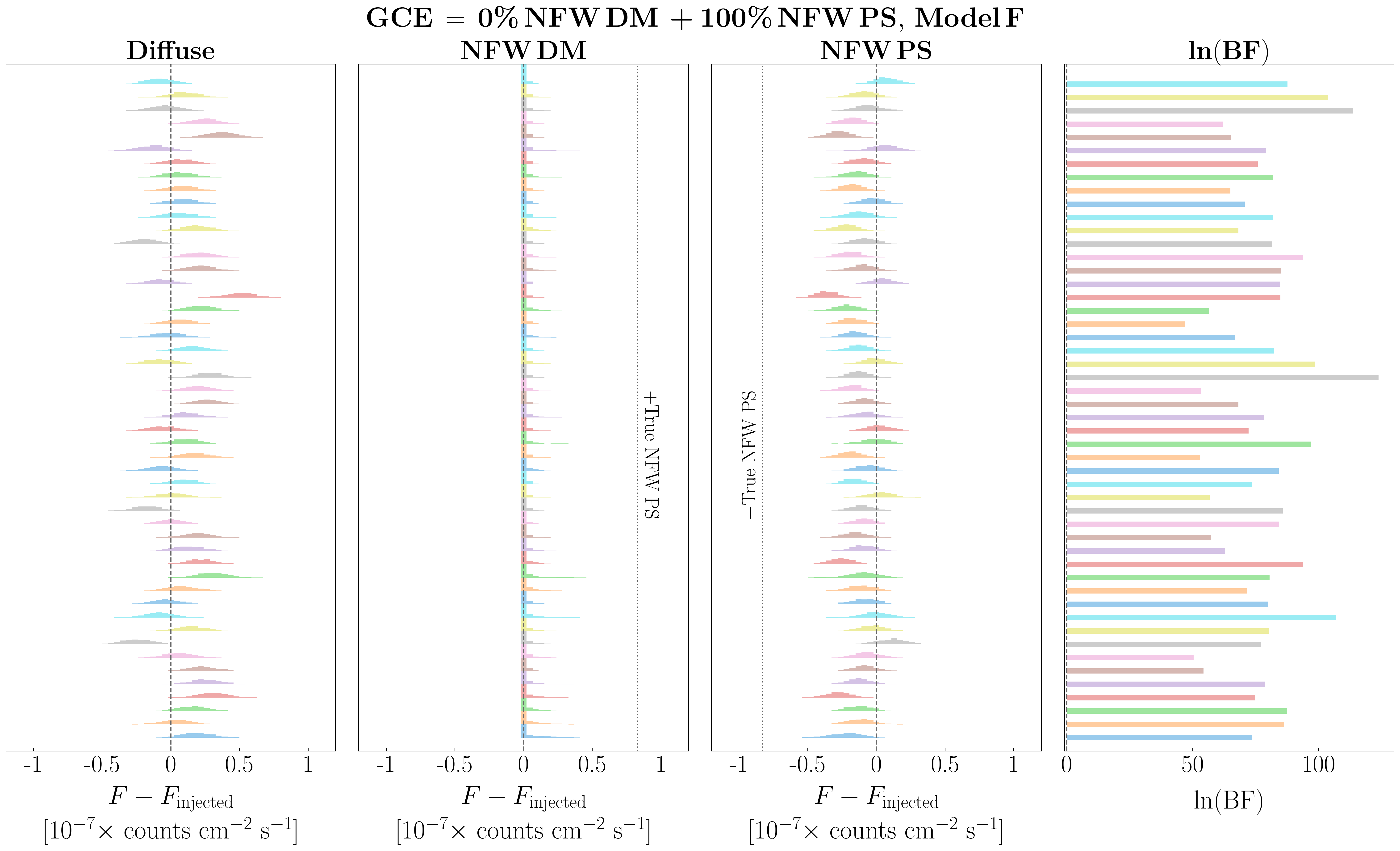}
\caption{Same as Fig.~\ref{fig:0DM_100PS_fplot}, but this time analyzed using Model F as the Galactic diffuse emission model. This corresponds to the case where the diffuse emission is mismodeled. Decisive evidence for a PS population is seen for each realization, even with the diffuse mismodeling. We emphasize that the overall scale of the BFs should not be compared to any results on \emph{Fermi} data as these maps are not intended to closely model an actual data realization.}
\label{fig:0DM_100PS_fplot_modelF}
\end{figure*}

\begin{figure*}[h]
\centering{}
\includegraphics[width=0.9\textwidth]{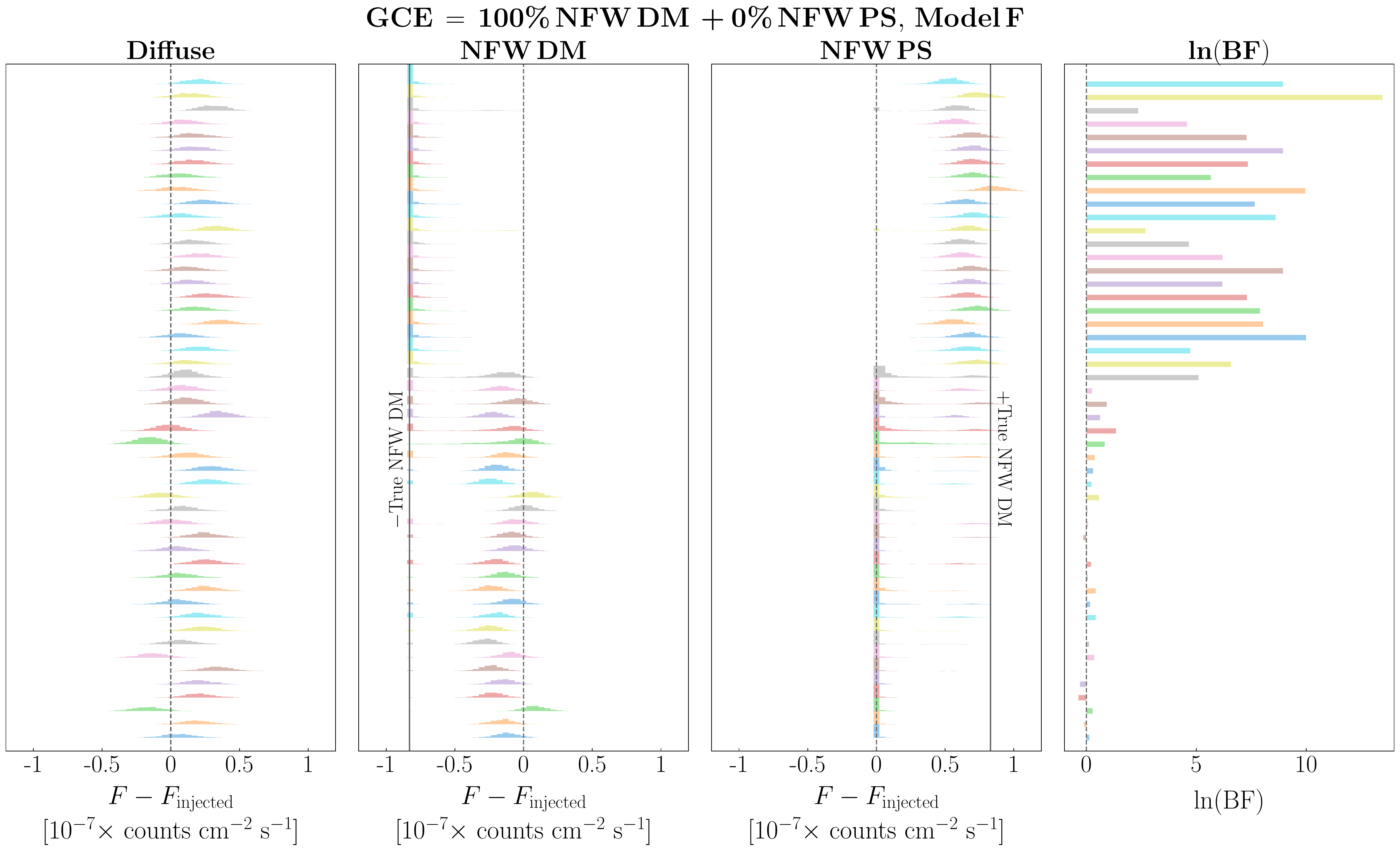}
\caption{Same as Fig.~\ref{fig:100DM_0PS_fplot}, but this time analyzed using Model F as the Galactic diffuse emission model (note the different scale in the rightmost panel from Fig.~\ref{fig:100DM_0PS_fplot}). Misattribution of the DM flux to PSs is significantly exacerbated in the presence of diffuse mismodeling, and evidence for a PS population can be erroneously inferred for a subset of the realizations due to residuals from the mismodeling. However, the misattribution is ameliorated when more DM signal is injected into the mock data, as demonstrated in the second column of Fig.~\ref{fig:signal_on_signal}. We emphasize that while the relative values of the BFs are useful in comparing the different scenarios and realizations studied here, their overall scale should not be compared to any results on \emph{Fermi} data as these maps are not intended to closely model an actual data realization.}
\label{fig:100DM_0PS_fplot_modelF}
\end{figure*}

\begin{figure*}[h]
\centering{}
\includegraphics[width=0.9\textwidth]{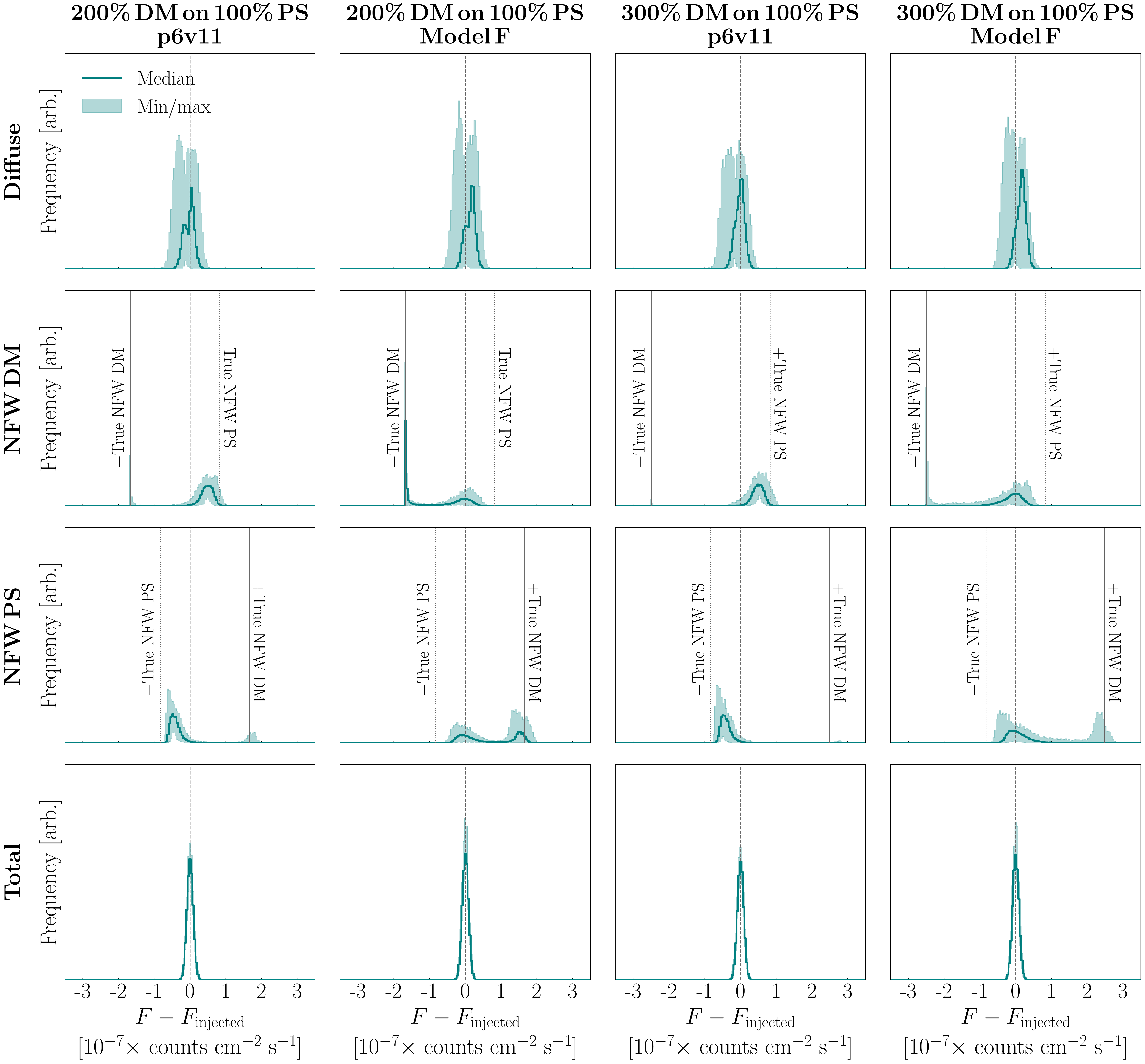}
\caption{Same as the right two columns of Fig.~\ref{fig:signal_on_signal}, but with even brighter DM signals injected on top of an existing GCE-strength soft PS signal. The simulated data maps consist of 10 distinct ``base'' maps in which the GCE is entirely accounted for by soft PSs, onto which an additional 200\% GCE flux (left two columns) or 300\% GCE flux (right two columns) DM signal is injected. The injected DM signal is Poisson fluctuated to generate 10 realizations for each base case. In the absence of diffuse mismodeling (first and third columns), the injection of increasingly bright DM signals reduces the bimodality of the DM and PS posterior flux distributions (see third column of Fig.~\ref{fig:signal_on_signal} for comparison). On average, the injected DM signal is fully recovered, and the DM template additionally picks up the flux contribution from ultrafaint PSs below the $1\sigma$ significance threshold. In the presence of diffuse mismodeling, the injection of brighter and brighter DM signals mitigates the absorption of the DM signal by the PS template. Note the different scale of the horizontal axes in this case compared to Fig.~\ref{fig:signal_on_signal}.}
\label{fig:more_signal_on_signal}
\end{figure*}

\begin{figure*}[t]
\centering{}
\includegraphics[width=1.0\textwidth]{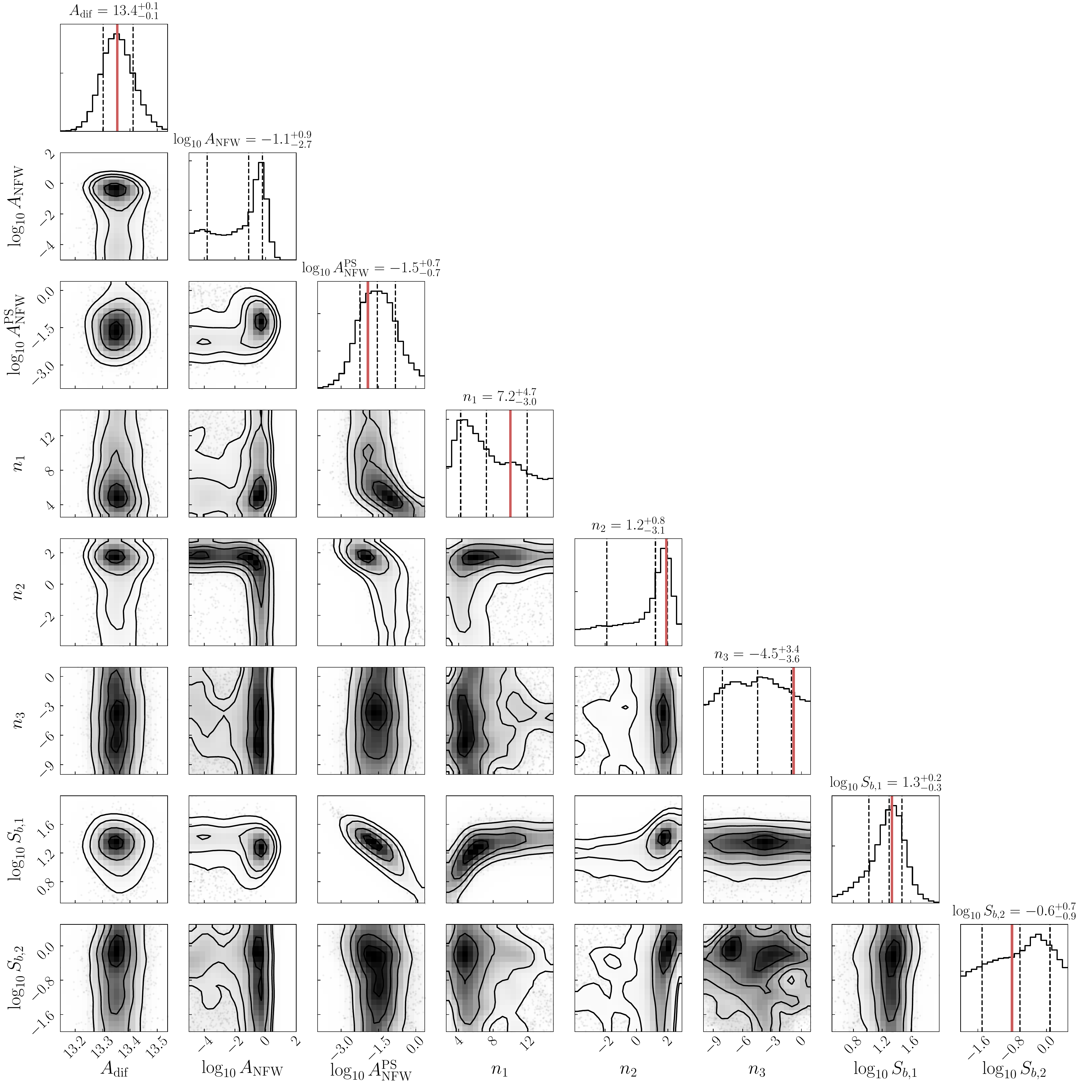}
\caption{An example triangle plot for an NPTF scan that passes the convergence criteria on $S_{b,1}$ described in Section~\ref{sec:NPTF_procedure}: the posterior distribution for $S_{b,1}$ is nicely converged within the prior range. The simulated data in this particular instance consists of 100\% soft PSs and diffuse emission. The templates used in this scan are: \emph{(i)} NFW PS, \emph{(ii)} NFW DM, and \emph{(iii)} Galactic diffuse emission (\texttt{p6v11}). Where applicable, the true simulated values are indicated on the 1d posterior distributions by thick solid red lines.}
\label{fig:corner_pass_PS}
\end{figure*}

\begin{figure*}[t]
\centering{}
\includegraphics[width=1.0\textwidth]{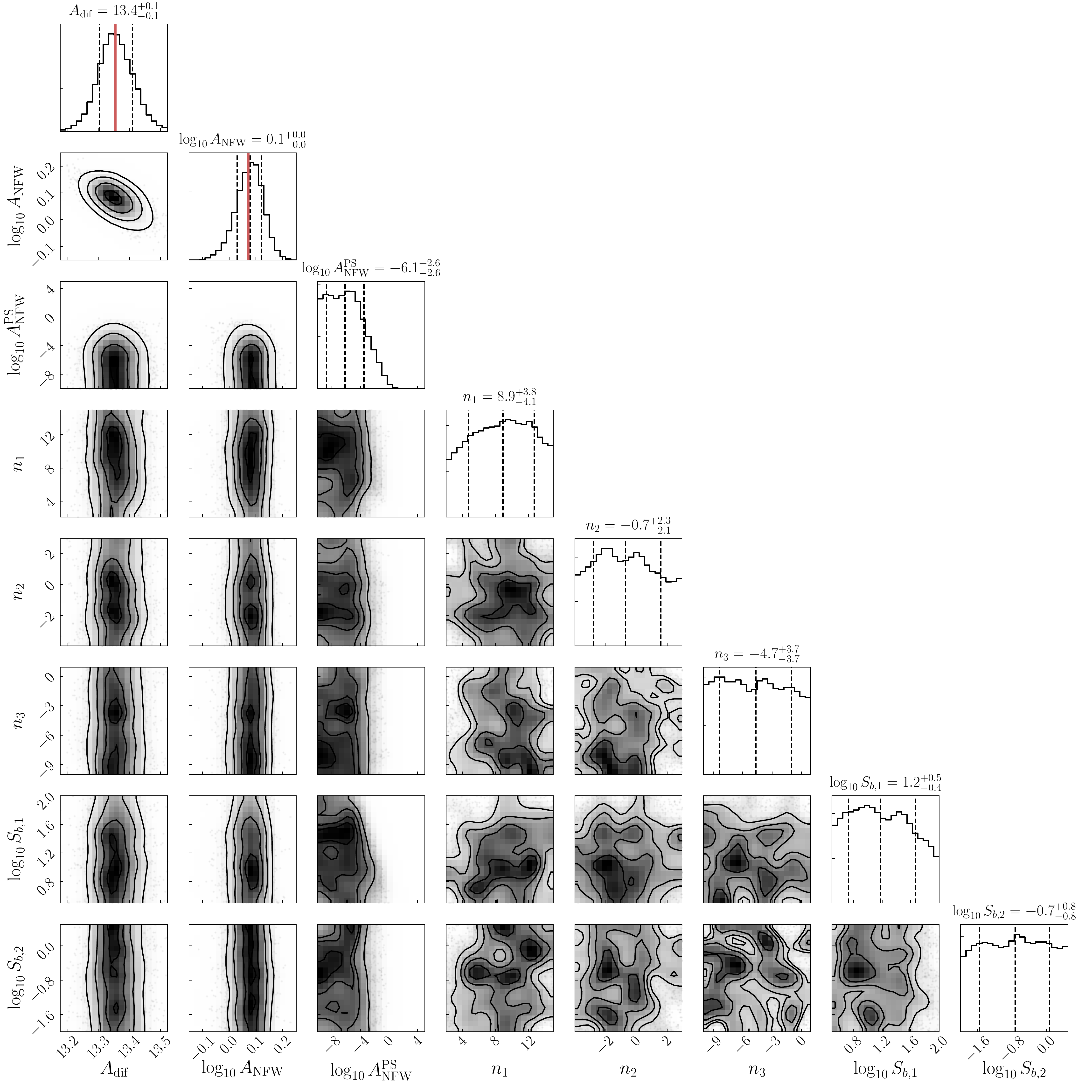}
\caption{Same as Fig.~\ref{fig:corner_pass_PS}, but where the simulated data consists of 100\% DM and diffuse emission. In this case, the posterior distribution for $S_{b,1}$ is unconstrained over the prior range. Where applicable, the true simulated values are indicated on the 1d posterior distributions by thick solid red lines.}
\label{fig:corner_pass_DM}
\end{figure*}

\begin{figure*}[t]
\centering{}
\includegraphics[width=1.0\textwidth]{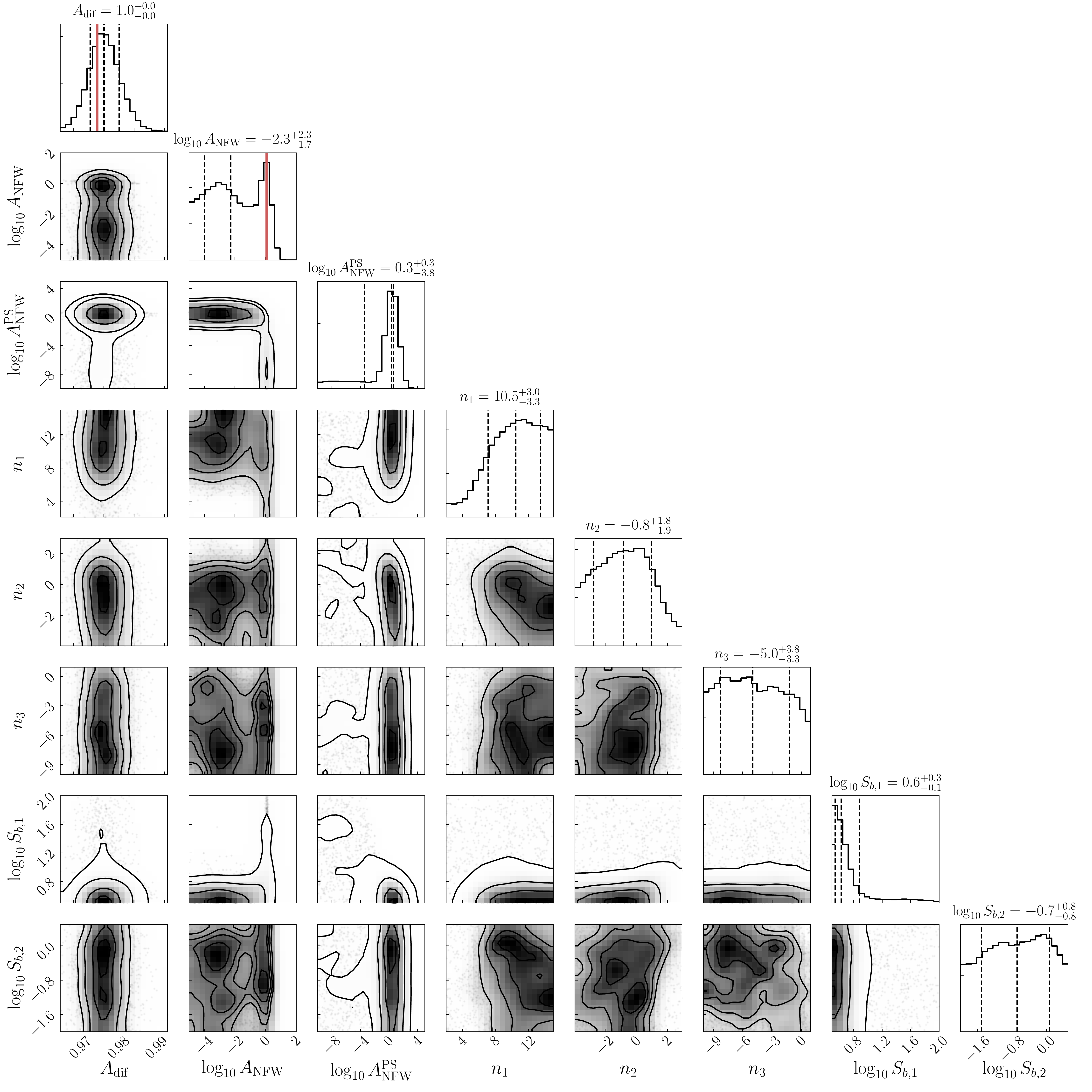}
\caption{An example triangle plot for an NPTF scan that fails the convergence criteria on $S_{b,1}$ described in Section~\ref{sec:NPTF_procedure}: the posterior distribution for $S_{b,1}$ is pushed against the lower prior edge. The simulated data in this particular instance consists of 100\% DM and diffuse emission, and there is diffuse mismodeling present in the scan. The templates used in the scan are: \emph{(i)} NFW PS, \emph{(ii)} NFW DM, and \emph{(iii)} Galactic diffuse emission (Model F). Where applicable, the true simulated values are indicated on the 1d posterior distributions by thick solid red lines---the ``true" diffuse model normalization here corresponds to the norm that yields equivalent flux to the true simulated (\texttt{p6v11}) flux. Scans like this are discarded from all results presented in this paper. Additionally, we note that implementing a lower flux cutoff in the source-count function (detailed in Appendix~\ref{sec:cutoff}) drastically reduces the number of such scans.}
\label{fig:corner_fail}
\end{figure*}

\begin{figure*}[h]
\centering{}
\includegraphics[width=0.9\textwidth]{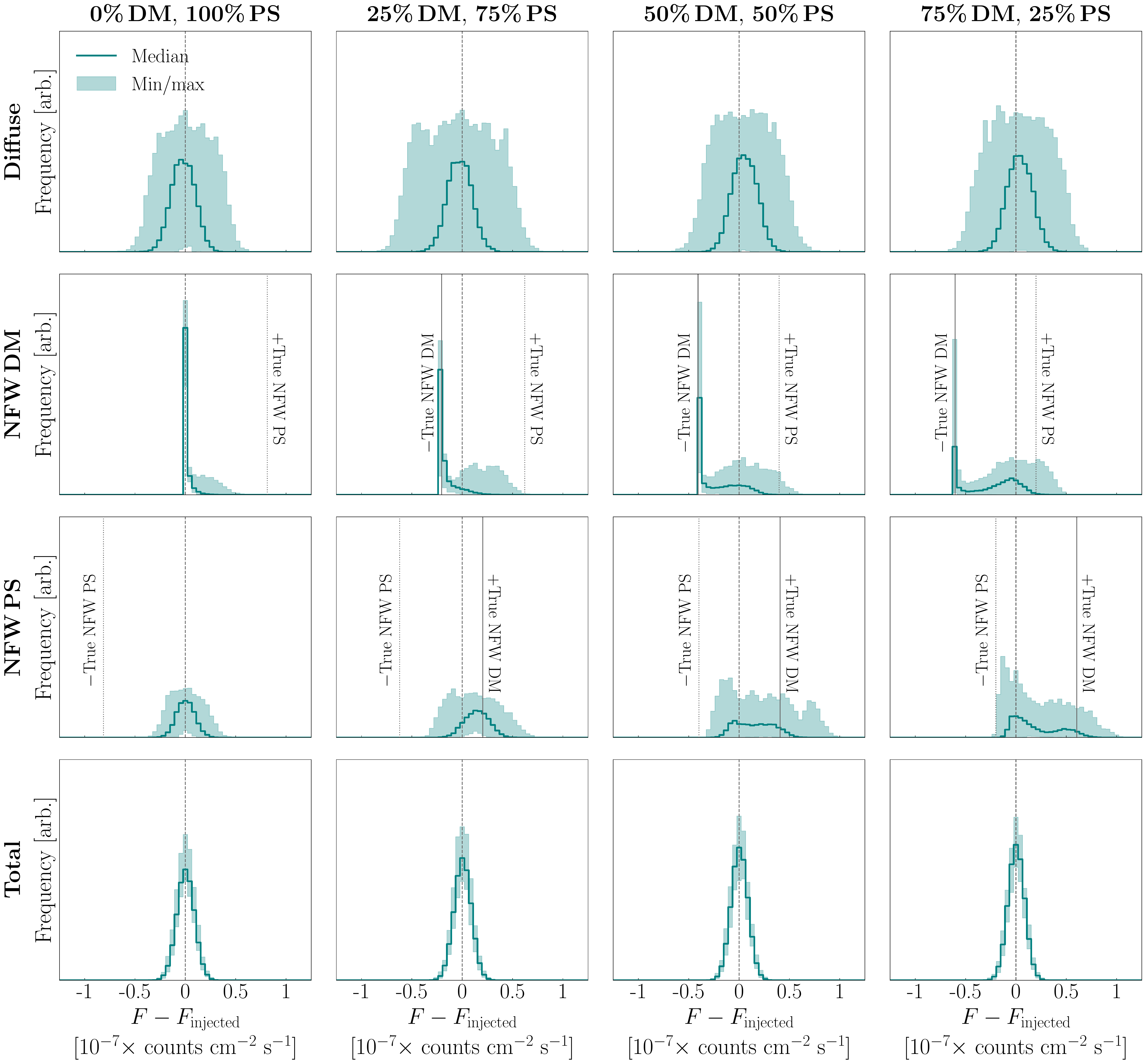}
\caption{Same as Fig.~\ref{fig:25_75_fullpanel}, except for the case of the hard PS source-count distribution. In this case, the PSs never get fully absorbed by the DM template, as is evidenced by the absence of a strong peak at ``$-$True NFW PS'' in the NFW PS posterior flux distributions. However, there is still a probability that the DM signal gets absorbed by the PS template.}
\label{fig:25_75_fullpanel_hardSC}
\end{figure*}

\begin{figure*}[h]
\centering
\includegraphics[width=0.9\textwidth]{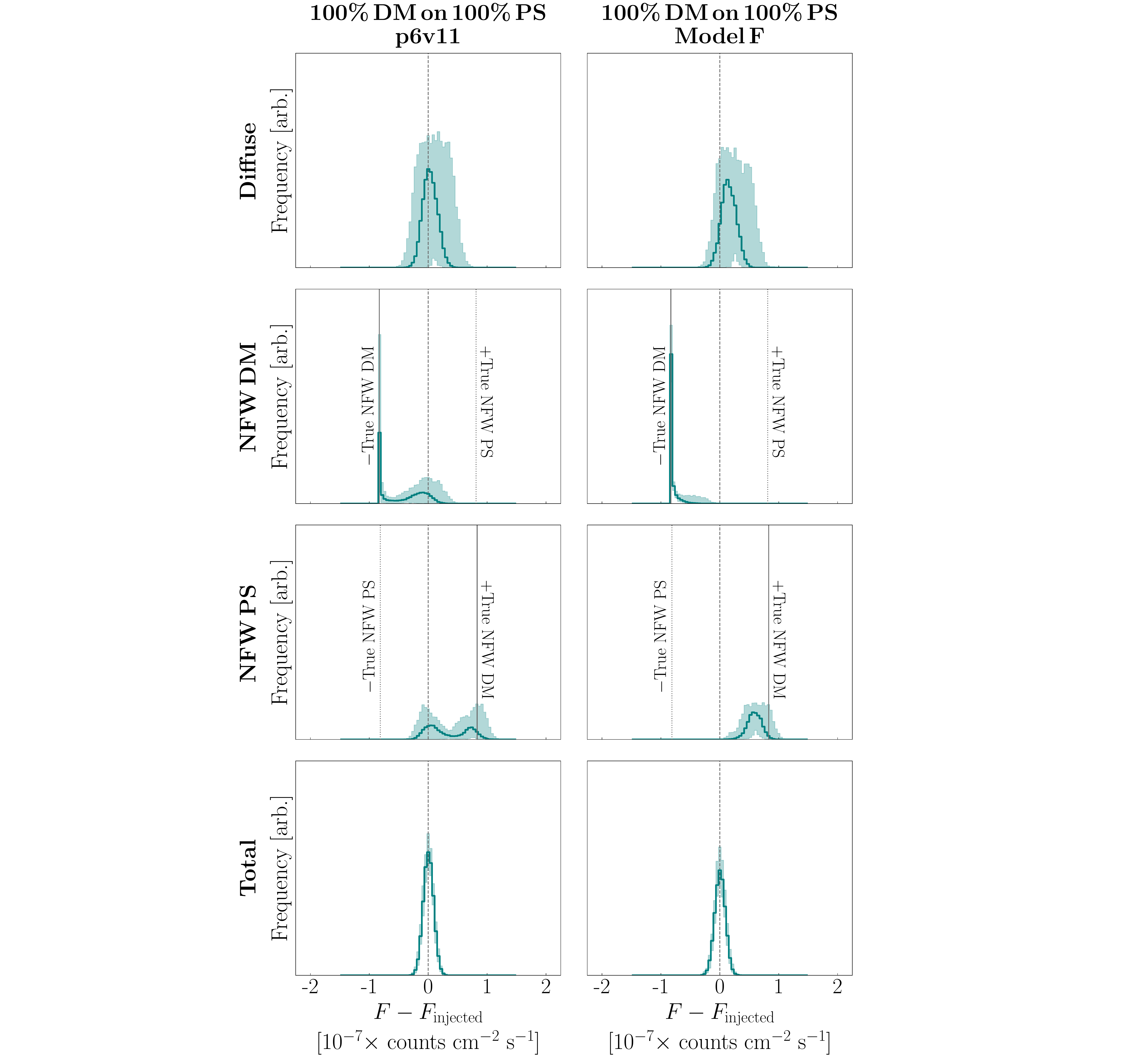} 
\caption{Same as Fig.~\ref{fig:signal_on_signal}, but for the case of the hard PS source-count distribution. Compared to case of the soft source-count distribution, there is less spread in the NFW DM and NFW PS flux posteriors. In particular, the PS flux is never absorbed by the DM template in these cases, while the DM template still may be absorbed by the PS template. The injected DM flux is consistently misatrributed to PSs in the presence of diffuse mismodeling (right column).}
\label{fig:signal_on_signal_hardSC}
\end{figure*}

In this Appendix, we provide a set of supplementary figures that are referenced and described in the main text. In Figs.~\ref{fig:0DM_100PS_fplot}--\ref{fig:corner_fail}, PSs (when relevant) correspond to the soft source-count distribution. Figs.~\ref{fig:25_75_fullpanel_hardSC}--\ref{fig:signal_on_signal_hardSC} show results for the hard source-count distribution.

\vspace{0.75in}

\clearpage

\setcounter{equation}{0}
\setcounter{figure}{0}
\setcounter{table}{0}
\renewcommand{\theequation}{B\arabic{equation}}
\renewcommand{\thefigure}{B\arabic{figure}}
\renewcommand{\thetable}{B\arabic{table}}

\section{Residuals from mismodeling}
\label{sec:residuals}

\begin{figure*}[h]
\centering{}
\includegraphics[width=0.95\textwidth]{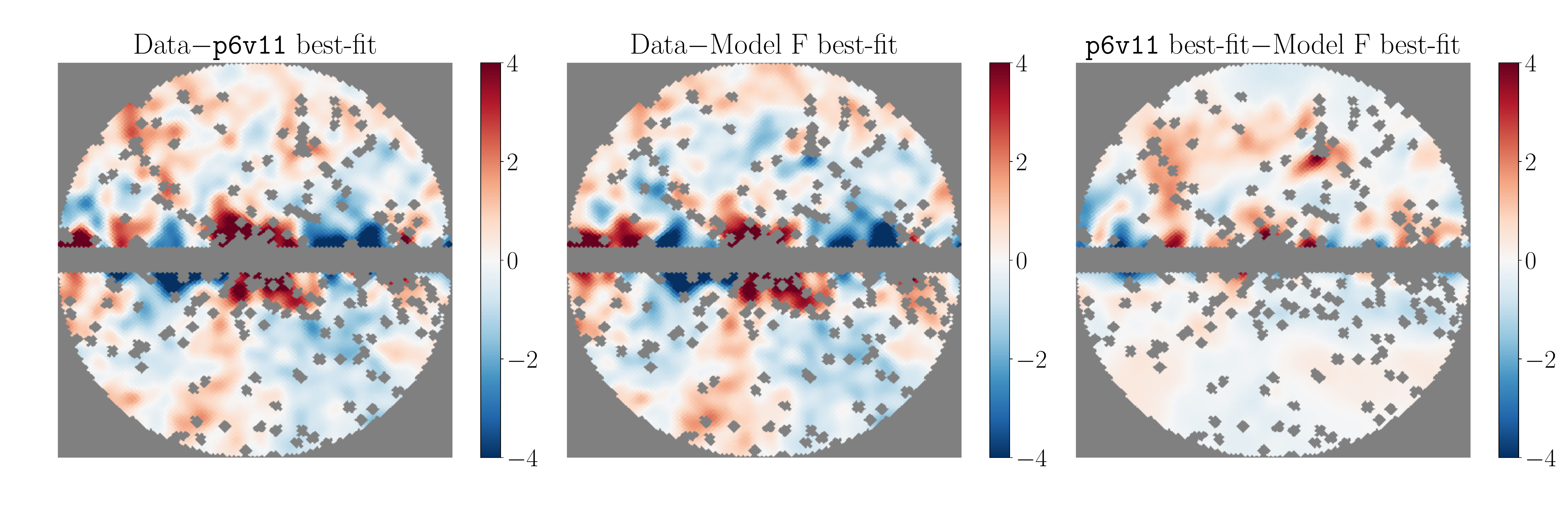}

\caption{Residual photon counts in our ROI from fitting the data with a model that includes \emph{(i)}~the Galactic diffuse emission, \emph{(ii)}~the \emph{Fermi} bubbles~\cite{2010ApJ...724.1044S}, \emph{(iii)}~isotropic emission, and \emph{{(iv)}}~resolved \emph{Fermi} 3FGL PSs~\cite{Acero:2015hja}. We show the residual sky maps for the cases where the Galactic diffuse emission is described by the \texttt{p6v11} model (left panel) or Model F (middle panel). We also show the difference between the best-fit sky maps obtained using each of the two models. For presentation purposes, we have smoothed each map by a Gaussian with $\sigma=1^\circ$.}
\label{fig:residual_maps}
\end{figure*}

This Appendix provides some relevant figures to give a sense of the degree of mismodeling we have considered in Sec.~\ref{sec:mismodeling}. In each case described within this Appendix, the model is purely Poissonian, and consists of \emph{(i)}~the Galactic diffuse emission, \emph{(ii)}~the \emph{Fermi} bubbles, \emph{(iii)}~isotropic emission, and \emph{{(iv)}}~resolved \emph{Fermi} 3FGL PSs. Figure~\ref{fig:residual_maps} shows the photon count residuals from fitting the $\emph{Fermi}$ data to such a model, where the diffuse emission template is either \texttt{p6v11} (left panel) or Model F (middle panel). To illustrate how the \texttt{p6v11} and Model F fits to data differ, we also show difference of the two best-fits (right panel). 

\begin{figure*}[h]
\centering{}
\includegraphics[width=3.5in]{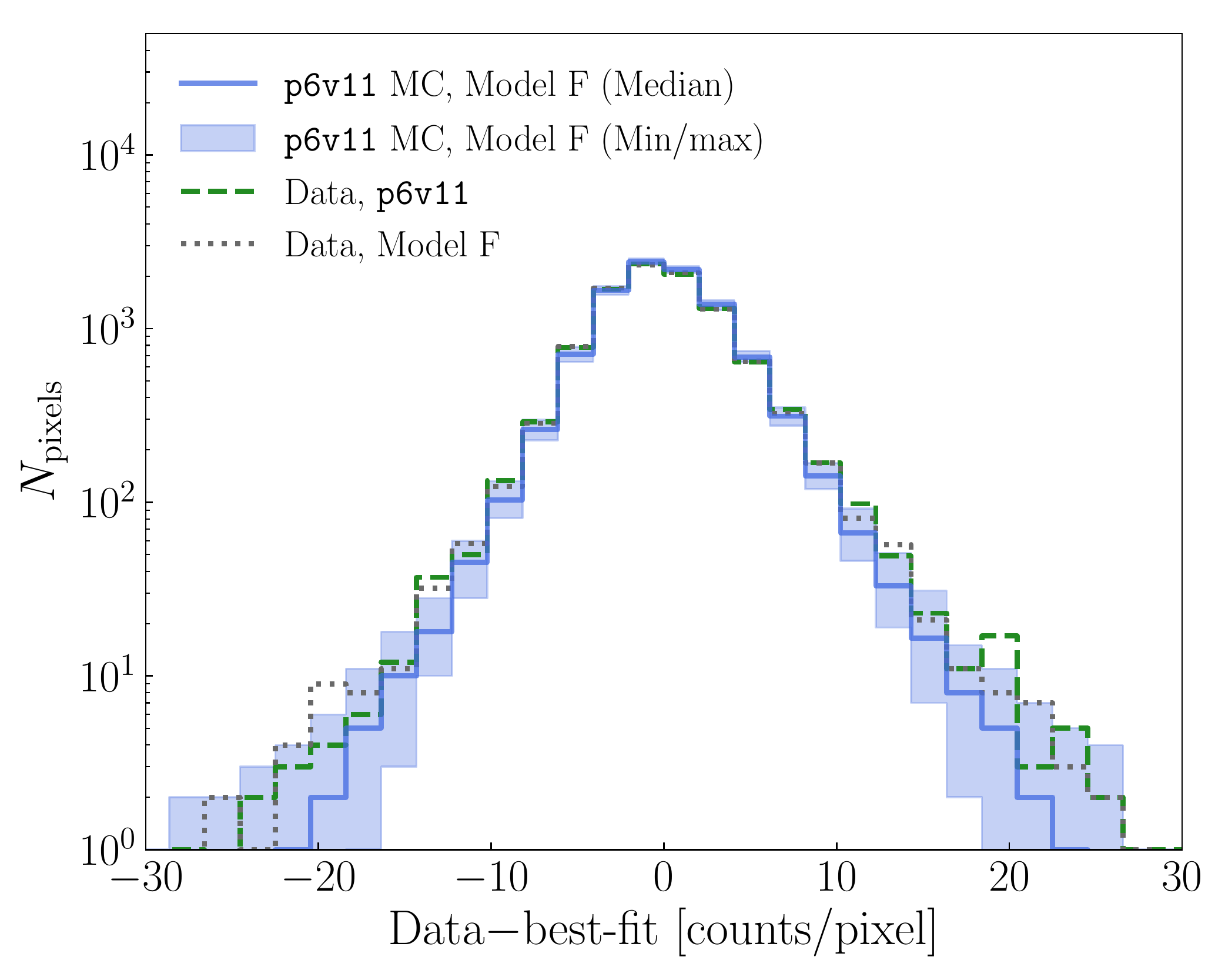}

\caption{Histogram of residual photon counts in our ROI. The green dashed line indicates the case where the $\emph{Fermi}$ data is analyzed using the \texttt{p6v11} diffuse model (this corresponds to the left panel of Fig.~\ref{fig:residual_maps}). The gray dotted line indicates the case where the $\emph{Fermi}$ data is analyzed using Model F (this corresponds to the middle panel of Fig.~\ref{fig:residual_maps}). For comparison, we generate 100 Monte Carlo~(MC) data maps consisting of the following components (best-fit from the $\emph{Fermi}$ data): \emph{(i)}~\texttt{p6v11} Galactic diffuse emission, \emph{(ii)}~the \emph{Fermi} bubbles~\cite{2010ApJ...724.1044S}, \emph{(iii)}~isotropic emission, and \emph{{(iv)}}~resolved \emph{Fermi} 3FGL PSs~\cite{Acero:2015hja}, which we analyze using Model F and Poissonian components  \emph{(ii)}--\emph{(iv)}. The median result is shown by the solid blue line, while the shaded blue bands span the minimum and maximum value in each bin over the 100 simulated maps. The three cases are roughly comparable, although on average, there tend to be fewer residuals with large magnitudes when fitting the \texttt{p6v11} simulated data with Model F than when fitting the data with either model.}
\label{fig:residual_hist}
\end{figure*}

For a more quantitative comparison, Fig.~\ref{fig:residual_hist} shows the histogram of the residual photon counts for the \texttt{p6v11} case (green dashed) and the Model F case (gray dotted). We have also generated 100 simulated data maps using the best-fit \texttt{p6v11} diffuse emission (along with the other Poissonian components), and analyzed the simulated maps using Model F to describe the diffuse emission. The median residual histogram from the 100 Monte Carlo realizations is shown in solid blue, and the shaded blue band spans the minimum/maximum across realizations. On average, the residuals from the latter scenario are somewhat smaller in magnitude than the residuals obtained on the real data. However, taking into account the variation across Monte Carlo realizations, the three are comparable. We therefore conclude that our method for simulating diffuse mismodeling in Sec.~\ref{sec:mismodeling} is a reasonable proxy for the typical degree of mismodeling on the real data.

\clearpage

\setcounter{equation}{0}
\setcounter{figure}{0}
\setcounter{table}{0}
\renewcommand{\theequation}{B\arabic{equation}}
\renewcommand{\thefigure}{B\arabic{figure}}
\renewcommand{\thetable}{B\arabic{table}}

\section{Source-count Function with Flux Cutoff }
\label{sec:cutoff}

\renewcommand{\thefigure}{C\arabic{figure}}
\renewcommand{\theHfigure}{C\arabic{figure}}
\setcounter{figure}{0}

The fact that emission from unresolved PSs is degenerate with Poissonian DM emission in the ultra-faint limit is a critical point for understanding the output of the NPTF analysis.  Here, we consider the scenario where the NPTF only identifies PSs that are bright enough to be distinguished from Poissonian emission, and does not attempt to distinguish fainter sources from DM below some flux cutoff.  In practice, this means that we use the source-count function from Eq.~\ref{eq:sourcecount2break} above some flux cutoff, but set it to zero below.  For illustration, we will consider a cutoff value that corresponds to a $1\sigma$ point-source significance within our setup.

\begin{figure*}[b]
\centering{}
\includegraphics[width=0.65\textwidth]{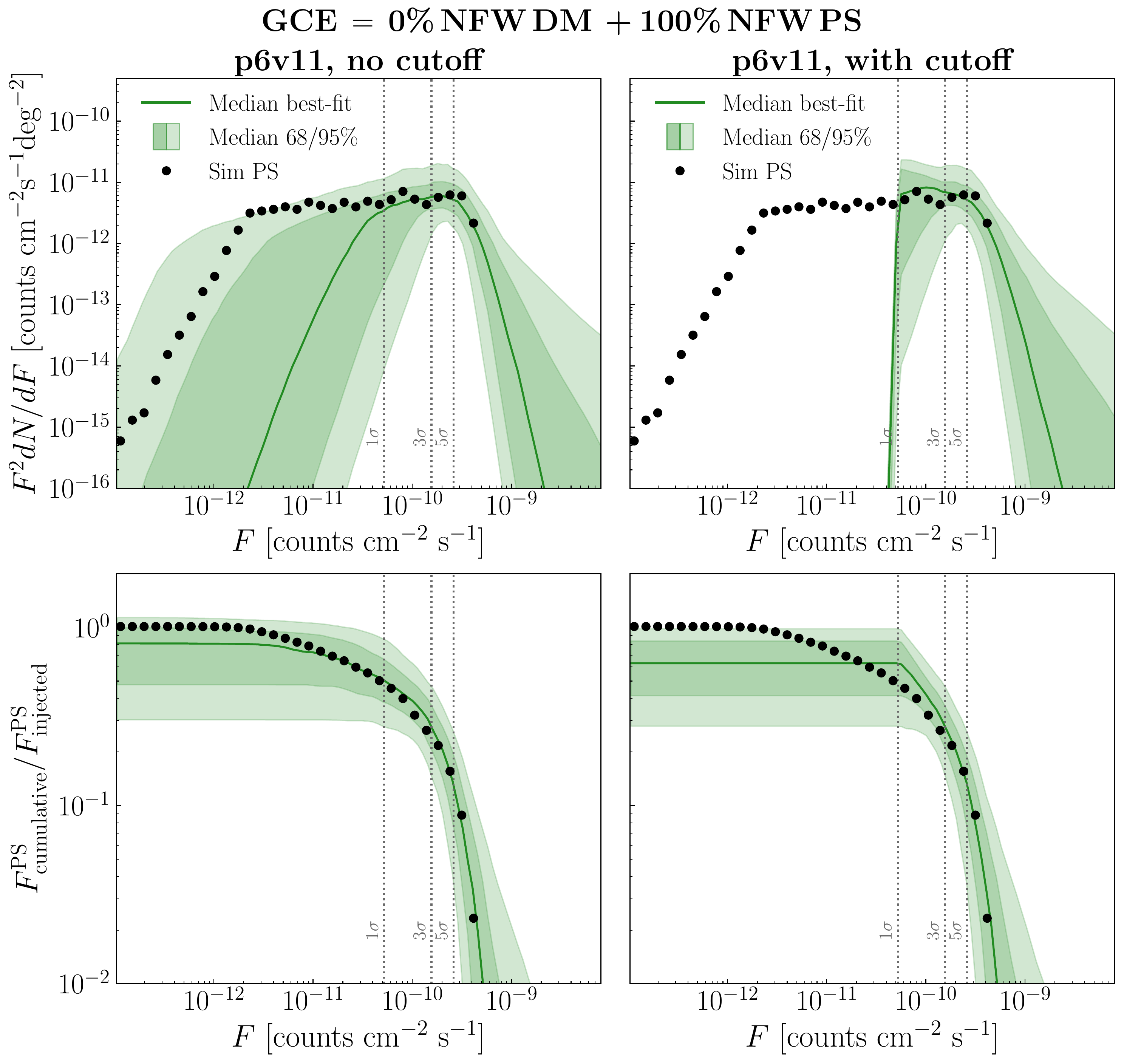}
\caption{Differential source-count distributions (top panels) and cumulative flux distributions, integrated above a given threshold (bottom panels), shown for simulations with PSs with a soft source-count distribution, accounting for 100\% of the GCE (no DM contribution).  The diffuse emission is not mismodeled in this case.  The solid lines indicate the median best-fit distributions recovered over 100 Monte Carlo realizations of the simulated data maps.  The bands show the median 68 and 95\% confidence bands over these 100 iterations.  We show a copy of the right column of  Fig.~\ref{fig:0DM100PS_fullpanel} (left) and the the corresponding result when a flux cutoff is added to the parameterization of the source-count function in the NPTF analysis (right).  The addition of the cutoff does not strongly affect the recovered source-count distribution.  However, a slight discrepancy (at the 68\% level) is introduced between the recovered and true cumulative flux distributions near the cutoff. }
\label{fig:0DM_100PS_cutoff_SCD_p6v11}
\end{figure*}

Figure~\ref{fig:0DM_100PS_cutoff_SCD_p6v11} shows the effect of the cutoff on the recovered source-count and cumulative flux distributions for the case where the simulated data consists of soft PSs and Galactic diffuse emission, modeled assuming \texttt{p6v11}.  The templates used in the model include: \emph{(i)}~NFW PS, \emph{(ii)}~NFW~DM, and \emph{(iii)}~\texttt{p6v11} Galactic diffuse emission (\emph{i.e.}, no mismodeling).  Note that, in this implementation, the NFW~DM template should be picking up both the true DM emission, as well as the flux from unresolved sources that fall below the flux cutoff.  The left column is a copy of  Fig.~\ref{fig:0DM100PS_fullpanel} (right panel), and the right column shows the corresponding result when the cutoff is implemented.  We see that the presence of the cutoff does not greatly affect the best-fit source-count function, though it does lead to a slight overestimate of the flux above the cutoff.  We do however find a significant reduction in the number of iterations where the the upper break ($S_{b,1}$) of the NFW~PS template is peaked at the lower edge of the prior range---cases that we discarded in the main analyses (see Sec.~\ref{sec:NPTF} for a discussion).  This suggests that the cutoff may help to regulate the anomalous cases where the source-count distribution is peaked in the regime where the NPTF loses sensitivity.

\begin{figure*}[t]
\centering{}
\includegraphics[width=0.65\textwidth]{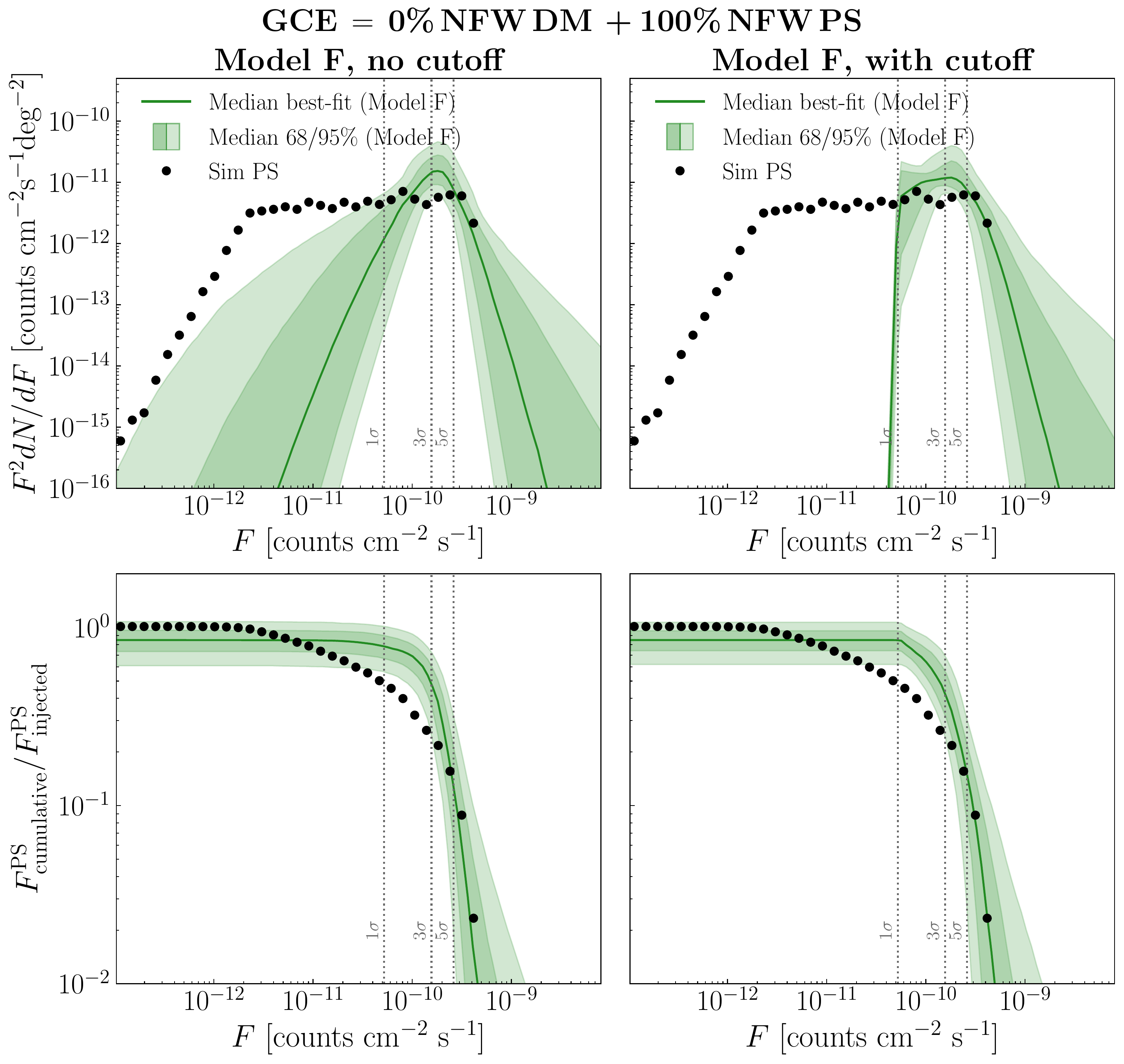}
\caption{Same as Fig.~\ref{fig:0DM_100PS_cutoff_SCD_p6v11}, except for the case where the diffuse emission is mismodeled.  We show a copy of the left column of  Fig.~\ref{fig:100PS_modelF} (left) and the the corresponding result when a flux cutoff is added to the parameterization of the source-count function in the NPTF analysis (right).  Here, we find that the cutoff helps to reduce the peak in the recovered source-count distribution near $\sim 2\times10^{-10}$~counts~cm$^{-2}$~s$^{-1}$.}
\label{fig:0DM_100PS_cutoff_SCD_modelF}
\end{figure*}

Figure~\ref{fig:0DM_100PS_cutoff_SCD_modelF} shows the corresponding result in the presence of diffuse mismodeling.  In this case, the addition of the cutoff in the source-count function regulates the excess above $\sim 10^{-10}$~counts~cm$^{-2}$~s$^{-1}$.  

\begin{figure*}[h]
\centering{}
\includegraphics[width=0.65\textwidth]{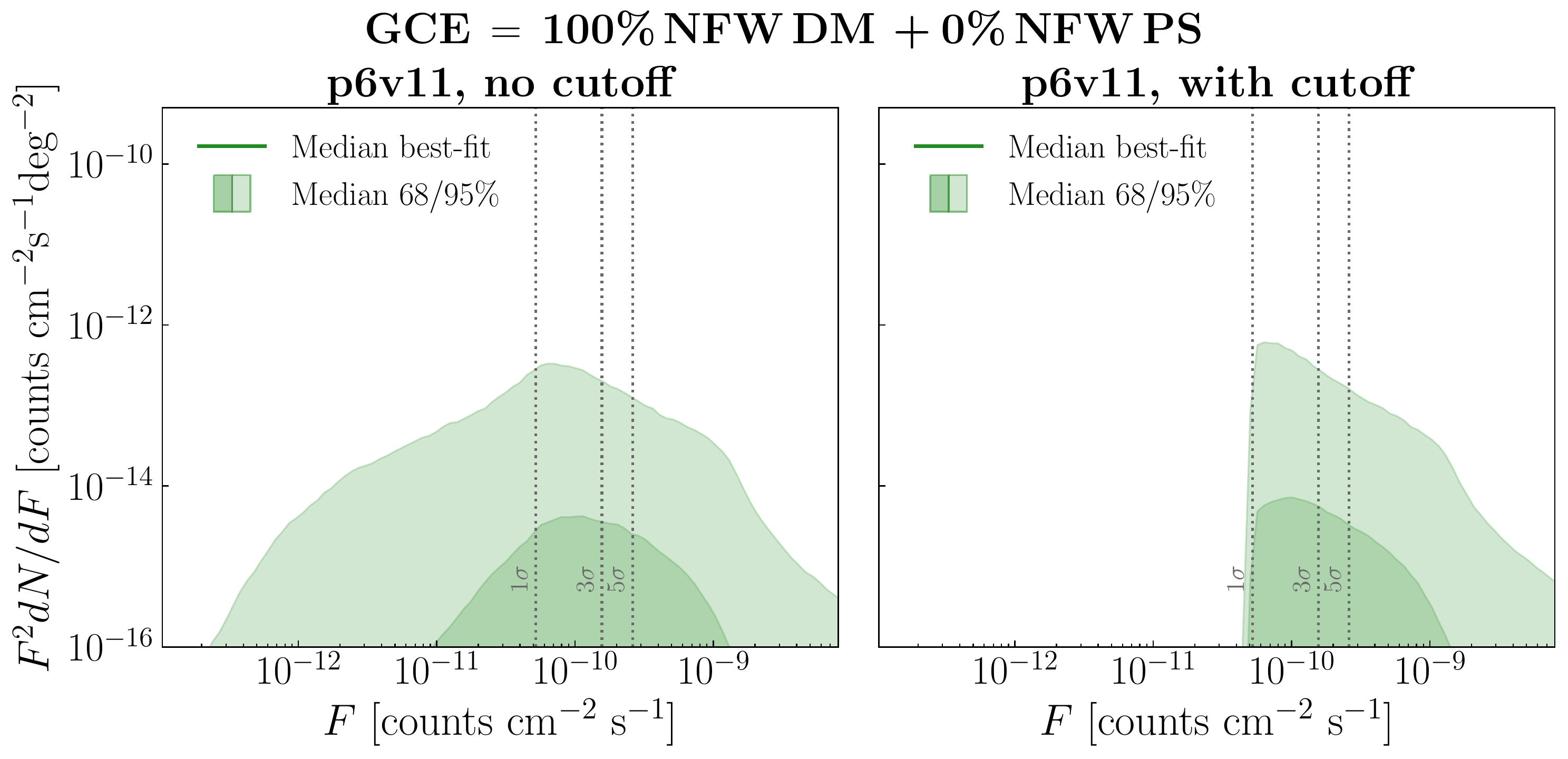}
\caption{Same as Fig.~\ref{fig:0DM_100PS_cutoff_SCD_p6v11}, except for the case where the GCE consists entirely of DM.}
\label{fig:100DM_0PS_cutoff_SCD_p6v11}
\end{figure*}

\begin{figure*}[h]
\centering{}
\includegraphics[width=0.65\textwidth]{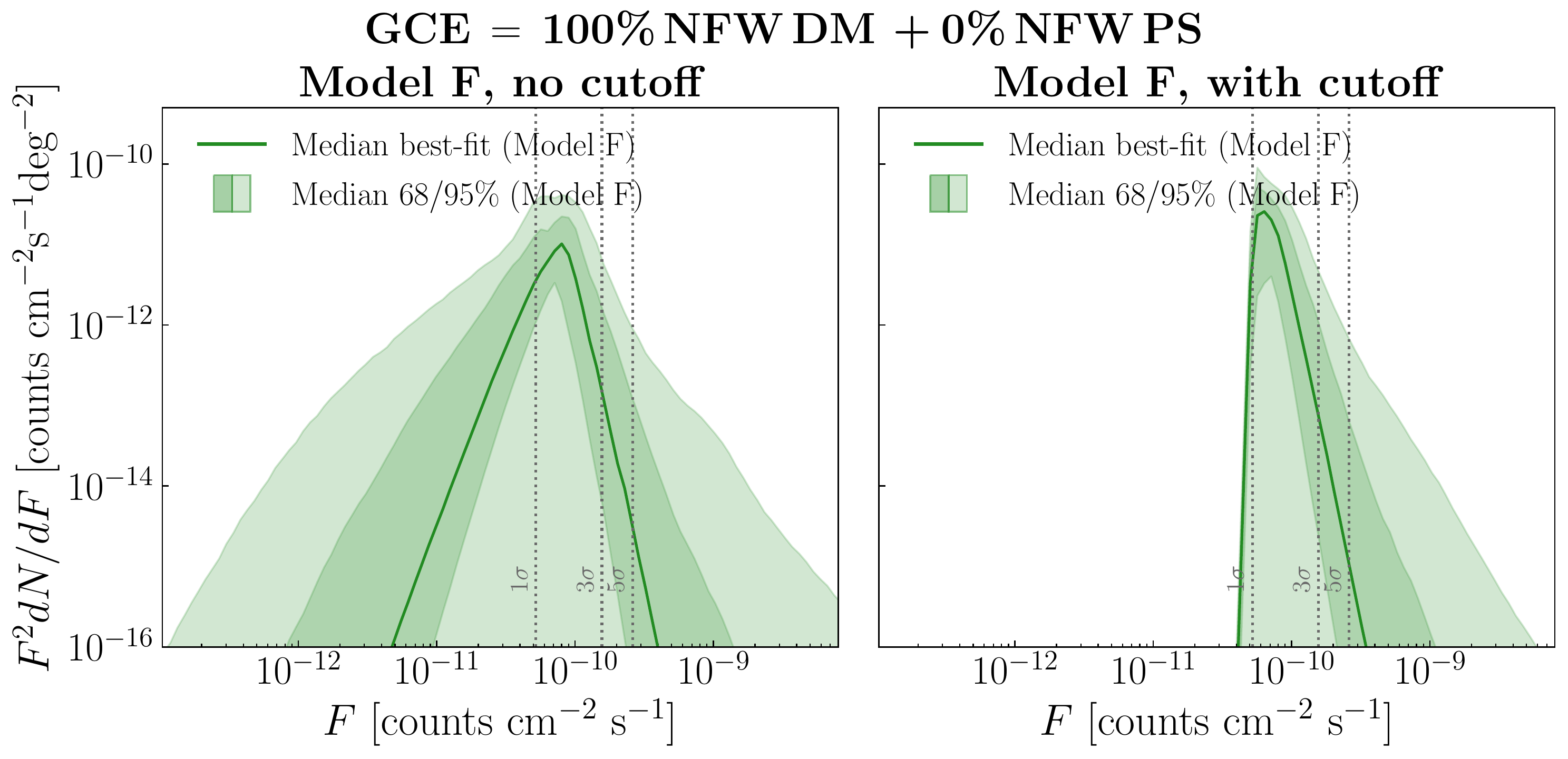}
\caption{Same as Fig.~\ref{fig:0DM_100PS_cutoff_SCD_modelF}, except for the case where the GCE consists entirely of DM.}
\label{fig:100DM_0PS_cutoff_SCD_modelF}
\end{figure*}

\begin{figure*}[h]
\centering{}
\includegraphics[width=0.65\textwidth]{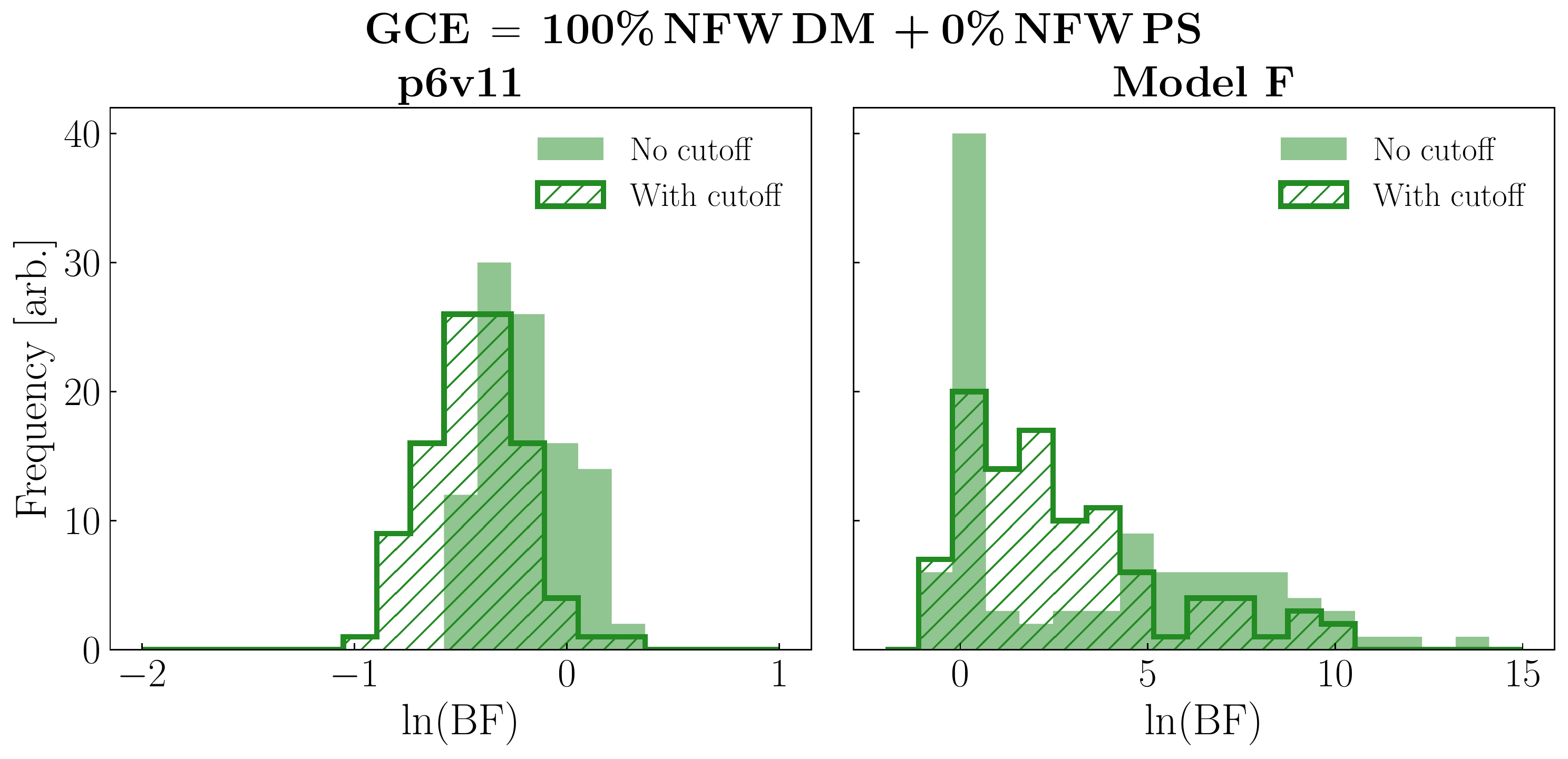}
\caption{The Bayes factors in preference for PSs when the GCE consists entirely of DM in the absence (left panel) or presence (right panel) of diffuse mismodeling.  In each case, solid(hatch)-shaded histogram corresponds to recovered source-count distributions without(with) a flux cutoff. In the \texttt{p6v11} analyses (left panel), the addition of the flux cutoff shifts the distribution of BFs to lower values while leaving the shape of the distribution unchanged. On the other hand, in the Model F analyses (right panel), the cutoff reduces the number of iterations with $\ln(\text{BF})\gtrsim5$ while increasing the number of iterations with $1\lesssim\ln(\text{BF})\lesssim5$.}
\label{fig:BF_100DM_0PS_modelF_cutoff}
\end{figure*}

We have also considered the effect of the flux cutoff when the NPTF is run on simulated data maps where the GCE consists entirely of DM.  Figures~\ref{fig:100DM_0PS_cutoff_SCD_p6v11} and~\ref{fig:100DM_0PS_cutoff_SCD_modelF} show the resulting source-count distributions when, respectively, the \texttt{p6v11} and Model~F templates are used in the analysis.  Again, the addition of the flux cutoff does not seem to have a strong effect on the source-count distribution. However, for the case of diffuse mismodeling, the addition of the cutoff does reduce the number of instances when the DM is misidentified as PSs with $\ln\text{(BF)} > 5$, as shown in Fig.~\ref{fig:BF_100DM_0PS_modelF_cutoff} (right panel). For comparison, we also show in the left panel of Fig.~\ref{fig:BF_100DM_0PS_modelF_cutoff} that the distribution of BFs in the absence of diffuse mismodeling is shifted to lower values and unchanged in shape with the addition of the cutoff. 

\begin{figure*}[h]
\centering{}
\includegraphics[width=0.9\textwidth]{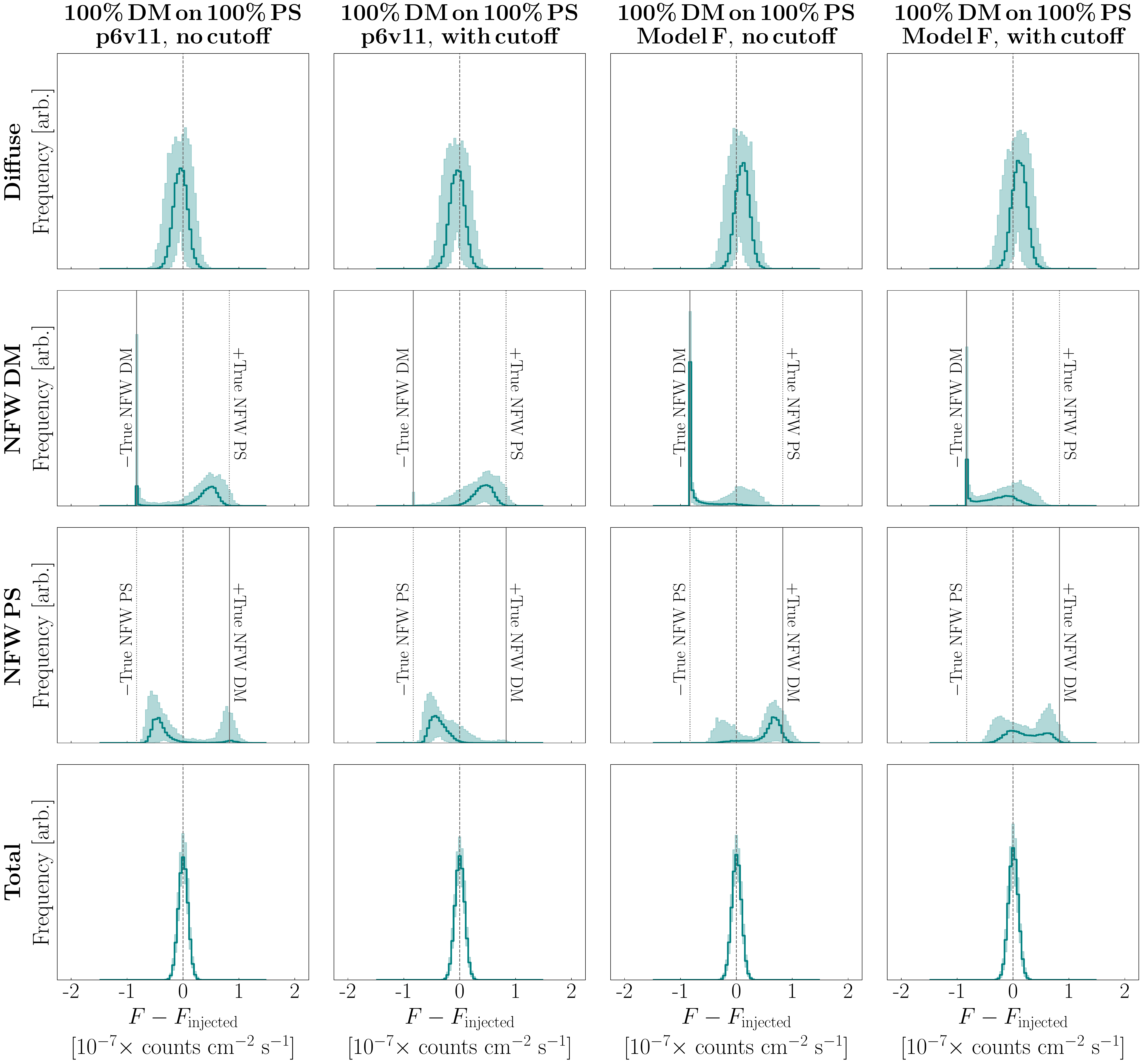}
\caption{Posterior flux distributions for the case where the simulated data is made up of the GCE, comprised by soft PSs, with an additional GCE-strength DM signal injected on top. The first and third columns are duplicates of the third and fourth columns of Fig.~\ref{fig:signal_on_signal}, respectively, and are provided here for comparison. The second and fourth columns show the corresponding results when a flux cutoff is applied to the source-count distribution in the NPTF analysis. In the absence of diffuse mismodeling (first and second column), the cutoff mitigates the bimodality of the posterior flux distributions. In the presence of diffuse mismodeling (third and fourth columns), the cutoff reduces---on average---the probability that the injected DM signal is absorbed by the PS template. This is demonstrated by reduced peak in the median DM posterior at ``$-$True NFW DM" in going from the third column to the fourth column.}
\label{fig:signal_on_signal_cutoff}
\end{figure*}

Lastly, we have studied the effect of the flux cutoff in the cases where the simulated data consists of the GCE, comprised entirely of soft PSs, with an additional GCE-strength DM signal injected on top. Figure~\ref{fig:signal_on_signal_cutoff} shows the resulting posterior flux distributions. We find that in the absence of diffuse mismodeling (left two columns), the cutoff slightly mitigates the bimodality of the flux posteriors, whereas in the presence of diffuse mismodeling (right two columns), the cutoff slightly reduces the probability for the injected DM signal to be absorbed by the PS template. We note that we have additionally tested the effect of the flux cutoff in the cases where the GCE consists of
25\% DM and 75\% PS, 50\% DM and 50\% PS, and 75\% DM and 25\% PS. In those cases, we have found that the cutoff has a negligible effect on the results,
and we therefore omit the corresponding figures. 

\end{document}